\documentclass[final,11pt]{article}
\pdfoutput=1
\usepackage{MRnotes}

\usepackage{comment}
\usepackage{mathtools}
\usepackage{upgreek}
\usepackage{empheq} 
\usepackage[many]{tcolorbox}

\usepackage[
    backend=biber,
    style=alphabetic,
    maxalphanames=10,
    maxnames=10
  ]{biblatex}

\usepackage[compat=1.1.0]{tikz-feynman}

\renewcommand{\arraystretch}{1.6}

 \newtheorem{claim}{Claim}[section]

\crefname{claim}{Claim}{Claims}
\crefname{remark}{Remark}{Remarks}
\crefname{example}{Example}{Examples}

\theoremstyle{definition}
\newtheorem{remark}{Remark}[section]

\definecolor{rust}{rgb}{0.8,0.2,0.2}

\newcommand{\prn}[1]{\left ( #1 \right )}
\newcommand{\prnbig}[1]{\Big( #1  \Big)}
\newcommand{\brk}[1]{\left [ #1 \right ]}
\newcommand{\brkbig}[1]{\Big [ #1  \Big]}
\newcommand{\cbrk}[1]{\left\{ #1 \right \} }

\newtcolorbox{coloremph}{colback=blue!5!white,
  boxrule=0pt,
  boxsep=1pt,
  left=2pt,right=2pt,top=2pt,bottom=2pt,
  oversize=2pt,
  sharp corners,
  before skip=\topsep,
  after skip=\topsep,}

\newcommand{\CFT}[1]{CFT$_{#1}$}

\renewcommand{\Re}{\mathfrak{R}}
\renewcommand{\Im}{\mathfrak{I}}

\newcommand{\bk}{\vb{k}}  
\newcommand{\Dz}{\mathbb{D}}  
\newcommand{\MZ}{\mathscr{Z}}
\newcommand{\Lk}{\Lambda_k}
\newcommand{\Gret}{K}
\newcommand{\holonorm}{\mathfrak{f}}
\newcommand{\markov}{\mathscr{M}}

\newcommand{\Tcl}{\mathtt{T}}
\newcommand{\wf}{\psi}
\newcommand{\xr}{x}
\newcommand{\acc}{\mathcal{E}}
\newcommand{\mexp}{\theta}
\newcommand{\cmexp}{\sigma}
\newcommand{\hcl}{\delta}
\newcommand{\intz}{\mathbb{Z}}
\newcommand{\sign}{\epsilon}
\newcommand{\signconv}{\kappa}
\newcommand{\connm}{\mathsf{C}}
\newcommand{\VB}{\mathcal{F}}
\newcommand{\drep}[2]{ \expval{ #1 , #2 } }
\newcommand{\ldiff}{\mathcal{L}}
\newcommand{\Fus}{\mathsf{F}}
\newcommand{\Fuscl}{\mathsf{F}^{\mathrm{cl}}}
\newcommand{\Fusclval}[5]{\frac{\Gamma\prn{1-2\, #1 \,#3} \Gamma\prn{2\, #2 \,#4}}{\Gamma\prn{\half - #1 \,#3 + #2 \,#4 \pm #5}}}

\newcommand{\CL}{\textup{S.C.}}
\newcommand{\VBcl}{\mathcal{W}}
\newcommand{\wfnorm}{\mathcal{N}}
\newcommand{\cmexpope}{\mathfrak{s}}

\newcommand{\xrp}{{x^\prime}}
\newcommand{\cmexpp}{\sigma^\prime}

\newcommand{\rp}{r_{+}}
\renewcommand{\rm}{r_{-}}
\newcommand{\rpm}{r_{\pm}}
\newcommand{\rcomplex}{r_{\mathrm{c}}}
\newcommand{\ri}{r_{i}}
\newcommand{\Qext}{Q_{\mathrm{ext}}}

\newcommand{\hor}{\mathrm{hor}}
\newcommand{\curv}{\mathrm{curv}}
\newcommand{\bdy}{\mathrm{bdy}}
\renewcommand{\real}{\mathbb{R}}
\newcommand{\freq}{\mathfrak{w}}
\newcommand{\mom}{\mathfrak{q}}

\newcommand{\ingo}{\mathrm{in}}
\newcommand{\outgo}{\mathrm{out}}
\newcommand{\nor}{\mathrm{nor}}
\newcommand{\nnor}{\mathrm{nnor}}

\newcommand{\nhalf}{\mathfrak{n}}

\newcommand{\prepot}{\mathscr{F}} 
\newcommand{\swd}{\lambda} 
\newcommand{\SW}{\textup{SW}} 
\newcommand{\Vir}{\textup{Vir}}

\newcommand{\xrapp}{\mathfrak{x}}
\newcommand{\asnum}{s}
\newcommand{\asc}{\mathsf{ASC}}
\newcommand{\scpara}{\Theta}

\newcommand{\dlogz}{\mathtt{d}_z}
\newcommand{\DDz}{\mathfrak{D}_z}
\newcommand{\frob}{\Psi}
\newcommand{\rec}{\mathtt{R}}
\newcommand{\chexp}{\lambda}
\newcommand{\chexpp}{\chexp_{+}}
\newcommand{\chexpm}{\chexp_{-}}
\newcommand{\chexppm}{\chexp_{\pm}}
\newcommand{\chexppn}{\chexp^{(n)}_{+}}
\newcommand{\chexpmn}{\chexp^{(n)}_{-}}
\newcommand{\chexppmn}{\chexp^{(n)}_{\pm}}
\newcommand{\chexpcrit}{\chexp_{*}}
\newcommand{\froblog}{\Phi}
\newcommand{\nlog}{\mathrm{nlog}}

\newcommand{\falloffc}{\mathcal{C}}
\newcommand{\coeff}{\mathrm{coeff}}
\newcommand{\zx}{\mathtt{z}_{_x}}

\newcommand{\pf}{\VB}
\newcommand{\cl}{\textup{cl}}
\newcommand{\cb}{\mathcal{W}}
\newcommand{\lop}{\mathfrak{l}}

\tikzset{bsty/.style={rectangle, inner sep=0pt, minimum height=0pt, minimum width=4pt,draw}}
\tikzset{tsty/.style={circle, fill, inner sep=0pt,minimum size=3pt,draw}}
\tikzset{disc/.style={snake it}}
\tikzset{conn/.style={dashed}}

\title{ 
Holographic thermal correlators and quasinormal modes from semiclassical Virasoro blocks  
}
\author{Hewei Frederic Jia, Mukund Rangamani}

\affiliation{Center for Quantum Mathematics and Physics (QMAP)\\
Department of Physics \& Astronomy, University of California, Davis, CA 95616 USA}

\emailAdd{fjia@ucdavis.edu}
\emailAdd{mukund@physics.ucdavis.edu}

\abstract{Motivated by its relevance for  thermal correlators in strongly coupled holographic CFTs, we refine and further develop a recent exact analytic approach to black hole perturbation problem, based on the semiclassical Virasoro blocks, or equivalently via AGT relation, the Nekrasov partition functions in the Nekrasov-Shatashvili limit. Focusing on asymptotically \AdS{5} black hole backgrounds, we derive new universal exact expressions for holographic thermal two-point functions, both for scalar operators and conserved currents. Relatedly, we also obtain exact quantization conditions of the associated quasinormal modes (QNMs). Our expressions for the holographic \CFT{4} closely resemble the well-known results for 2d thermal CFTs on $\mathbb{R}^{1,1}$. 
This structural similarity stems from the locality of fusion transformation for Virasoro blocks. We provide numerical checks of our quantization conditions for QNMs. Additionally, we discuss the application of our results to understand specific physical properties of QNMs, including their near-extremal and asymptotic limits. The latter is related to a certain large-momentum regime of semiclassical Virasoro blocks dual to Seiberg-Witten prepotentials. }

\newpage

\begin{document}
\maketitle


\section{Introduction}\label{sec:intro}

Recently, an exact analytic approach to black hole perturbations has been developed, based on the semiclassical Virasoro block. Equivalently, one can view this as an application of techniques used to analyze 4d supersymmetric field theories, in particular the  Nekrasov partition function (in the so called Nekrasov-Shatashvili limit)  to the black hole context. Our aim here is to exploit these developments to achieve the following:
\begin{itemize}[wide,left=0pt]
\item  First, to provide some further insights into the  connection, refining the general method, and extending it to incorporate some physically motivated generalizations. 
\item Second, to utilize this framework to analyze the matter and graviton fluctuations around asymptotically AdS black hole backgrounds. We are motivated here by the AdS/CFT correspondence, and, as we shall demonstrate, these techniques can be applied to study thermal correlators of generic operators and conserved currents of strongly coupled CFTs. 
\end{itemize}

In the rest of the introduction, we elaborate on the physical motivations and technical background, and summarize the results we obtain. 

\subsection{Background}

\paragraph{Quasinormal modes and holographic real-time thermal correlators:} Perturbation around black hole spacetime, due to the existence of horizon, features characteristic damped oscillations known as quasinormal modes (QNMs). The study of QNMs is an important subject of long history due to their relevance for distinct branches of physics, ranging from gravitational wave astronomy to gauge/gravity duality. We refer to the comprehensive reviews~\cite{Kokkotas:1999bd,Berti:2009kk,Konoplya:2011qq} for more background; see also~\cite{Hatsuda:2021gtn} for a review specialized to the exact WKB method. 

In the gauge/gravity context where black holes are generically dual to thermal field theory (at least above the deconfinement transition in compact volume), QNMs are related to the relaxation to thermal equilibrium~\cite{Horowitz:1999jd}. The retarded thermal two-point function, is meromorphic, and has poles at QNMs. This observable can be extracted from the wave equations associated with black hole perturbation using the prescription of~\cite{Son:2002sd}. In fact, up to an overall constant, the two-sided thermal two-point function is entirely fixed by QNMs~\cite{Dodelson:2023vrw}. 

Recently, a systematic approach for computing real-time holographic thermal correlators has since been developed. This not only justifies the prescription of~\cite{Son:2002sd}, which is aimed at  thermal two-point functions, but also provides a framework for computing arbitrary higher-point correlators. 
In general, real-time correlators in QFT should be computed using the Schwinger-Keldysh formalism. In the holographic context, the Schwinger-Keldysh contour is represented as a complexified two-sheeted geometry introduced in~\cite{Glorioso:2018mmw} (for important prior  work, see~\cite{Skenderis:2008dg,Skenderis:2008dh}). This prescription was analyzed and refined further in~\cite{Chakrabarty:2019aeu,Jana:2020vyx,Loganayagam:2022zmq,Loganayagam:2022teq,Loganayagam:2024mnj}. Specifically, these works argued for the efficacy of the contour integral prescription demonstrating that one could compute correlation functions using Witten diagrams on a complexified spacetime, dubbed the \emph{grSK geometry}. (for related work see 
~\cite{Loganayagam:2020iol,Loganayagam:2020eue,Chakrabarty:2020ohe}). The upshot of this analysis is that a general correlator is obtained as a single integral over the exterior region of the black hole. The integrand is a multiple discontinuity of a non-analytic function, obtained by suitable convolution of a set of Green's functions in the grSK geometry. Employing this new formulation, it is demonstrated in~\cite{Jana:2020vyx,Loganayagam:2022zmq} that, higher-point thermal correlators are also meromorphic -- they have poles at QNMs or anti-QNMs.\footnote{The anti-QNMs are the poles of the advanced Green's function, and are complex conjugates of the QNMs. We also note that higher-point functions could additionally have poles at (kinematic) Matsubara frequencies.}  In this work, we will primarily focus on thermal two-point functions, which are amenable to analysis via the prescription of~\cite{Son:2002sd}, though we shall comment briefly on higher-point functions.

\paragraph{Holographic open quantum systems:} Apart from their intrinsic interest, a secondary motivation for analyzing holographic thermal correlators is to further our understanding of effective field theories of open quantum dynamics. 

The study of open quantum systems, viz., systems coupled to an environment, has wide-ranging relevance in many physical problems. The key question here is to integrate out the environmental degrees of freedom, and to derive the (non-unitary) effective dynamics of the open system. The basic paradigm for analyzing such systems was explained in~\cite{Feynman:1963fq}, and has been extensively applied in 1d quantum mechanical systems  (see~\cite{Breuer:2002pc} for an overview). The analysis in the quantum field theoretic context, however, has been less developed, see~\cite{Avinash:2017asn}.

In general, thermal correlators form the basic data for constructing the open EFT of a probe coupled to a thermal environment. As thermal correlators in generic QFT are difficult to compute, holographic environment, whose thermal correlators admit dual gravitational descriptions, provides a valuable avenue for explicitly determining the open EFT~\cite{Jana:2020vyx,Loganayagam:2022zmq}. More concretely, the idea is to use a strongly coupled holographic field theory as a thermal environment, and probe it with a quantum system. Integrating out the holographic degrees of freedom, one obtains the open effective dynamics of the probe quantum system. An exact description of holographic thermal correlators will therefore be a useful tool in understanding such open quantum systems. 

\paragraph{Black hole wave equations in AdS and Heun-type opers:} Despite their physical significance and long history of study, QNMs and the associated thermal two-point functions have long resisted exact analytic understanding. In a way that aligns with our subsequent discussions, the reason can be explained as follows. The differential operator governing the (complexified) radial wave equations, for a mode with definite frequency $\omega$ and momentum $k$, on black hole background can generally be recast to certain $SL(2,\mathbb{C})$ opers\footnote{The notion of opers refers in general to a class of connections, which may also be identified with differential operators. We will use the terminology synonymously with the latter interpretation. } on $\mathbb{P}^1$, viz.,
\begin{equation}
    \partial^2_z + \Tcl(z|\omega,k)
\end{equation}
with meromorphic $\Tcl(z)dz^2$. The poles/punctures in the quadratic differential corresponds to special locations in black hole geometry, in particular including asymptotic boundary and horizon. 

An immediate consequence of the definition of QNMs and the prescription in~\cite{Son:2002sd} is that, solving QNMs and thermal two-point functions amounts to understanding certain global property of such oper, viz., the \emph{connection problem} between asymptotic boundary and horizon. This turns out to be rather non-trivial beyond the well-understood three-puncture hypergeometric oper: standard exact results on QNMs, in particular the case of BTZ black hole in \AdS{3}~\cite{Birmingham:2001pj}, indeed all correspond to opers of this type. 

We will focus on asymptotically AdS black hole backgrounds, where the associated oper generically has regular punctures~\cite{Loganayagam:2022teq}, viz., $\Tcl(z)dz^2$ with double poles. In more elementary term, differential equations associated with such opers are referred to as Fuchsian equations. Other asymptotics will generically involve opers with irregular punctures, viz., $\Tcl(z)dz^2$ with poles beyond second order. We refer to opers with more than three regular punctures as Heun-type opers.

\paragraph{Heun-type opers and semiclassical Virasoro blocks:}The new feature of Heun-type opers compared to the hypergeometric opers is the presence of \emph{accessory parameters}. It has been known since~\cite{Zamolodchikov:1986two} that accessory parameters in Heun-type opers are directly related to semiclassical Virasoro blocks; see also~\cite{Litvinov:2013sxa}. There exists a long list of applications of this relation to various topics in theoretical and mathematical physics, ranging from \AdS{3}/CFT$_2$~\cite{Hartman:2013mia,Fitzpatrick:2014vua} to isomonodromic deformation~\cite{Teschner:2017rve}.

A recent breakthrough was made in~\cite{Bonelli:2022ten} to utilize, among other things, this relation to solve the connection coefficients of Heun-type opers in terms of semiclassical Virasoro blocks.\footnote{Connection coefficients in presence of irregular punctures are also derived in~\cite{Bonelli:2022ten}.} The original derivation in~\cite{Bonelli:2022ten} involves using crossing symmetry of Liouville correlators and the DOZZ formula. The resulting expression involves a degenerate fusion matrix and the semiclassical Virasoro block. Subsequently, it was clarified in~\cite{Consoli:2022eey,Lisovyy:2022flm} that the result of~\cite{Bonelli:2022ten} directly follows from examining the fusion transformation of degenerate Virasoro block in the semiclassical limit. 

\paragraph{Semiclassical Virasoro block and black hole perturbation:}The result of~\cite{Bonelli:2022ten} relating connection coefficients of Heun-type opers to semiclassical Virasoro blocks, for previously explained reasons, is immediately applicable to obtain new exact analytic characterizations of QNMs and thermal two-point functions in \AdS{d>3}. There already exists a considerable amount of literature on applying semiclassical Virasoro blocks to black hole perturbation problems based on the relation with connection coefficients, either in AdS or flat spacetime asymptotics. Moreover, there have also emerged, in recent years, other new exact analytic approach based on distinct methods such as quantum Seiberg-Witten period~\cite{Aminov:2020yma}, ODE/IM correspondence~\cite{Fioravanti:2021dce}, exact WKB~\cite{Hatsuda:2021gtn}, some of which also involve semiclassical Virasoro block for reasons besides the relation with connection coefficients. We refer to~\cite{Aminov:2023jve} for a quite extensive list of references.

While previous works such as~\cite{Dodelson:2022yvn} have analyzed thermal two-point functions (for scalar operators) in \AdS{5} using the semiclassical Virasoro block approach, we will provide a new and considerably simpler exact analytic description. As we describe, previous work in this context utilizes what we refer to as the $t$-channel connection formula. However, a simpler $s$-channel connection formula for the \emph{same} connection coefficients exists.  The  explicit dependence on the choice of channel of the semiclassical Virasoro block has remained rather underappreciated in the literature. As will be made clear later, the $t$-channel expression does have its merit that the $t$-channel OPE limit corresponds to physical limits in the black hole perturbation problem, while the $s$-channel one does not.

These considerations motivate our analysis. We were inspired by exploiting this technology to compute conserved current correlators of holographic CFTs. Thanks to large $N$, these results are universal for all holographic CFTs. For instance, the  energy-momentum tensor correlators obtained holographically remain valid for all $d$-dimensional CFTs with a dual \AdS{d+1} gravitational description. We will focus on $d=4$ for simplicity. A few additional comments are in order:

\paragraph{Additional comments}
\begin{itemize}[wide,left=0pt]
    \item A large part of the existing literature on applying semiclassical Virasoro block to black hole perturbation involves using the gauge theory side of AGT correspondence~\cite{Alday:2009aq}, which relates semiclassical Virasoro blocks to Nekrasov instanton partition functions in the Nekrasov-Shatashvili limit. We would like to emphasize that the connection coefficients approach to black hole perturbation problems can be purely, and arguably more simply, understood in CFT terms, without the need for AGT relation. For example, the gauge-theoretic derivation of BPZ equation, which is crucial for deriving the connection coefficients for Heun-type opers, was only understood rather recently in~\cite{Nekrasov:2017gzb,Jeong:2018qpc}. We have therefore chosen to phrase our discussions exclusively in CFT language. The only exception to this is the analysis in~\cref{sec:SW} on the WKB regime of QNMs.
    \item We would like to clarify explicitly that there exists two distinct notions of CFTs in our discussions: one being the \CFT{4} whose thermal correlators are holographically dual to black hole perturbations in \AdS{5}, the other being the auxiliary \CFT{2} whose semiclassical Virasoro blocks are used to analyze the Heun-type opers arising from perturbation equations in \AdS{5}. Which notion of CFT is referred to in our  subsequent discussions should be clear from context.
\end{itemize}

\subsection{Summary of results}

We give a brief summary of our salient results, with pointers to the appropriate part of the paper for further details. 

\paragraph{Universal exact descriptions:} Our central results are universal exact analytic characterizations of QNMs and holographic thermal two-point functions in \AdS{5}, described in terms of the auxiliary data of spherical four-point semiclassical Virasoro blocks, in both $s$-channel (cf.~\eqref{eq-Gret-s} and~\eqref{eq-QNM-quantization}) and $t$-channel (cf.~\eqref{eq-Gret-t} and~\eqref{eq-QNM-quantization-t}). These are  different descriptions of the same physical objects. The universality of the results refers to their applicability to a wide range of examples in \AdS{5}.\footnote{Here we refer to distinct types of perturbations (scalar, electromagnetic, gravitational) on different asymptotically \AdS{5} backgrounds, both uncharged and charged black holes with planar or spherical horizon topology.} While technically these results were derived assuming the absence of logarithmic singularity at \AdS{5} boundary, viz., for perturbations dual to \CFT{4} operators with non-integer conformal dimensions, the answers for logarithmic cases can be extracted via a suitable continuation procedure (cf.~\cref{claim:QNM-log-quant},~\cref{claim:TwoPt-log} and~\cref{remark:logCFT}).

As alluded to previously, while the $t$-channel expressions have appeared in the earlier literature in certain examples. The $s$-channel expressions, having a considerably simpler analytic structure, are new to the best of our knowledge. In particular, they resemble the formal structure of the answers for BTZ black hole in \AdS{3} dual to thermal correlator for holographic \CFT{2}.\footnote{The form of thermal 2-point function in \CFT{2} on $\mathbb{R}^{1,1}$ is independent of the central charge, and thus holds even for non-holographic CFTs.}

Certain examples do require extension of the aforementioned results. In particular, the extension involves adding one more regular puncture (non-scalar perturbations on charged planar black hole background; cf.~\cref{subsec:five-punc}), or taking into account punctures with trivial monodromy (scalar channel of metric perturbation; cf.~\cref{subsec:app-sing}). The latter corresponds to certain heavy degenerate operator in the auxiliary \CFT{2}. The aforementioned $s$-channel exact expressions admit straightforward extension to these cases, thanks to the locality of fusion transformation of the (degenerate) Virasoro blocks.

\paragraph{Specific properties:} Besides the universal exact expressions, we also discuss their specific physical aspects illustrated in various examples, including
\begin{itemize}[wide,left=0pt]
    \item \emph{Computing QNMs using $s$-channel exact quantization condition:} The simplicity of the analytic structure of $s$-channel expressions facilitates explicit computation of QNMs using semiclassical Virasoro block data. This is illustrated in \cref{ex:QNM-expansion-BB} for different types of perturbations on planar Schwarzschild-\AdS{5} black hole background. 
    \item \emph{WKB regime of QNMs from Seiberg-Witten limit of $s$-channel exact quantization condition.} The WKB regime of QNMs, where QNMs become equally spaced, is also more transparent from the $s$-channel exact quantization condition. It corresponds to a specific large-momenta  limit of semiclassical Virasoro block, which is related to Seiberg-Witten prepotential from gauge theory side of AGT relation. This is discussed in~\cref{sec:SW}.
    \item \emph{Physical limits of QNMs from the OPE limit of $t$-channel exact quantization condition.} Compared to their $s$-channel counterpart, the $t$-channel exact quantization condition does have the advantage that its OPE limit corresponds to physical limits in the black hole perturbation problems. In particular, in~\cref{ex:near-extremal-QNM}, we obtain new analytic expressions for purely decaying QNMs in near-extremal planar black hole.
\end{itemize}

\subsection{Outline}

The outline of the paper is as follows. In~\cref{sec:waveeqs} we introduce the black hole perturbation problem, and the holographic computation of thermal observables. We take the opportunity here to give an algorithm for the real-time computations. We then turn in~\cref{sec-connection-formula-CFT} to a broad overview of connection formulae for Heun type opers using 2d CFT techniques, specifically developing the relation to the semiclassical Virasoro blocks. These results are used in~\cref{sec:exactAdS5} to derive properties of thermal correlation functions in holographic CFTs. In~\cref{sec:SW} we explain how the connection with supersymmetric field theory observables can aid our understanding of the WKB asymptotics. The reader interested in explicit results for energy-momentum tensor correlators of 4d CFTs can find them compiled in~\cref{sec:emcorr}. We conclude in~\cref{sec:discuss} with a discussion of open questions and future directions. 

\section{Black holes and holographic CFTs}\label{sec:waveeqs}

Consider a CFT in $d$ spacetime dimensions, $\Sigma_{d-1} \times \mathbb{R}$. We will primarily focus on the cases where the spatial manifold $\Sigma_{d-1}$ is either the round sphere $\mathbf{S}^{d-1}$ or flat Euclidean space $R^{d-1}$, corresponding to placing the CFT on the Einstein Static Universe, or Minkowski spacetime, respectively. We work at finite temperature $T = \frac{1}{\beta}$. We will also allow ourselves the freedom to turn on chemical potentials for conserved global currents, should the CFT admit any.  

Focusing on holographic CFTs, which have a large number of degrees of freedom, and a sparse low-lying spectrum, the thermal ensemble at high enough temperatures is described by a black hole in the dual gravitational theory. For thermal equilibrium, we will be interested in static black holes with \AdS{d+1} asymptotics. Working in ingoing coordinates, the line element takes the form
\begin{equation}\label{eq:staticbh}
ds^2 = -r^2 \, f(r)\,dv^2 + 2\, dv\, dr + r^2\, ds^2_{\Sigma_{d-1}}\,, 
\end{equation}
where $ds^2_{\Sigma_{d-1}}$ is the round metric on $\mathbf{S}^{d-1}$ or the flat Euclidean metric on $\mathbb{R}^{d-1}$, depending on the case of interest. The black hole horizon is at $\rp$, the  largest real root of $f(r)$, $f(\rp) =0$. The temperature of the black hole and the dual CFT is 
\begin{equation}\label{eq:Tdef}
T = \frac{\rp^2\, f'(\rp)}{4\pi}\,.
\end{equation}
In the absence of any chemical potentials, viz., in the canonical ensemble, the duals are the \SAdS{d+1} black holes. The explicit metric functions in the coordinates chosen above are 
\begin{equation}\label{eq:fSchw}
\begin{split}
f(r)  
= \begin{dcases}
    \frac{\lads^2}{r^2} \left[1 + \frac{r^2}{\lads^2} - \frac{\rp^{d-2}}{r^{d-2}}
         \left( 1+ \frac{\rp^2}{\lads^2} \right) \right]\,, &\qquad  \Sigma_{d-1} = \mathbf{S}^{d-1} \,,\\ 
    1 - \frac{\rp^d}{r^d} \,, &\qquad  \Sigma_{d-1} = \mathbb{R}^{d-1}\,.
\end{dcases}
\end{split}
\end{equation}
The ingoing coordinates are regular at the future horizon, and the radial derivative adapted to this choice is 
\begin{equation}\label{eq:Dzdef}
\Dz_+   \equiv  r^2f\, \pdv{r} + \pdv{v} \,.
\end{equation}  
%

\subsection{Quasinormal modes and thermal Green's functions}\label{sec:QNMs}

We would like to compute real-time thermal correlators using the holographic description. A single trace, gauge invariant CFT operator $\mathcal{O}(x)$, where $x$ coordinatizes $\mathbb{R} \times \Sigma_{d-1}$ is dual to a field $\Phi(r,x)$. We are suppressing the representation labels of the operator for simplicity. The field $\Phi(r,x)$ has a classical action in the \AdS{} black hole background, with a kinetic term, and a potential, which characterizes its interactions (both with itself and other fields).  The basic ingredient for our analysis will be the linear wave equation arising from the kinetic term, which schematically takes the form 
\begin{equation}\label{eq:waveeqn}
\mathfrak{D} \Phi  + V(r) \, \Phi = 0 \,, \qquad 
\mathfrak{D} = - \frac{1}{\sqrt{-g}\, e^{\chi(r)}} \partial_A \left( \sqrt{-g}\, e^{\chi(r)}\, g^{AB}\, \partial_B \right) .
\end{equation}
Here $V(r)$ and $\chi(r)$ are some auxiliary functions that depend on the nature of the kinetic operator for the field $\Phi$. We will outline some examples below in~\cref{sec:bhwave}.

The CFT correlation functions of interest will be  real-time thermal correlators. Per se, these should be evaluated using the Schwinger-Keldysh contour, which  computes the generating function  $\Tr(U[J_{\text{R}}]\, \rho_\beta\, (U[J_{\text{L}}])^\dag )$. Such observables probe the thermal state and can be viewed as characterizing the retarded response, and associated stochastic fluctuations. Sources $J_{\text{R}}$ and $J_{\text{L}}$ are inserted in the forward and backward evolution segments of the contour. We give a quick overview of implementing this in holography in~\cref{sec:algorithm}. For the present, we will give a somewhat simpler characterization (which we justify below).

The wave equation~\eqref{eq:waveeqn} has two linearly independent solutions. For non-extremal black holes, the solutions near the horizon are ingoing or outgoing. Using the Killing symmetry $\partial_v$ to pass to the frequency domain,\footnote{We work in the frequency domain with wavefunctions having temporal dependence $e^{-i\omega v}$. } the local solutions, near the horizon $r\sim \rp$, are  $\Phi(r) = c_1 + c_2\, (r-\rp)^{i\beta\omega }$. The constant mode is the ingoing solution that is analytic at the horizon. At the asymptotic AdS boundary, we have as usual normalizable and non-normalizable modes, the latter corresponding to  the sources we can turn on for the boundary operator. 

The asymptotically normalizable solution to the wave equation with ingoing boundary conditions at the horizon exists only for a discretuum of frequencies. These are the quasinormal modes (henceforth QNMs) of the black hole. One way to extract them is to examine the retarded 2-point function of the dual boundary operator (spatial dependence suppressed for simplicity), 
\begin{equation}\label{eq:Kdef}
\Gret(\omega) = \expval{\mathcal{O}(\omega) \, \mathcal{O}(-\omega)}_\beta\,.
\end{equation}
In holographic theories $\Gret(\omega)$ is meromorphic, with poles at the quasinormal frequencies. 

A prescription for computing $\Gret(\omega)$ was given in~\cite{Son:2002sd}. The essential idea was to phrase the calculation in terms of a connection problem for~\eqref{eq:waveeqn}. Consider turning on the non-normalizable asymptotic mode for $\Phi$ (to activate a source for $\mathcal{O}$) and imposing ingoing boundary condition at the horizon. The solution defines the ingoing boundary-bulk Green's function $G_{\text{in}}(r,\omega)$. The 2-point function is determined in terms of the ratio of normalizable and non-normalizable modes of $G_{\text{in}}$, viz.,  
\begin{equation}\label{eq:KfromG}
\begin{split}
\Gret(\omega) =   \frac{\textrm{Normalizable}(G_{\text{in}})}{\textrm{Non-normalizable}(G_{\text{in}})}.
\end{split}
\end{equation}  
This prescription, which was argued for in~\cite{Herzog:2002pc} has the virtue of being simple, but does not readily generalize to computing higher point functions. The more natural way to proceed is to directly implement the real-time Schwinger-Keldysh computation in gravity. As noted above, this has been achieved recently~\cite{Glorioso:2018mmw} (see also~\cite{vanRees:2009rw}). For completeness, we give a brief synopsis of this development, and argue for how to recover~\eqref{eq:Ginpoles}.

\subsection{Real-time observables in thermal holographic CFTs}\label{sec:algorithm}

%
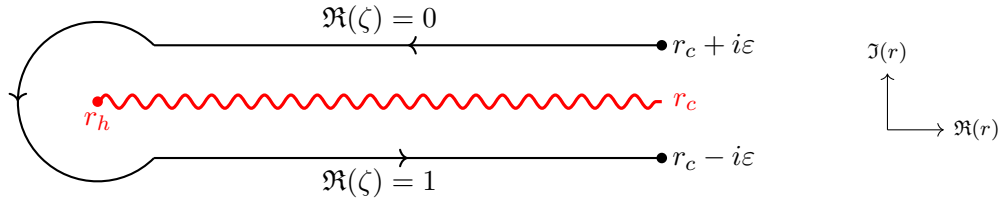
\begin{figure}[t!]
\begin{center}
\begin{tikzpicture}[scale=0.75]
\draw[thick,color=red,fill=red] (-5,0) circle (0.45ex);
\draw[thick,color=black,fill=black] (5,1) circle (0.45ex);
\draw[thick,color=black,fill=black] (5,-1) circle (0.45ex);
\draw[very thick,snake it, color=red] (-5,0) node [below] {$r_h$} -- (5,0) node [right] {$r_c$};
\draw[thick,color=black, ->-] (5,1)  node [right] {$r_c+i\varepsilon$} -- (0,1) node [above] {$\Re(\zeta) =0$} -- (-4,1);
\draw[thick,color=black,->-] (-4,-1) -- (0,-1) node [below] {$\Re(\zeta) =1$} -- (5,-1) node [right] {$r_c-i\varepsilon$};
\draw[thick,color=black,->-] (-4,1) arc (45:315:1.414);
\draw[thin, color=black,  ->] (9,-0.5) -- (9,0.5) node [above] {$\scriptstyle{\Im(r)}$};
\draw[thin, color=black,  ->] (9,-0.5) -- (10,-0.5) node [right] {$\scriptstyle{\Re(r)}$};  
\end{tikzpicture}
\caption{ The complex $r$ plane, with the locations of the two boundaries and the horizon marked. The grSK contour is a codimension-1 surface in this plane (drawn at fixed $v$). The direction of the contour is as indicated counter-clockwise, encircling the branch point at the horizon.}
\label{fig:grsk}
\end{center}
\end{figure}

The Schwinger-Keldysh contour is implemented in gravity by a complex two-sheeted geometry obtained from~\eqref{eq:staticbh}. Following~\cite{Jana:2020vyx} introduce the complexified (mock) tortoise coordinate 
\begin{equation}\label{eq:zetadef}
d\zeta = \frac{2}{i\,\beta}\, \frac{dr}{r^2f}\,.
\end{equation}
The one form $\frac{dr}{f(r)}$ has a branch-point at the horizon/ We avoid this by working on a keyhole contour encircling the horizon, see~\cref{fig:grsk}.

The algorithm for computation of the CFT correlators on this geometry can be summarized as follows:    
\begin{itemize}[wide,left=0pt]
\item First, we obtain the ingoing boundary to bulk propagator $G_{\text{in}}(r,\omega)$ for~\eqref{eq:waveeqn}. It satisfies 
\begin{equation}\label{eq:}
\lim_{r\to \infty} G_{\text{in}} =1 \,, \qquad \dv{G_{\text{in}}}{r} \bigg|_{\rp}  =0 \,.
\end{equation}	

\item  Next, we uplift the fluctuations of fields in the gravitational theory to the grSK contour. The bulk dynamics is then prescribed by a contour integral
\begin{equation}\label{eq:bulkS}
S_{\text{bulk}} = \oint d\zeta \int d^dx \, \sqrt{-g} \; \mathcal{L}[g_{AB}, \Phi]\,,
\end{equation}	

\item $G_{\text{in}}(r,\omega)$is regular at the horizon, and therefore is insensitive to the branch cut in $\zeta$. Therefore, the ingoing grSK propagator $G_{\text{in}}(\zeta,\omega)$ is obtained by replacing $r\to r(\zeta)$.

\item The solution for $\Phi$ with L and R Schwinger-Keldysh sources on the respective boundaries is obtained by exploiting the time-reversal involution of the static solution. One finds 
\begin{equation}\label{eq:phiLR}
\begin{split}
\Phi(\zeta,\omega) 
&= 
    G_{\text{in}}(\zeta,\omega) \,  \bigg( (1+n_\omega)\, J_{\text{R}} -n_\omega \, J_{\text{L}} \bigg) - G_{\text{in}}(\zeta,-\omega) \, e^{\beta\, \omega\, (1-\zeta)}\,n_\omega \bigg( J_{\text{R}} - J_{\text{L}} \bigg) \,, \\ 
&\equiv 
 - G_{\text{in}}(\zeta,\omega)\, J_F + G_{\text{in}}(\zeta,-\omega)\, e^{\beta\omega\, (1-\zeta)}\, J_P\,.
\end{split}
\end{equation}
Here $n_\omega = (e^{\beta\omega}-1)^{-1}$ is the Bose-Einstein function. In the second line, we introduced the retarded/advanced sources $J_F$ and $J_P$ as linear combinations of the R and L sources. 
\item Finally, the propagation of the field between two bulk spacetime points is captured by the bulk-bulk propagator ($\mathcal{N}(\omega)$ is a normalization factor)
\begin{equation}\label{eq:Gbulk}
\begin{split}
G_{\text{bulk}}(\zeta,\zeta';\omega)
&=  
   \mathcal{N}(\omega)\;  e^{\beta\omega\,\zeta'} 
   \bigg[\Theta(\zeta - \zeta') \, G_L(\zeta, k) \, G_R(\zeta', k) 
   +\Theta(\zeta'- \zeta) \, G_L(\zeta', k)\, G_R(\zeta, k)  \bigg] \,.
\end{split}
\end{equation} 
Here $\Theta(\zeta-\zeta')$ is a contour-ordered step function along the contour depicted in Fig.~\ref{fig:grsk}, while 
\begin{equation}\label{eq:GLRdef}
\begin{split}
G_R(\zeta,\omega)
&=
        e^{\beta\,\omega}\, n_\omega \left(G_{\text{in}}(\zeta,\omega) - e^{-\beta\omega\,\zeta} \, G_{\text{in}}(\zeta,-\omega)\right) ,
    \\ 
G_L(\zeta,\omega)
&=
     -n_\omega \left(G_{\text{in}}(\zeta,k) - e^{\beta\omega\, (1-\zeta)}\, G_{\text{in}}(\zeta,-\omega) \right) .
\end{split}
\end{equation}  
The functions $G_R$ and $G_L$ have source on the right ($\zeta =1$) and left ($\zeta =0$) boundaries, respectively, and are normalizable at the other end. 
\end{itemize}

Real-time correlators are computed by Witten diagrams on the grSK geometry. Any such diagram comprises a set of propagators, vertex functions, and radial-ordering step-functions. This constructs the integrand for the contour integral over the contour depicted in~\cref{fig:grsk}. The non-analytic parts of the integrand come from the factors $e^{-\beta\omega\,\zeta}$ and $\Theta(\zeta - \zeta')$. Accounting for these, the final result for an $n$-point function can be expressed as an integral of  (multiple) discontinuities of the original integrand in the region exterior of the black hole (in the original radial coordinate $r \in [\rp,\infty)$, cf.~\cite{Jana:2020vyx,Loganayagam:2022zmq,Loganayagam:2024mnj} for details.\footnote{As noted in~\cite{Chakrabarty:2018dov,Loganayagam:2022zmq} one may also have localized contributions at the horizon in certain situations.}  

The essential data necessary for real-time observables is the ingoing boundary to bulk Green's function for $\Phi$. Since this Green's function has a source at the boundary, and is infalling at the horizon, it should be meromorphic, with poles at the quasinormal frequencies. Factoring out the contribution from the quasinormal modes, one can write 
\begin{equation}\label{eq:Ginpoles}
\begin{split}
 G_{\text{in}}(\zeta,\omega) = \Gret(\omega)\,  \widetilde{G}_{\text{in}}(\zeta,\omega)\,.
\end{split}
\end{equation}  
The function $\Gret(\omega)$ is the boundary 2-point introduced in~\eqref{eq:Kdef}. This can be seen by computing the corresponding Witten diagram. The contour integral picks out the discontinuity proportional to $J_F\, J_P$ (the other combinations vanish as they have no discontinuities consistent with Schwinger-Keldysh and KMS constraints). Since the field is on-shell in the bulk, the result simply reduces to the product of the boundary value of the field and its conjugate momentum. The latter is the normalizable mode of the field. Accounting for the fact that $G_{\text{in}}$ was defined with a unit boundary source, we recover~\eqref{eq:KfromG}. This recovers the prescription of~\cite{Son:2002sd} quoted above. 

For purposes of computing the 2-point function $\Gret(\omega)$, it suffices to phrase the calculation as a connection problem.  We will use this language, since it allows us to reframe the question of computing the thermal real-time correlation functions using semiclassical conformal blocks of an auxiliary two-dimensional CFT. However, the grSK perspective will be useful, since the ingoing bulk-boundary propagator itself can be reinterpreted as a conformal block. Thus, all correlation functions can in principle determined from the data of this auxiliary two-dimensional CFT.

\section{Connection formulae for Heun-type opers from semiclassical Virasoro blocks}\label{sec-connection-formula-CFT}

The wave equations of interest in \AdS{d+1} black hole backgrounds are generically Fuchsian differential equations~\cite{Aminov:2020yma,Loganayagam:2022teq}.  These equations are characterized by having regular singular points or punctures. By a suitable choice of coordinates, one may bring the equations to the canonical form of Heun-type opers. We will assume this has been achieved for the present discussion. Some well-known examples are the hypergeometric equation that arises in the case of $d=2$, i.e., for wave equations in the BTZ background. Our focus will be on the case $d=4$, relevant to 4d CFTs such as $\mathcal{N}=4$ SYM. In this case, the equations of interest (for neutral black holes) turn out to have 4 punctures. Such equations are usually referred to as Heun's equation. For the case of a charged black hole, we will also encounter an equation with 5 punctures, which we shall also comment on.

We revisit the large-$c$ CFT derivation of the connection formula for Heun's equation, largely following the presentation of \cite{Lisovyy:2022flm}. As explained in \cite{Lisovyy:2022flm}, the generalization to equations with more singular points is immediate. In CFT parlance, this follows from the locality of fusion transformation. We therefore primarily focus on \AdS{5} black holes, and the associated Heun's equations, for simplicity in presentation. 

The connection between~\eqref{eq:waveeqn} and 2d CFTs can be motivated as follows. 
Consider a degenerate Virasoro block, involving the insertion of $n$-heavy operators and one light degenerate operator. Here, heavy/light refers to conformal weights being $\order{c^1}$ or $\order{c^0}$, respectively, in the large-$c$ limit. The BPZ equation satisfied by such degenerate blocks reduces to Fuchsian equation in the large-$c$ limit. The heavy operators correspond to the punctures. The fusion of the light degenerate operator with different heavy operators corresponds to different Frobenius solutions. The change of basis of the Frobenius solutions is then directly related to the degenerate fusion transformation of these degenerate Virasoro blocks, from which one directly reads off the connection formula. Eventually, the connection matrix depends on degenerate fusion matrix and classical Virasoro block of the heavy operators. 

The above derivation can be carried out by starting with degenerate Virasoro block in \emph{any} channel and then performing degenerate fusion transformation. The resulting connection matrix will be apparently different for different choices of channel. As we shall see, for the connection problem in the Heun case (4 punctures), there will be a distinction between $s$- and $t$- channel connection formulae. The $s$-channel one takes a simpler form than the $t$-channel one (the reason being the former is derived from using one fusion transformation, whereas the latter a composition of two). 

This important point, to the best of our knowledge, has not been explicitly explained or emphasized in literature. We will use the simpler $s$-channel connection formula for application to QNMs and thermal 2-point functions of 4d holographic CFTs. The $t$-channel connection formula is equivalent to the one employed in~\cite{Dodelson:2022yvn}. Details of this will become clear in the sequel.

\paragraph{Notation:} The following convenient notation will be used throughout our subsequent discussions:
\begin{equation}
    \Gamma(\alpha \pm \beta) \coloneqq \Gamma(\alpha + \beta) \, \Gamma(\alpha - \beta)\,.
\end{equation}

\subsection{Elements of 2d CFT}\label{sec:Liouville}
We first give a brief review of elements of 2d CFT relevant for our subsequent discussions. We refer to, e.g., \cite{DiFrancesco:1997nk,Ribault:2014hia} for more comprehensive background.

\paragraph{Liouville parametrization:}The Liouville parametrization of central charge and conformal weights is, in general, convenient for understanding representations of Virasoro algebra. The central charge is parametrized by
\begin{equation}
    c = 1 + 6\, Q^2 \,, \qquad Q = b + \frac{1}{b} \,.
\end{equation}
Primary operators $V_P$ are labeled by Liouville momenta $P$, and have conformal weight
\begin{equation}
   h =  \frac{Q^2}{4} - P^2  \,.
\end{equation}

\paragraph{Degenerate representations:}The Verma module atop $V_P$ is obtained by acting with Virasoro raising generators $L_{-n}$ with $n \geq 1$.  For generic $P$, such a  module is non-degenerate in the sense that no null vectors appear in the module. Degenerate modules occur for a discrete set of momenta characterized by two integers 
\begin{equation}\label{eq:degenrep}
  P_{\drep{r}{s}} = \frac{1}{2} \left(r \, b +  \frac{s}{b}\right)  \,.
\end{equation}
The corresponding degenerate primary is $V_{\drep{r}{s}}$, and it has conformal weight
\begin{equation}\label{eq:dengenh}
  h_{\drep{r}{s}} = \frac{1}{4} \left[ \left( b + \frac{1}{b} \right)^2 -
  \left(r \, b +  \frac{s}{b}\right)^2 \right] .
\end{equation}
The fusion rules for the degenerate operators with a generic primary operator with Liouville momentum $P$ is given by 
\begin{equation}\label{eq:Lfusion}
    V_{\drep{r}{s}} \times V_P = \sum\limits_{m=-\frac{r-1}{2}}^{\frac{r-1}{2}}\, \sum\limits_{n=-\frac{s-1}{2}}^{\frac{s-1}{2}}\, V_{P+m\,b + \frac{n}{b}} \,.\
\end{equation}
\paragraph{Virasoro blocks and their semiclassical limits:}The spherical four-point Virasoro block in $s$-channel is defined via a series expansion in cross-ratio:
\begin{equation}
\begin{split}
    &\VB^s\prn{x|\vb*{P},P_\sigma} \coloneqq x^{h_\cmexp - h_0 - h_\xr} \prn{1 + \sum^\infty_{k=1} \VB^s_k\prn{\vb*{P},P_\sigma} x^k}, \\
    &\VB^s_1\prn{\vb*{P},P_\sigma} = \frac{\prn{h_\cmexp - h_0 + h_\xr} \,\prn{h_\cmexp - h_\infty + h_1}}{2 \,h_\cmexp}, \ \;\; \mathrm{etc}.
\end{split}
\end{equation}
The expansion coefficients, $\VB^s_k\prn{\vb*{P},P_\sigma}$, are entirely fixed by Virasoro algebra, and their precise definitions can be found in standard references such as~\cite{DiFrancesco:1997nk,Ribault:2014hia}. The relation between conformal weights and Liouville momenta is understood. We use $\vb*{P}$ to collectively denote $\{P_i\}_{i= 0, \xr, 1, \infty}$. The expansion coefficients also admit closed-form combinatorial expression from instanton counting thanks to the AGT relation~\cite{Alday:2009aq}. A standard graphical notation is
\begin{equation}
    \VB^s(x) = 
    \begin{tikzpicture}[scale= .5, baseline = .5ex]
        \coordinate (inf) at (-2,-2);
        \coordinate (one) at (-2,2);
        \coordinate (inf-one) at (0,0);
        \coordinate (xr-zero) at (4,0);
        \coordinate (midint) at (2,0);
        \coordinate (xr) at (6,2);
        \coordinate (zero) at (6,-2);
        \draw (inf) -- (inf-one);
        \draw  (one) -- (inf-one);
        \draw (inf-one) -- (xr-zero);
        \draw (xr-zero) -- (xr);
        \draw (xr-zero) -- (zero);
        \node [below left] at (inf) {\scriptsize $P_\infty$};
        \node [above left] at (one) {\scriptsize $P_1$};
        \node [below] at (midint) {\scriptsize $P_\cmexp$};
        \node [above right] at (xr) {\scriptsize $P_\xr$};
        \node [below right] at (zero) {\scriptsize $P_0$};
    \end{tikzpicture} 
\end{equation}
An alternative $t$-channel expansion, corresponding to a different pair-of-pants decomposition of the four-punctured sphere, can be performed for the same set of external momenta $\vb*{P}$. This is graphically represented by
\begin{equation}
    \VB^t(x) = 
    \begin{tikzpicture}[scale= .5, baseline = -5.5ex]
        \coordinate (xr) at (2,2);
        \coordinate (one) at (-2,2);
        \coordinate (one-xr) at (0,0);
        \coordinate (inf-zero) at (0,-4);
        \coordinate (midint) at (0,-2);
        \coordinate (inf) at (-2,-6);
        \coordinate (zero) at (2,-6);
        \draw (one) -- (one-xr);
        \draw  (xr) -- (one-xr);
        \draw (one-xr) -- (inf-zero);
        \draw (inf-zero) -- (inf);
        \draw (inf-zero) -- (zero);
        \node [below left] at (inf) {\scriptsize $P_\infty$};
        \node [above left] at (one) {\scriptsize $P_1$};
        \node [left] at (midint) {\scriptsize $P_\cmexp$};
        \node [above right] at (xr) {\scriptsize $P_\xr$};
        \node [below right] at (zero) {\scriptsize $P_0$};       
    \end{tikzpicture}
\end{equation}
The blocks in the two channels are related by
\begin{equation}
    \VB^t(\xr) =  \VB^s(1-\xr) \bigg|_{P_0 \leftrightarrow P_1}\,.
\end{equation}
Now consider the following semiclassical limit:
\begin{equation}\label{eq-quasiclassical-limit}
    \CL: b \to 0, \quad P_i,\, P_\sigma \to \infty, \quad b \,P_i \to \theta_i, \quad b\,P_\sigma \to \sigma.
\end{equation}
The semiclassical Virasoro block $\VBcl(x)$,\footnote{This notation comes from the gauge theory side of AGT relation, where semiclassical Virasoro block is related to effective twisted superpotential. We find that it does more justice to the significance of the object, compared to the frequently used notation $f(x)$.} in either channel, is then defined by
\begin{equation}
    \VBcl\prn{x|\vb*{\mexp},\cmexp} \coloneqq \lim_{\CL} \,b^2\, \log\VB\prn{x|\vb*{P},P_\sigma}.
\end{equation}
The limit is understood to be taken term by term in the cross-ratio expansion of $\log\VB(x)$. For $s$-channel, the explicit expansion reads:
\begin{equation}
\begin{split}
    &\VBcl^s(x) = \prn{\hcl_\cmexp - \hcl_0 - \hcl_\xr} \log\xr + \sum^\infty_{k=1} \VBcl^s_k\prn{\vb*{\mexp},\cmexp} x^k \,, \\
    &\VBcl^s_1\prn{\vb*{\mexp},\cmexp} = \frac{\prn{\hcl_\cmexp - \hcl_0 + \hcl_\xr} \prn{\hcl_\cmexp - \hcl_\infty + \hcl_1}}{2 \,\hcl_\cmexp}, \;\; \mathrm{etc}.
\end{split}
\end{equation}
where $\hcl_i = \frac{1}{4} - \mexp_i^2$ is (semi)classical conformal weight, and by convention $\mexp_\cmexp = \cmexp$.
The existence of the defining large-$c$ limit of semiclassical Virasoro block is non-trivial beyond $k=1$: it requires cancellation between all the $\order{c^2}, \dots, \order{c^k}$ terms in the $x^k$ term of $\log\VB^s(x)$ expansion. The exponentiation of Virasoro block is originally conjectured in~\cite{Zamolodchikov:1986two} and usually motivated by saddle-point argument in Liouville theory; a proof is recently claimed in~\cite{Besken:2019jyw}.

The definitions of Virasoro blocks and their semiclassical limits also extend to cases with more punctures, which will also be relevant in our subsequent discussions.

\paragraph{Degenerate fusion transformation and its locality property:} The infinite-dimensional fusion transformation relating generic Virasoro blocks at different channels is, in general, known in closed form~\cite{Ponsot:1999uf}. For our purpose, we only need the simplest two-dimensional case involving Virasoro block with the insertion of a degenerate operator $V_{\drep{2}{1}}$, whose fusion with $V_P$ results in $V_{P\pm \frac{b}{2}}$. In graphical notation, such degenerate fusion transformation is represented as:
\begin{equation}\label{eq-deg-block-fusion-transformation}
    \begin{tikzpicture}[scale= .9, baseline = 2.5ex]
        \draw (0,0) -- (1,0); 
        \draw (1,0) -- (3.5,0);
        \draw (1,0) -- (1,2.5);
        \draw [dashed] (2.5,0) -- (2.5,1.5);

        \node[left] at (0,0) {$\dots$};
        \node[right] at (3.5,0) {$\dots$};
        \node[above] at (1,2.5) {$\vdots$};

        \node[below right] at (0,0) {\scriptsize$P^{\prime\prime}$};
        \node[below right] at (1,0) {\scriptsize$P + \frac{\sign \,b}{2}$};
        \node[below right] at (2.5,0) {\scriptsize$P$};
        \node[below left] at (1,2.5) {\scriptsize$P^\prime$};

        \node[above] at (2.5,1.5){\scriptsize$P_{\drep{2}{1}}$};
    \end{tikzpicture}
    = \sum_{\sign^\prime = \pm} \Fus_{\sign\sign^\prime}\prn{P,P^\prime,P^{\prime\prime}}
    \begin{tikzpicture}[scale= .9, baseline = 2.5ex]
        \draw (0,0) -- (1,0); 
        \draw (1,0) -- (3.5,0);
        \draw (1,0) -- (1,2.5);
        \draw [dashed] (1,1.5) -- (2.5,1.5);

        \node[left] at (0,0) {$\dots$};
        \node[right] at (3.5,0) {$\dots$};
        \node[above] at (1,2.5) {$\vdots$};

        \node[below right] at (0,0) {\scriptsize$P^{\prime\prime}$};
        \node[left] at (1,1) {\scriptsize$P^\prime + \frac{\sign^\prime \,b}{2}$};
        \node[below right] at (2.5,0) {\scriptsize$P$};
        \node[below left] at (1,2.5) {\scriptsize$P^\prime$};

        \node[right] at (2.5,1.5){\scriptsize$P_{\drep{2}{1}}$};
    \end{tikzpicture}
\end{equation}
with 
\begin{equation}\label{eq:Fdegfusion}
    \Fus_{\sign\sign^\prime}\prn{P,P^\prime,P^{\prime\prime}} 
= 
    \frac{\Gamma\prn{1-2 \,\sign\, b\, P} \Gamma\prn{2 \,\sign^\prime \,b\, P^\prime}}{ \Gamma\prn{\half - \sign \,b\, P+ \sign^\prime\, b\, P^\prime \,\pm  b\, P^{\prime\prime}}} \,.
\end{equation}
The crucial property is the \emph{locality} of such transformation~\cite{Moore:1988qv,Moore:1989vd}: the omitted dots above could represent arbitrary conformal block diagrams with additional external operators/punctures. The degenerate block in the case with three non-degenerate external operators is the well-known hypergeometric block, and the above transformation is equivalent to the classically known connection formulae for hypergeometric functions. As will be seen with more details later, the locality of such transformation allows using the CFT method to obtain connection formulae beyond the hypergeometric case.

The only necessary ingredients we haven't introduced thus far are the BPZ equation satisfied by the degenerate Virasoro block, and the relation between accessory parameters in Heun-type opers and semiclassical Virasoro blocks. These will be introduced along with the derivation of connection coefficients.

For much of the discussion, we will be interested in a five-point degenerate Virasoro block where four of the operators $V_{P_i}$ are generic, and the fifth corresponds to the degenerate representation $V_{\drep{2}{1}}$. This will turn out to be sufficient for understanding thermal 2-point functions of scalar operators, and tensor and vector polarizations of the energy-momentum tensor in 4d holographic CFTs. The scalar polarization of the energy-momentum tensor will, however, require
an additional insertion of $V_{\drep{1}{3}}$. The reason for this will be the presence of an apparent singular point in the associated differential equation~\cite{Loganayagam:2022teq}.

\subsection{Prelude: The connection problem for Heun-type opers and subtleties in logartihmic cases}\label{subsec:conn-prob-frobenius}

For concreteness, consider Heun's equation, which is the most general second order ODE with four punctures. The generalization of subsequent discussions to generic Heun-type opers with more punctures is immediate. For Heun's equation, the location of punctures can always be set to $\{0,\xr,1,\infty \}$ via Möbius transformation. The equation in normal form reads
\begin{equation}\label{eq-heun-normal-form}
\begin{split}
     &\wf''(z) + \Tcl(z)\, \wf(z) = 0 \,,\\
     &\Tcl(z) = \frac{\delta_0}{z^2} + \frac{\delta_1}{(z-1)^2} + \frac{\delta_\xr}{(z-\xr)^2} +  \frac{\delta_{\infty} - \delta_0 - \delta_1 - \delta_\xr}{z\,(z-1)} + \frac{(\xr-1) \,\acc}{z\,(z-1)\,(z-\xr)} \,,\\
     & \delta_i = \frac{1}{4} - \mexp^2_i\,.
\end{split}
\end{equation}
The characteristic exponents at each puncture are $\half \pm \mexp_i$, and $\acc$ is the accessory parameter. 

We focus on the connection problem between the punctures at $0$ and $\xr$. It will be clear that the CFT method can be used to derive the connection formula between any other pair of punctures. Assume $\mexp_0, \mexp_\xr \not\in \frac{\intz}{2}$ so that there exists unique normalized Frobenius bases
\begin{equation}
\begin{split}
\wf^{[0]}_{\sign}(z) 
&= 
    z^{\half-\epsilon \,\mexp_0} \, \prn{ 1 + \sum_{k=1}^\infty \wf^{[0]}_{\sign,k}\; z^k} ,\\
\wf^{[\xr]}_{\sign}(z) 
&= 
    (z-\xr)^{\half-\epsilon \,\mexp_\xr} \prn{ 1 + \sum_{k=1}^\infty \wf^{[\xr]}_{\sign,k} \; (z-\xr)^k}, \qquad \sign=\pm\,.
\end{split}
\end{equation}
The connection problem is the problem of finding the connection matrix $\connm$ such that
\begin{equation}\label{eq:conn-prob-def}
    \wf^{[0]}_{\sign}(z) = \sum_{\sign^\prime = \pm} \, \connm_{\sign \sign^\prime}\;  \wf^{[\xr]}_{\sign^\prime}(z).
\end{equation}

While the case of exponent being half-integer appears rather non-generic mathematically, for many physical examples in black hole perturbation problems we will encounter $\mexp_{\xr} \in \intz/2$, which is related to the fall-off behavior at AdS boundary. In the half-integer cases, one of the local solutions is logarithmic and there are additional subtleties on the appropriate generalization of the connection problem. While eventually, we will claim that the connection coefficient in the logarithmic cases relevant for defining QNMs can be obtained via a suitable continuation of the CFT answer in the non-logarithmic case, we find it important to first give a direct mathematical characterization of the logarithmic cases. This requires a more careful examination of the local Frobenius expansion.

\subsubsection{Details on Frobenius expansion: subtleties in logarithmic cases}

Here we give more details on the local Frobenius expansion, in particular in the logarithmic cases. We follow the presentation in~\cite{iwasaki1991from}, adapting to our set-up and convention. We would like to find local Frobenius solutions to Fuchsian equation $\prn{\partial^2_z + \Tcl(z)} \psi(z)$ at $z=0$, where $\Tcl(z)$ has a double pole. It is convenient to rewrite the equation in the following manner:
\begin{equation}\label{eq:heun-dlogz}
        0  = \brk{z^2\partial^2_z + z^2\, \Tcl(z)} \psi(z) = \brk{\dlogz^2 - \dlogz +  z^2\, \Tcl(z) } \psi(z) \equiv \DDz \psi(z).
\end{equation}
Here $\dlogz \coloneqq z \partial_z$, and we have used $z^k \partial^k_z = \dlogz\, (\dlogz -1) \,\cdots\, (\dlogz - k +1)$. Consider the normalized Frobenius expansion,
\begin{equation}
    \frob(z) = z^\lambda \prn{ 1 + \sum^\infty_{k=1} \frob_k\, z^k}.
\end{equation}
One then reads off from~\eqref{eq:heun-dlogz}:
\begin{equation}
    \begin{split}
        \DDz\frob(z) &= z^\lambda \,\sum^\infty_{k=0} \rec_k\prn{\frob_0, \cdots,\frob_k|\lambda}\, z^k
        \\
        \rec_k\prn{\frob_0, \cdots,\frob_k|\lambda} &= \frob_k \,(\lambda+k)\,(\lambda+k-1) + \sum^k_{p=0} \frob_p \,\lop_{p-k}.
    \end{split}
\end{equation}
Here the $\lop$ operators are expansion coefficients in $z^2\, \Tcl(z) = \sum^\infty_{k=0} \,\lop_{-k}\,z^k$. In particular, $\lop_0 = \frac{1}{4} - \mexp^2$ with sign convention $\Re(\mexp) \geq 0$. Also $\frob_0 = 1$ by convention. The coefficients $\rec_k$ can be more conveniently written as
\begin{equation}
\begin{split}
    \rec_k\prn{\frob_0, \cdots,\frob_k|\lambda}  &= \frob_k\, f_k(\chexp)
  + \sum^{k-1}_{p=0} \frob_p\, \lop_{p-k}  \\
    f_k(\chexp) &= \prod_{\sign = \pm} \brkbig{\lambda -\prn{ \lambda_{\sign} - k} }.
\end{split}
\end{equation}
Here $\lambda_{\pm} = \half \pm \mexp$, with $f_0(\chexp) = 0$ being the characteristic equation. Now define
\begin{equation}\label{eq:frobenius-recursion-def}
    \begin{split}
        \frob_k(\lambda) &\coloneqq \text{solution to recursion relation } \; \rec_k\prn{\frob_0, \cdots,\frob_k|\lambda} =0\,, \quad k \geq 1 \\
        \frob(z|\lambda) &\coloneqq z^\lambda \prn{ 1 + \sum^\infty_{k=1} \frob_k(\lambda) \,z^k}.
    \end{split}
\end{equation}
Eventually we will set $\lambda = \lambda_{\pm}$. But when $\chexppm = \frac{1\pm n}{2} \equiv \chexppmn$ with $n \in \mathbb{Z}_{>0}$ (viz., $\mexp = n/2$), the recursive solutions $\frob_k(\chexpmn)$ are not well-defined at all $k$. In particular, the coefficient $f_n(\chexpm)$ for $\frob_n$ in $\rec_n\prn{\frob_0, \cdots,\frob_n|\chexpmn}$ vanishes due to $\chexpmn  = \chexppn -n$. In this case, define $\frob_k\prn{\chexpmn}$ in the following way. With the coefficients $\frob_0\prn{\chexpmn}, \dots, \frob_{n-1}\prn{\chexpmn}$ still defined from solving the recursion relation, choose $\frob_n\prn{\chexpmn}$ arbitrarily. Then  $\frob_{k>n}\prn{\chexpmn}$ can be solved for recursively starting from this arbitrary choice. This then defines the associated $\frob\prn{z|\chexpmn}$.

One then has
\begin{equation}\label{eq:frob-sols-generic-exp}
    \DDz\frob(z|\chexp) = z^\chexp\prn{f_0(\chexp) + z^n \delta_{\chexp,\chexpmn} \sum^{n-1}_{p=0} \frob_p\prn{\chexpmn} \lop_{p-n}}\,, \quad n \in \intz_{>0}.
\end{equation}
Therefore, for all $\mexp$, $\frob(z|\chexpp)$ is one of the solutions:
\begin{equation}
    \DDz\frob(z|\chexpp) = 0.
\end{equation}

\paragraph{Case I ($\mexp \notin \intz/2$):} In this case there is no ambiguity.  The two independent solutions are
\begin{equation}
    \psi_{\pm}(z) = \frob(z|\chexp_{\pm}) = z^{\chexp_{\pm}} \prn{ 1 + \sum^\infty_{k=1} \frob_k(\chexp_{\pm}) \,z^k}.
\end{equation}

To find the other solution in the remaining cases, consider taking derivative w.r.t.~$\chexp$ on both sides of~\eqref{eq:frob-sols-generic-exp} at $\chexp = \chexpp$, which is allowed since the discontinuities are only at $\chexpmn$. This gives:
\begin{equation}
\begin{split}
    \DDz  \froblog(z|\chexpp)
    &= 
        z^{\chexpp}\, f^\prime_0(\chexpp) \\
    \froblog(z|\chexpp) 
    &\coloneqq 
    \partial_{\chexp} \frob(z|\chexp)\big|_{\chexp = \chexpp} 
    = \;\log(z) \,\frob(z|\chexpp) + z^{\chexpp}\, \sum^\infty_{k=1} \frob^\prime_k(\chexpp)\, z^k.
\end{split}  
\end{equation}

\paragraph{Case II ($\mexp = 0$):} In this case $\chexp_{\pm}= \chexpcrit = 1/2$. Since $f_0^\prime(\chexpcrit) = 0$, $\froblog(z|\chexpcrit)$ is another independent solution. A basis of solutions is therefore given by
\begin{equation}
    \begin{split}
        \psi_{_{\nlog}}(z) 
        &= \frob\prn{z|\chexpcrit} \\
        \psi_{_{\log}}(z) 
        &= \froblog(z|\chexpcrit).
    \end{split}
\end{equation}

\paragraph{Case III ($\mexp = \frac{n}{2}$, $n \in \mathbb{Z}_{>0}$):} 
In this case, one first observes that 
\begin{equation}
    \DDz  \frob\prn{z|\chexpmn} = z^{\chexppn}\, \sum^{n-1}_{p=0} \frob_p\prn{\chexpmn} \lop_{p-n}.
\end{equation}
Therefore another independent solution is given by a suitable linear combination of $\frob\prn{z|\chexpmn}$ and $\froblog\prn{z|\chexppn}$. Explicitly, a basis of solution is:
\begin{equation}
    \begin{split}
        \psi_{_{\nlog}}(z) 
        &= \frob\prn{z|\chexppn} \\
        \psi_{_{\log}}(z) 
        &=  \frob\prn{z|\chexpmn} - \frac{\sum^{n-1}_{p=0} \frob_p\prn{\chexpmn}\, \lop_{p-n}}{f^\prime_0(\chexppn)}\; \froblog\prn{z|\chexppn}\,.
    \end{split}
\end{equation}

\begin{remark}\label{remark:ambiguity-integern}
    Recall that only the $\frob_{k<n}\prn{\chexpmn}$ coefficients in the Frobenius expansion $\frob\prn{z|\chexpmn}$ are unambiguously determined, whereas the coefficients $\frob_{k>n}\prn{\chexpmn}$ depend on the arbitrary choice of $\frob_{n}\prn{\chexpmn}$. This arbitrary choice corresponds to shifting $\psi_{\log}(z)$ by an arbitrary multiple of $\psi_{\nlog}(z)$. Therefore, there is an ambiguity in the non-logarithmic part of $\psi_{\log}(z)$, whereas the logarithmic part of $\psi_{\log}(z)$, which only involves $\froblog\prn{z|\chexppn}$ and $\frob_{k<n}\prn{\chexpmn}$, has no ambiguity.
\end{remark}

\begin{remark}
    The apparent singularity condition $\asc_n$ is thus given by 
 \begin{equation}
        \asc_n = \sum^{n-1}_{p=0} \frob_p\prn{\chexpmn}\, \lop_{p-n} = 0 \,,
 \end{equation}
    with, $\frob_p\prn{\chexpmn}$ understood to be defined via~\eqref{eq:frobenius-recursion-def}. This is an alternative to the method discussed in~\cref{app-apparent-singularity}. The first few expressions for $\asc_n$ are listed in~\cref{tab-apparent-singularity-conditions}.
\end{remark}

\subsubsection{Connection coefficients in logarithmic cases}

We will be interested in the situation where $\mexp_0$ is generic while $\mexp_{\xr} \in \intz/2$. In the logarithmic cases, we seek generalization of~\eqref{eq:conn-prob-def} to 
\begin{equation}
    \wf^{[0]}_{\sign}(z) = \connm_{\sign\,{\log}}\, \wf^{[\xr]}_{\log}(z) + \connm_{\sign\,{\nlog}} \, \wf^{[\xr]}_{\nlog}(z)\,.
\end{equation}
\paragraph{Case II ($\mexp_{\xr} = 0$):} In this case, both $\wf^{[\xr]}_{\log}(z)$ and $\wf^{[\xr]}_{\nlog}(z)$ are unambiguously defined. So the connection coefficients $\connm_{\sign\,{\log}}$ and $\connm_{\sign\,{\nlog}}$ have no ambiguity.
\paragraph{Case III ($\mexp_{\xr} = \frac{n}{2}$, $n \in \mathbb{Z}_{>0}$):} In this case, as discussed in~\cref{remark:ambiguity-integern}, there is an ambiguity of shifting the non-logarithmic part of $\wf^{[\xr]}_{\log}(z)$ by arbitrary multiple of $\wf^{[\xr]}_{\nlog}(z)$. Therefore, only the connection coefficient $\connm_{\sign\,{\log}}$ is unambiguously defined, with $\connm_{\sign\,{\nlog}}$ subject to arbitrary shift. 

Therefore, in both logarithmic cases, the connection coefficient $\connm_{\sign\,{\log}}$ is unambiguously defined. It is only this type of connection coefficient that will appear in the definition of QNMs. 

\subsubsection{Fall-off coefficients}\label{subsec:fall-off-coeff}

While the notion of connection coefficients is sufficient for defining QNMs, it will be necessary to use \emph{fall-off coefficients} to define the thermal two-point function in one of the logarithmic cases due to the aforementioned ambiguity in connection coefficient. The fall-off coefficients are defined as follows in the three cases.
\paragraph{Case I ($\mexp_{\xr} \notin \intz/2$):} 
\begin{equation}
    \falloffc_{\sign\,\nor/\nnor} \ \coloneqq\ \wf^{[0]}_{\sign}(z)\bigg|_{\coeff\brk{\prn{\zx}^{\chexppm}}} = \connm_{\sign\mp}.
\end{equation}
Here $\coeff(\cdot)$ denotes the coefficient of $(\cdot)$ in the expansion near $z=x$, with $\zx = z-x$.
\paragraph{Case II ($\mexp_{\xr} = 0$):}
\begin{equation}
    \begin{split}
        \falloffc_{\sign\,\nor} &\coloneqq \wf^{[0]}_{\sign}(z)\bigg|_{\coeff\brk{(\zx)^{\chexpcrit}}} = \connm_{\sign\,\nlog}\,, \\
        \falloffc_{\sign\,\nnor} &\coloneqq \wf^{[0]}_{\sign}(z)\bigg|_{\coeff\brk{ (\zx)^{\chexpcrit}\log\zx } } = \connm_{\sign\,\log}\,.
    \end{split}
\end{equation}
\paragraph{Case III ($\mexp_{\xr} = \frac{n}{2}$, $n \in \mathbb{Z}_{>0}$):}
\begin{equation}
    \begin{split}
        \falloffc_{\sign\,\nor} &\coloneqq \wf^{[0]}_{\sign}(z)\bigg|_{\coeff\brk{\prn{\zx}^{\chexpp}}} = \connm_{\sign\,{\nlog}} + \frob_{n}\prn{\chexpmn} \connm_{\sign\,{\log}}\,,\\
        \falloffc_{\sign\,\nnor} &\coloneqq \wf^{[0]}_{\sign}(z)\bigg|_{\coeff\brk{\prn{\zx}^{\chexpm}}} = \connm_{\sign\,{\log}}\,.
    \end{split}
\end{equation}
Note that, despite the appearance, there is no ambiguity in $\falloffc_{\sign\,\nor}$ by definition.

\subsection{The connection formula in the \texorpdfstring{$s$}{s}-channel expansion}

We now proceed to the derivation of the $s$-channel connection formula, closely following~\cite{Lisovyy:2022flm}. The relevant object is the five-point Virasoro block with four non-degenerate vertex operators $V_{P_i}$ and one degenerate vertex operator $V_{\drep{2}{1}}$. Without loss of generality, we can place the four non-degenerate vertex operators at $0$, $x$, $1$, and $\infty$. We will label the momenta by these loci for simplicity.  The degenerate vertex operator will be inserted at $z$. 

\paragraph{The fusion transformation and BPZ equation:} 
Consider the five-point Virasoro blocks with insertion of degenerate representation $V_{\drep{2}{1}}(z)$ in the following channels:
\begin{equation}
    \VB^{s,[0]} _\sign(z,\xr) \coloneqq 
    \begin{tikzpicture}[scale= .5, baseline = .5ex]
        \coordinate (inf) at (-2,-2);
        \coordinate (one) at (-2,2);
        \coordinate (inf-one) at (0,0);
        \coordinate (xr-zero) at (4,0);
        \coordinate (midint) at (2,0);
        \coordinate (xr) at (6,2);
        \coordinate (zero) at (6,-2);
        \draw (inf) -- (inf-one);
        \draw  (one) -- (inf-one);
        \draw (inf-one) -- (xr-zero);
        \draw (xr-zero) -- (xr);
        \draw (xr-zero) -- (zero);
        \node [below left] at (inf) {\scriptsize $P_\infty$};
        \node [above left] at (one) {\scriptsize $P_1$};
        \node [below] at (midint) {\scriptsize $P_\cmexp$};
        \node [above right] at (xr) {\scriptsize $P_\xr$};
        \node [below right] at (zero) {\scriptsize $P_0$};
        
        \coordinate (deg-fusion) at ($ 0.3*(xr-zero) + 0.7*(zero)$);
        \coordinate (deg) at ($ (deg-fusion) + (2,2) $);
        \coordinate (deg-fusion-label) at ($0.5*(xr-zero) + 0.5*(deg-fusion)$);
        \draw [dashed] (deg) -- (deg-fusion);
        \node [above right] at (deg) {\scriptsize $P_{\drep{2}{1}}(z)$}; 
        \node [below left] at (deg-fusion-label) {\scriptsize $P_0 + \frac{\sign\, b}{2}$};
    \end{tikzpicture}
\end{equation}
\begin{equation}
    \VB^{s,[\xr]} _\sign(z,\xr) \coloneqq 
    \begin{tikzpicture}[scale= .5, baseline = .5ex]
        \coordinate (inf) at (-2,-2);
        \coordinate (one) at (-2,2);
        \coordinate (inf-one) at (0,0);
        \coordinate (xr-zero) at (4,0);
        \coordinate (midint) at (2,0);
        \coordinate (xr) at (6,2);
        \coordinate (zero) at (6,-2);
        \draw (inf) -- (inf-one);
        \draw  (one) -- (inf-one);
        \draw (inf-one) -- (xr-zero);
        \draw (xr-zero) -- (xr);
        \draw (xr-zero) -- (zero);
        \node [below left] at (inf) {\scriptsize $P_\infty$};
        \node [above left] at (one) {\scriptsize $P_1$};
        \node [below] at (midint) {\scriptsize $P_\cmexp$};
        \node [above right] at (xr) {\scriptsize $P_\xr$};
        \node [below right] at (zero) {\scriptsize $P_0$};
        
        \coordinate (deg-fusion) at ($ 0.3*(xr-zero) + 0.7*(xr)$);
        \coordinate (deg) at ($ (deg-fusion) + (2,-2) $);
        \coordinate (deg-fusion-label) at ($0.5*(xr-zero) + 0.5*(deg-fusion)$);
        \draw [dashed] (deg) -- (deg-fusion);
        \node [below right] at (deg) {\scriptsize $P_{\drep{2}{1}}(z)$}; 
        \node [above left] at (deg-fusion-label) {\scriptsize $P_\xr + \frac{\sign \,b}{2}$};
    \end{tikzpicture}
\end{equation}
Here $\epsilon = \pm 1$ to succinctly account for the fusion relations~\eqref{eq:Lfusion}. We will use $P_\sigma$ to denote the momentum in the internal leg. This sets up our notation for the blocks in question.

By considering the OPE limits, $z\to 0$ and $z\to x$, respectively, we can simplify the above, and recover a leading behavior,
\begin{equation}
\label{eq-deg-block-ope-s}
    \VB^{s,[0]} _\sign(z,\xr) \quad \xrightarrow{z \to 0} \quad z^{\frac{1+b^2}{2} - \sign\, b\, P_0} 
        \begin{tikzpicture}[scale= .5, baseline = .5ex]
        \coordinate (inf) at (-2,-2);
        \coordinate (one) at (-2,2);
        \coordinate (inf-one) at (0,0);
        \coordinate (xr-zero) at (4,0);
        \coordinate (midint) at (2,0);
        \coordinate (xr) at (6,2);
        \coordinate (zero) at (6,-2);
        \draw (inf) -- (inf-one);
        \draw  (one) -- (inf-one);
        \draw (inf-one) -- (xr-zero);
        \draw (xr-zero) -- (xr);
        \draw (xr-zero) -- (zero);
        \node [below left] at (inf) {\scriptsize $P_\infty$};
        \node [above left] at (one) {\scriptsize $P_1$};
        \node [below] at (midint) {\scriptsize $P_\cmexp$};
        \node [above right] at (xr) {\scriptsize $P_\xr$};
        \node [below right] at (zero) {\scriptsize $P_0 + \frac{\sign\, b}{2}$};
    \end{tikzpicture}
\end{equation}
\begin{equation}
    \VB^{s,[\xr]} _\sign(z,\xr)\quad  \xrightarrow{z \to \xr} \quad  (z-\xr)^{\frac{1+b^2}{2} - \sign \,b \,P_\xr} 
        \begin{tikzpicture}[scale= .5, baseline = .5ex]
        \coordinate (inf) at (-2,-2);
        \coordinate (one) at (-2,2);
        \coordinate (inf-one) at (0,0);
        \coordinate (xr-zero) at (4,0);
        \coordinate (midint) at (2,0);
        \coordinate (xr) at (6,2);
        \coordinate (zero) at (6,-2);
        \draw (inf) -- (inf-one);
        \draw  (one) -- (inf-one);
        \draw (inf-one) -- (xr-zero);
        \draw (xr-zero) -- (xr);
        \draw (xr-zero) -- (zero);
        \node [below left] at (inf) {\scriptsize $P_\infty$};
        \node [above left] at (one) {\scriptsize $P_1$};
        \node [below] at (midint) {\scriptsize $P_\cmexp$};
        \node [above right] at (xr) {\scriptsize $P_\xr + \frac{\sign \,b}{2}$};
        \node [below right] at (zero) {\scriptsize $P_0 $};
    \end{tikzpicture}
\end{equation}
We already see the features of interest, viz., there are two scaling behaviors in this OPE limit as $z\to\{0,x\}$, which should be reminiscent of the local Frobenius basis of Heun's equation.

To make the connection precise, let us further recall that the degenerate blocks in the two channels are related by the degenerate fusion matrix,
\begin{equation}\label{eq-deg-block-fusion-transformation-s}
\begin{split}
\VB^{s,[0]} _\sign(z,\xr) 
&= 
    \sum_{\sign^\prime = \pm} \,\Fus_{\sign\sign^\prime}\prn{P_0,P_x,P_\sigma} \,\VB^{s,[x]} _{\sign^\prime}(z,\xr) \,,
\end{split}
\end{equation}
where the fusion matrix $\Fus_{\sign\sign^\prime}$ is as defined in~\eqref{eq-deg-block-fusion-transformation}.
The final piece of data we need is that these degenerate blocks satisfy a BPZ decoupling equation (in any channel). This reads
\begin{equation}\label{eq-BPZ}
    \prnbig{b^{-2} \,\partial^2_z + \ldiff_{-2}} \VB(z,\xr) = 0\,,
\end{equation}
with the differential operator $\ldiff_{-2}$ being given by  
\begin{equation}\label{eq:L2op}
\begin{split}
\ldiff_{-2} 
&= 
    \frac{h_0}{z^2} + \frac{h_1}{(z-1)^2} + \frac{h_\xr}{(z-\xr)^2} + \frac{h_\infty - h_{\drep{2}{1}} - h_0 - h_1 - h_\xr}{z\,(z-1)} \\
&\qquad \quad
    + \frac{\xr\, (\xr -1)}{z\,(z-1)\,(z-\xr)}\partial_\xr 
     - \prn{ \frac{1}{z} + \frac{1}{z-1}} \partial_z \,.
\end{split}
\end{equation}

We now have all the ingredients necessary for relating the connection matrix of Heun's differential equation to these degenerate conformal blocks. The final step involves taking a suitable semiclassical limit. 

\paragraph{The connection formula from  semiclassical limit:}   Recall the previously introduced semiclassical limit where we scale $b\to0$,  and let the momenta diverge as $b^{-1}$, cf.~\eqref{eq-quasiclassical-limit}.
The notation in~\eqref{eq-quasiclassical-limit} is suggestively intended to make the connection with~\eqref{eq-heun-normal-form} transparent. 
In such semiclassical limit, there are two important simplifications: Firstly, the degenerate blocks are expected to obey heavy-light factorization. Secondly, the heavy part is expected to exponentiate in the central charge $c$. Altogether, this implies
\begin{equation}\label{eq-deg-block-CL-s}
    \VB^{s,[i]} _\sign(z,\xr)
    \;\;  \xrightarrow{\CL} \;\; 
    \wfnorm^{[i]}_\sign \, \wf^{[i]}_{\sign}(z) \;\exp\brk{b^{-2} \,\VBcl^s(x)}, \quad i=0,x \,.
\end{equation}
Here $\VBcl^s$ is the classical Virasoro block in $s$-channel, viz.,   
\begin{equation}
     \VBcl^s(x) \coloneqq \begin{tikzpicture}[scale= .5, baseline = .5ex]
        \coordinate (inf) at (-2,-2);
        \coordinate (one) at (-2,2);
        \coordinate (inf-one) at (0,0);
        \coordinate (xr-zero) at (4,0);
        \coordinate (midint) at (2,0);
        \coordinate (xr) at (6,2);
        \coordinate (zero) at (6,-2);
        \draw (inf) -- (inf-one);
        \draw  (one) -- (inf-one);
        \draw (inf-one) -- (xr-zero);
        \draw (xr-zero) -- (xr);
        \draw (xr-zero) -- (zero);
        \node [below left] at (inf) {\scriptsize $\mexp_\infty$};
        \node [above left] at (one) {\scriptsize $\mexp_1$};
        \node [below] at (midint) {\scriptsize $\cmexp$};
        \node [above right] at (xr) {\scriptsize $\mexp_\xr$};
        \node [below right] at (zero) {\scriptsize $\mexp_0$};
    \end{tikzpicture} 
\end{equation}
The normalization factor $\wfnorm^{[i]}_\sign$ comes from considering the OPE limits~\eqref{eq-deg-block-ope-s} and taking into account the shift in conformal weights due to fusion with the light degenerate operator, 
\begin{equation}
 \wfnorm^{[i]}_\sign = \exp\brk{\frac{\sign}{2} \,\partial_{\theta_i} \VBcl^s(x)} \,.
\end{equation}

We are now in a position to state the connection formula. Note that the wave functions $\wf^{[i]}_\sign(z)$, with exponents $\half - \sign\, \theta_i$ in the semiclassical limit, can be identified with the normalized Frobenius solutions of Heun's equation. This is straightforward to see from the  OPE limit~\eqref{eq-deg-block-ope-s}.  Moreover, one recovers Heun's equation~\eqref{eq-heun-normal-form} to leading order, $\order{b^{-2}}$, upon substituting the classical limit~\eqref{eq-deg-block-CL-s} into BPZ equation~\eqref{eq-BPZ}.  In this limit,  the accessory parameter is identified to be related to the classical Virasoro block, viz.,
\begin{equation}\label{eq:Zamrel}
    \acc = x \,\partial_x \VBcl^s(x).
\end{equation}
This is sometimes referred to as the \emph{Zamolodchikov relation}.

The connection formula for $\wf^{[i]}_\sign(z)$ can then be derived from taking the semiclassical limit of \eqref{eq-deg-block-fusion-transformation-s}. The degenerate fusion matrix has an obvious classical limit using~\eqref{eq:Fdegfusion},
\begin{equation}
    \Fus_{\sign\sign^\prime}\prn{P_0,P_x,P_\cmexp} 
    \quad\xrightarrow{\CL} \quad \Fuscl_{\sign\sign^\prime}\prn{\mexp_0,\mexp_x,\cmexp} 
    = \frac{\Gamma\prn{1-2 \,\sign\, \mexp_0} \, \Gamma\prn{2\, \sign^\prime \,\mexp_\xr}}{\Gamma\prn{\half - \sign\, \mexp_0 + \sign^\prime \,\mexp_\xr \pm \cmexp}}.
\end{equation}
From this expression, we arrive at our sought for $s$-channel connection formula 
\begin{empheq}[box=\fbox]{equation}\label{eq:schannel}
\begin{split}
\connm_{\sign \sign^\prime} 
&=  
    \Fuscl_{\sign\sign^\prime}\prn{\mexp_0,\mexp_x,\cmexp} \,\frac{\wfnorm^{[x]}_{\sign^\prime}}{\wfnorm^{[0]}_{\sign}} \\
&= 
    \frac{\Gamma\prn{1-2 \,\sign\, \mexp_0} \Gamma\prn{2 \,\sign^\prime \,\mexp_\xr}}{\Gamma\prn{\half - \sign\, \mexp_0 + \sign^\prime\, \mexp_\xr \pm \cmexp}} \exp\brk{\half \prn{\sign^\prime\, \partial_{\mexp_x} - \sign \,\partial_{\mexp_0} } \VBcl^s(x) } \,.
\end{split}
\end{empheq}
The key point to note is that the Liouville momentum $\sigma$ appearing in the connection matrix is only defined implicitly through the Zamolodchikov relation~\eqref{eq:Zamrel}. The connection matrix is clearly meromorphic; the poles arise from the Gamma functions. To understand the structure of this formula, we need to have a better handle on $\sigma$. We therefore turn to take a closer look at the Zamolodchikov relation.

\paragraph{Inverting the Zamolodchikov relation using semiclassical Virasoro block:} 
Since the Zamolodchikov relation gives us a link between the accessory parameter and the classical Virasoro block, we can use it to obtain  $\cmexp$ as an expansion in $\xr$. To this end, introduce
\begin{equation}
    \cmexp = \cmexpope + \sum^\infty_{k=1} \cmexp_k \,\xr^k \,, \qquad  \cmexp^{(n)} \coloneqq \cmexpope + \sum^n_{k=1} \cmexp_k \,x^k \,,
\end{equation}
where $\cmexpope$  is the leading Liouville momentum. It is fixed by the leading OPE term in the classical Virasoro block
\begin{equation}
    \cmexpope^2 = \mexp^2_0 + \mexp^2_x - \acc - \frac{1}{4} \,.
\end{equation}
From the Zamolodchikov relation, we further have,
\begin{equation}
    \cmexp^2 = \cmexpope^2 + \sum^\infty_{k=1}\, k \,\VBcl_k(\vb*{\mexp},\cmexp) \,x^{k} \,.
\end{equation}
This can be exploited to obtain the following recursion relation for $\cmexp_k$:
\begin{equation}
    \cmexp_n = (2\cmexpope)^{-1} \prn{ \sum^n_{k=1} \,k\, \VBcl_k\prn{\vb*{\mexp},\cmexp^{(n-1)}} \, x^k \bigg|_{\text{coeff}[x^n]} - \sum^{n-1}_{k=1}\, \cmexp_k \,\cmexp_{n-k} } ,
\end{equation}
where $\big|_{\text{coeff}[x^n]}$ denotes taking coefficient of $x^n$. For instance, first two terms read
\begin{equation}\label{eq:sigmasol12}
\begin{split}
    \cmexp_1 &= 
        \frac{\VBcl^s_1(\vb*{\mexp},\cmexpope)}{2 \,\cmexpope} \,, \\
    \cmexp_2 &= 
        -\frac{\VBcl^s_1(\vb*{\mexp},\cmexpope)^2 - 8 \,\cmexpope^2 \,\VBcl^s_2(\vb*{\mexp},\cmexpope) - 2 \,\cmexpope \,\VBcl^s_1(\vb*{\mexp},\cmexpope) \,\partial_\cmexp \VBcl^s_1(\vb*{\mexp},\cmexpope)}{8\, \cmexpope^3} \,.
\end{split}
\end{equation}

The upshot is that the internal classical Liouville momentum $\cmexp$ is entirely fixed by external Liouville momenta, accessory parameter and Virasoro block data. We will use this in our analysis below to deduce properties of the thermal 2-point function. 

To wrap up our discussion of the $s$-channel connection formula, let us record the behavior in the OPE limit $\xr \to 0$. We simply retain the leading terms in the above to find 
\begin{equation}
\begin{split}
    \cmexp  & \to 
        \cmexpope \,, \\
    \connm_{\sign \sign^\prime} 
    & \to 
        \Fusclval{\sign}{\sign^\prime}{\mexp_0}{\mexp_\xr}{\cmexpope} \xr^{-\sign\, \mexp_0 +\,\sign^\prime\, \mexp_\xr}\,.
\end{split}
\end{equation}

The $s$-channel connection formula~\eqref{eq:schannel} is a particularly simple expression, especially when we compare with the $t$-channel expression, which we turn to next.

\subsection{The connection formula in the \texorpdfstring{$t$}{t}-channel expansion}

While we already have the result using the $s$-channel conformal blocks, we can equivalently derive an expression in the $t$-channel.  The $t$-channel expression will be useful for comparing with the literature, and does have its own merits in application to black hole perturbation problems. The logic, however, is similar to the above, where we make use of the semiclassical limit of a degenerate Virasoro block.

To obtain the result we seek, we again consider the following three five-point degenerate Virasoro blocks:
\begin{equation}
    \VB^{t,[0]}_\sign (z,x)\coloneqq 
        \begin{tikzpicture}[scale= .5, baseline = -5.5ex]
        \coordinate (xr) at (2,2);
        \coordinate (one) at (-2,2);
        \coordinate (one-xr) at (0,0);
        \coordinate (inf-zero) at (0,-4);
        \coordinate (midint) at (0,-2);
        \coordinate (inf) at (-2,-6);
        \coordinate (zero) at (2,-6);
        \draw (one) -- (one-xr);
        \draw  (xr) -- (one-xr);
        \draw (one-xr) -- (inf-zero);
        \draw (inf-zero) -- (inf);
        \draw (inf-zero) -- (zero);
        \node [below left] at (inf) {\scriptsize $P_\infty$};
        \node [above left] at (one) {\scriptsize $P_1$};
        \node [left] at (midint) {\scriptsize $P_\cmexp$};
        \node [above right] at (xr) {\scriptsize $P_\xr$};
        \node [below right] at (zero) {\scriptsize $P_0$};
        
        \coordinate (deg-fusion) at ($ 0.3*(inf-zero) + 0.7*(zero)$);
        \coordinate (deg) at ($ (deg-fusion) + (2,2) $);
        \coordinate (deg-fusion-label) at ($0.5*(inf-zero) + 0.5*(deg-fusion)$);
        \draw [dashed] (deg) -- (deg-fusion);
        \node [above right] at (deg) {\scriptsize $P_{\drep{2}{1}}(z)$}; 
        \node [right] at (inf-zero) {\scriptsize $P_0 + \frac{\sign\, b}{2}$};
    \end{tikzpicture}
\end{equation}
\begin{equation}
    \VB^{t,[\sigma]}_\sign (z,x) \coloneqq 
    \begin{tikzpicture}[scale= .5, baseline = -5.5ex]
        \coordinate (xr) at (2,2);
        \coordinate (one) at (-2,2);
        \coordinate (one-xr) at (0,0);
        \coordinate (inf-zero) at (0,-4);
        \coordinate (midint) at (0,-2);
        \coordinate (inf) at (-2,-6);
        \coordinate (zero) at (2,-6);
        \draw (one) -- (one-xr);
        \draw  (xr) -- (one-xr);
        \draw (one-xr) -- (inf-zero);
        \draw (inf-zero) -- (inf);
        \draw (inf-zero) -- (zero);
        \node [below left] at (inf) {\scriptsize $P_\infty$};
        \node [above left] at (one) {\scriptsize $P_1$};
        \node [above left] at ($(midint) + (0,0.5)$) {\scriptsize $P_\cmexp$};
        \node [above right] at (xr) {\scriptsize $P_\xr$};
        \node [below right] at (zero) {\scriptsize $P_0$};

        \coordinate (deg) at ($ (midint) + (2.5,0) $);
        \draw [dashed] (deg) -- (midint);
        \node [right] at (deg) {\scriptsize $P_{\drep{2}{1}}(z)$}; 
        \node [below left] at ($(midint) - (0,0.5)$) {\scriptsize $P_\cmexp + \frac{\sign \,b}{2}$};
    \end{tikzpicture}
\end{equation}
\begin{equation}
    \VB^{t,[x]}_{\sign,\sign^\prime} (z,x) \coloneqq 
    \begin{tikzpicture}[scale= .5, baseline = -5.5ex]
        \coordinate (xr) at (2,2);
        \coordinate (one) at (-2,2);
        \coordinate (one-xr) at (0,0);
        \coordinate (inf-zero) at (0,-4);
        \coordinate (midint) at (0,-2);
        \coordinate (inf) at (-2,-6);
        \coordinate (zero) at (2,-6);
        \draw (one) -- (one-xr);
        \draw  (xr) -- (one-xr);
        \draw (one-xr) -- (inf-zero);
        \draw (inf-zero) -- (inf);
        \draw (inf-zero) -- (zero);
        \node [below left] at (inf) {\scriptsize $P_\infty$};
        \node [above left] at (one) {\scriptsize $P_1$};
        \node [left] at (midint) {\scriptsize $P_\cmexp + \frac{\sign b}{2}$};
        \node [above right] at (xr) {\scriptsize $P_\xr$};
        \node [below right] at (zero) {\scriptsize $P_0$};
        
        \coordinate (deg-fusion) at ($ 0.3*(one-xr) + 0.7*(xr)$);
        \coordinate (deg) at ($ (deg-fusion) + (2,-2) $);
        \draw [dashed] (deg) -- (deg-fusion);
        \node [below right] at (deg) {\scriptsize $P_{\drep{2}{1}}(z)$}; 
        \node [right] at (one-xr) {\scriptsize $P_\xr + \frac{\sign^\prime b}{2}$};
    \end{tikzpicture}
\end{equation}
These degenerate blocks are related by the following fusion transformations
\begin{equation}
\begin{split}
    &\VB^{t,[0]}_\sign = \sum_{\signconv = \pm}  \Fus_{\sign \signconv}\prn{P_0,P_\cmexp,P_\infty} \VB^{t,[\cmexp]}_\signconv \,,\\
    &\VB^{t,[\cmexp]}_\signconv = \sum_{\sign^\prime = \pm} \Fus_{\Bar{\signconv}\sign^\prime}\prn{P_\cmexp + \frac{\signconv\, b}{2}, P_x, P_1}\, \VB^{t,[x]}_{\signconv,\sign^\prime} \,.
\end{split}
\end{equation}
In the second fusion, the convention relating the blocks is $\Bar{\signconv} = - \signconv$. Putting the two together, we can relate the fusion between the degenerate operator and $P_0$ to that between the operator and $P_x$. All told, this gives
\begin{equation}
\label{eq-deg-block-fusion-transformation-t}
    \VB^{t,[0]}_\sign = \sum_{\signconv,\sign^\prime = \pm} \,
    \Fus_{\sign \signconv}\prn{P_0,P_\cmexp,P_\infty} \,\Fus_{\Bar{\signconv}\sign^\prime}\prn{P_\cmexp + \frac{\signconv\, b}{2}, P_x, P_1} \VB^{t,[x]}_{\signconv,\sign^\prime}\,.
\end{equation}

The connection formula we seek is derived by taking the semiclassical limit of~\eqref{eq-deg-block-fusion-transformation-t}. In the semiclassical limit, exploiting the heavy-light factorization and exponentiation of Virasoro lock, we obtain
\begin{equation}
\begin{split}
\VB^{t,[0]}_\sign (z,x) 
&\;\; \xrightarrow{\CL} \;\; 
    \wfnorm^{[0]}_\sign\, \wf^{[0]}_{\sign}(z)\;
    \exp\brk{b^{-2} \,\VBcl^t(x)} \,,\\
\VB^{t,[x]}_{\signconv,\sign^\prime}(z,x) 
&\;\; \xrightarrow{\CL}\;\; 
    \wfnorm^{[x]}_{\signconv,\sign^\prime} \,
    \wf^{[x]}_{\sign^\prime}(z) \;\exp\brk{b^{-2}\, \VBcl^t(x)} \,.
\end{split}   
\end{equation}
Now $\VBcl^t$ is the $t$-channel semiclassical Virasoro block, 
\begin{equation}
\VBcl^t(x) \coloneqq 
    \begin{tikzpicture}[scale= .5, baseline = -5.5ex]
        \coordinate (xr) at (2,2);
        \coordinate (one) at (-2,2);
        \coordinate (one-xr) at (0,0);
        \coordinate (inf-zero) at (0,-4);
        \coordinate (midint) at (0,-2);
        \coordinate (inf) at (-2,-6);
        \coordinate (zero) at (2,-6);
        \draw (one) -- (one-xr);
        \draw  (xr) -- (one-xr);
        \draw (one-xr) -- (inf-zero);
        \draw (inf-zero) -- (inf);
        \draw (inf-zero) -- (zero);
        \node [below left] at (inf) {\scriptsize $\mexp_\infty$};
        \node [above left] at (one) {\scriptsize $\mexp_1$};
        \node [left] at (midint) {\scriptsize $\cmexp$};
        \node [above right] at (xr) {\scriptsize $\mexp_\xr$};
        \node [below right] at (zero) {\scriptsize $\mexp_0$};       
    \end{tikzpicture}
\end{equation}
The normalization factors are determined by considering OPE limits similar to the $s$-channel case, and are given by
\begin{equation}
\wfnorm^{[0]}_\sign = \exp\brk{\frac{\sign}{2} \, \partial_{\mexp_0} \VBcl^t(x)}, \qquad \wfnorm^{[x]}_{\signconv,\sign^\prime} = \exp\brk{\half \prn{\signconv \,\partial_\cmexp + \sign^\prime\, \partial_{\mexp_x}} \VBcl^t(x)} .
\end{equation}

The degenerate blocks $\VB(z,x)$ satisfy the BPZ equation in any channel. Therefore, for the same reason as in the $s$-channel case, the wave functions $\wf^{[i]}_{\sign}$ can be identified as normalized Frobenius solutions of Heun's equation. The accessory parameter is still identified via Zamolodchikov relation
\begin{equation}
    \acc = x \,\partial_x \VBcl^t(x)\,.
\end{equation}

Armed with this data, one then reads off the $t$-channel connection formula by taking the classical limit of~\eqref{eq-deg-block-fusion-transformation-t}. This results in 
\begin{empheq}[box=\fbox]{equation}\label{eq:tchannel}
\begin{split}
\connm_{\sign \sign^\prime} 
&= 
    \sum_{\signconv = \pm} \Fuscl_{\sign \signconv}\prn{\mexp_0,\cmexp,\mexp_\infty}\,  \Fuscl_{\Bar{\signconv}\sign^\prime}\prn{\cmexp,\mexp_\xr,\mexp_1}\, \frac{\wfnorm^{[x]}_{\signconv,\sign^\prime}}{\wfnorm^{[0]}_\sign} \,,\\
&= 
    \sum_{\signconv = \pm} \Fusclval{\sign}{\signconv}{\mexp_0}{\cmexp}{\mexp_\infty} \Fusclval{\Bar{\signconv}}{\sign^\prime}{\cmexp}{\mexp_\xr}{\mexp_1} \\
&\qquad \qquad    
    \times \exp\brk{\half \prn{\signconv \,\partial_\cmexp + \sign^\prime \,\partial_{\mexp_x} - \sign\, \partial_{\mexp_0}} \VBcl^t(x)} .
\end{split}
\end{empheq}
Once again, the internal classical Liouville momentum $\cmexp$ appearing in the connection matrix is defined via the Zamolodchikov relation. This is a slightly more involved expression than the $s$-channel counterpart~\eqref{eq:schannel} because we had to convolve two successive fusion transformations. 

Finally, let us record the behavior in the $t$-channel OPE limit $x \to 1$. Now the internal Liouville momentum simplifies to 
\begin{equation}\label{eq:topecmexp}
    \cmexp \to \cmexpope, \qquad \cmexpope^2 = \mexp^2_1 + \mexp^2_x + (1-\xr)\,\acc - \frac{1}{4} \,.
\end{equation}
We note that $(1-\xr)\acc$ is finite as per the convention in~\eqref{eq-heun-normal-form}. 

When $\Re(\cmexpope) \neq 0$, the sum over signs $\signconv$ in $\connm_{\sign \sign^\prime}$ simplifies. The leading OPE part of $t$-channel block $\VBcl^t(x)$ gives a factor $(1-x)^{-\signconv \cmexpope}$. Choosing w.l.o.g.~a sign convention such that $\Re(\cmexpope)>0$, the leading OPE contribution comes from the $\signconv = +$ term. One finds  
\begin{equation}
    \connm_{\sign \sign^\prime} \;\to\; \dfrac{\Gamma(1-2\, \sign\, \mexp_0) \Gamma(2\, \cmexpope)}{ \Gamma\prn{\half - \sign \,\mexp_0 + \cmexpope \pm \mexp_\infty}} \,
    \dfrac{\Gamma(1+2\,\cmexpope)\Gamma(2 \,\sign^\prime \,\mexp_\xr)}{ \,\Gamma\prn{\half + \cmexpope + \sign^\prime \,\mexp_\xr \pm \mexp_1}} (1-\xr)^{-\cmexpope \,+ \,\sign^\prime\, \mexp_\xr} + \order{(1-x)^\cmexpope} \,.
\end{equation}

\subsection{The connection formula in the five-punctured case}

We will also have occasion to need the generalization of~\eqref{eq-heun-normal-form} to five punctures. In this case, the normal form for the function $\Tcl(z)$ is 
\begin{equation}\label{eq:T5punctures}
\begin{split}
 \Tcl(z) &= 
    \frac{\hcl_0}{z^2} + \frac{\hcl_1}{(z-1)^2} + \frac{\hcl_\xr}{(z-\xr)^2} +  \frac{\hcl_\xrp}{(z-\xrp)^2} + \frac{\hcl_{\infty} - \hcl_0 - \hcl_1 - \hcl_{\xr} - \hcl_\xrp}{z\,(z-1)}  \\ 
& \qquad \quad 
    + \frac{(\xr-1)\, \acc_\xr}{z\,(z-1)\,(z-\xr)} + \frac{(\xrp-1) \,\acc_\xrp}{z\,(z-1)\,(z-\xrp)}\,.
\end{split}
\end{equation}
We are still interested in the connection problem between $0, \xr$, which is as defined earlier. As explained in~\cite{Lisovyy:2022flm}, thanks to locality of fusion transformation, the derivation of $s$-channel connection formula in the previous section can be immediately generalized to yield
\begin{empheq}[box=\fbox]{equation}
\begin{split}
\connm_{\sign \sign^\prime} 
 = 
    \dfrac{\Gamma\prn{1-2 \,\sign\, \mexp_0} \Gamma\prn{2\, \sign^\prime \,\mexp_\xr}}{\Gamma\prn{\half - \sign \,\mexp_0 + \sign^\prime\, \mexp_\xr \pm \cmexp}} \exp\brk{\half \prn{\sign^\prime\, \partial_{\mexp_x} - \sign \,\partial_{\mexp_0} } \VBcl(\xr,\xrp) } 
\end{split}
\end{empheq}
The semiclassical Virasoro block encountered here is 
\begin{equation}
    \VBcl(\xr,\xrp) \coloneqq 
    \begin{tikzpicture}[scale= .5, baseline = .5ex]
        \coordinate (inf) at (-2,-2);
        \coordinate (one) at (-2,2);
        \coordinate (inf-one) at (0,0);
        \coordinate (xr-zero) at (4,0);
        \coordinate (midint) at (2,0);
        \coordinate (xr) at (6,2);
        \coordinate (zero) at (6,-2);
        \draw (inf) -- (inf-one);
        \draw  (one) -- (inf-one);
        \draw (inf-one) -- (xr-zero);
        \draw (xr-zero) -- (xr);
        \draw (xr-zero) -- (zero);
        \node [below left] at (inf) {\scriptsize $\mexp_\infty$};
        \node [above left] at (one) {\scriptsize $\mexp_\xrp$};
        \node [above right] at (xr) {\scriptsize $\mexp_\xr$};
        \node [below right] at (zero) {\scriptsize $\mexp_0$};
        \node [below right] at ($(midint) + (0.5,-.2)$) {\scriptsize $\cmexp$};
        \node [below left] at ($(midint) - (.5,0)$) {\scriptsize $\cmexpp$};
        \coordinate (heavy-deg) at ($(midint) + (0,2)$);
        \draw (midint) -- (heavy-deg);
        \node [above] at (heavy-deg) {\scriptsize $\mexp_1$};
    \end{tikzpicture},
\end{equation}
Now the accessory parameters $\cmexp,\cmexpp$ are defined implicitly by
\begin{equation}
    \acc_\xr = \xr\, \partial_\xr \VBcl(\xr,\xrp), \qquad 
    \acc_\xrp = \xrp \,\partial_\xrp \VBcl(\xr,\xrp).
\end{equation}
%

\subsection{The connection formula for an equation with an apparent singularity}

We have thus far examined Fuchsian equation with four or five punctures. In order to have a complete characterization of the  energy-momentum tensor correlation functions, we need one more ingredient: an analysis of equations with an apparent singular point or puncture. We recall that an apparent puncture is a local where the indicial exponents of the two Frobenius solutions are integers. Naively, one expects a logarithmic branch of solutions. If this is absent, owing to a fine-tuning of the parameters, then we have an apparent puncture. Such a locus is, in fact, an ordinary point, with  Taylor series solutions and no monodromy. The wave equation for the scalar polarization of the energy-momentum tensor  in a neutral black hole background will turn out to be of this form. We will therefore generalize our considerations to include this case.

\paragraph{Apparent singularities and higher degenerate representations:} Consider the analog of an equation with five punctures, as discussed above. One of these punctures, which we designate as $\xrapp$, will be the apparent puncture. The normal form of the equation in this case will be as in~\eqref{eq:T5punctures}, which with some relabeling we write as 
\begin{equation}\label{eq-heun-as}
\begin{split}
\Tcl(z) 
&= 
    \frac{\hcl_0}{z^2} + \frac{\hcl_1}{(z-1)^2} + \frac{\hcl_\xr}{(z-\xr)^2} +  \frac{\hcl_\xrapp}{(z-\xrapp)^2} + \frac{\hcl_{\infty} - \hcl_0 - \hcl_1 - \hcl_{\xr} - \hcl_\xrapp}{z\,(z-1)} \\ 
&\qquad \quad
    + \frac{(\xr-1) \,\acc_\xr}{z\,(z-1)\,(z-\xr)} + \frac{(\xrapp-1) \,\acc_\xrapp}{z\,(z-1)\,(z-\xrapp)}\,.
\end{split}   
\end{equation}
The new feature is that we let $\mexp_\xrapp \in \intz /2$.

Generically, when the exponent at a singularity satisfies $\mexp_\xrapp \in \intz/2$, the local solution has logarithmic singularity and the local monodromy matrix takes a Jordan form. The special non-generic case where such singularity with $\mexp_\xrapp \in \intz/2$ has trivial monodromy is called apparent singularity. The condition for apparent singularity involves an algebraic relation between the weights and accessory parameters. The case with $\mexp_\xrapp = 3/2$ is the one relevant for the scalar polarization (sound channel) of the stress tensor.

To illustrate the apparent singularity condition explicitly, consider the following expansion of stress-tensor around $\xrapp$:
\begin{equation}
    \Tcl(z) = \sum^\infty_{m=0} \lop_{-m} \,(z-\xrapp)^{m-2} \,, 
    \qquad 
    \lop_{-m} \coloneqq \Res_{z = \xrapp} \brk{(z-\xrapp)^{1-m} \Tcl(z)}\,.
\end{equation}
The apparent singularity conditions for $\mexp_\xrapp = \asnum/2$ can be expressed in terms of these expansion coefficients $\lop_{-m}$. Explicit forms for the first few values of $\asnum$ are given in~\cref{tab-apparent-singularity-conditions}.
\begin{table}[H]
    \centering
    \begin{tabular}{|c|c|}
      \hline
       $\asnum$  &  $\asc_\asnum$ (=0)\\
       \hline\hline
       2 & $\lop^2_{-1} + \lop_{-2}$ \\
       \hline 
       3 & $\frac{1}{4}\, \lop^3_{-1} + \lop_{-1} \,\lop_{-2} 
       + \lop_{-3}$ \\
       \hline
       4 & $\frac{1}{36} \,\lop^4_{-1} + \frac{5}{18}\, \lop^2_{-1}\, \lop_{-2} + \frac{1}{4}\, \lop^2_{-2} + \frac{2}{3} \,\lop_{-1}\,\lop_{-3} + \lop_{-4}$ \\
       \hline      
    \end{tabular}
    \caption{The apparent singularity conditions (ASC) for $\mexp_\xrapp = \asnum/2$ tabulated for the first few non-trivial values of $\asnum$.}
    \label{tab-apparent-singularity-conditions}
\end{table}
To derive the connection formula in the presence of such an apparent singularity in the large-$c$ CFT approach, we need to know the corresponding heavy CFT operator at this locus. The answer is the following:
\begin{empheq}[box=\fbox]{equation}
\text{Apparent singularity with} \ \mexp_\xrapp = \frac{\asnum}{2} \longleftrightarrow \text{Insertion of} \ V_{\drep{1}{\asnum}}(\xrapp)
\end{empheq}
To justify this, a shortcut is to consider the fusion rule between the light probe $V_{\drep{2}{1}}$ and the heavy operator $V_{\drep{1}{\asnum}}$ \cite{Gaiotto:2011nm}:
\begin{equation}
    V_{\drep{2}{1}} \times V_{\drep{1}{\asnum}} = V_{\drep{2}{\asnum}}.
\end{equation}
As only one fusion is allowed, it must be that the monodromy is trivial. Alternatively, one can justify this by showing that the BPZ equation for $V_{\drep{1}{\asnum}}$ in the semiclassical limit reproduces the apparent singularity condition $\asc_\asnum$; see \cref{app-apparent-singularity} for more details.

\paragraph{Degenerate blocks with apparent punctures:} Now that we identified the presence of an apparent singularity with a higher degenerate operator, we can examine the BPZ equation. To do so  consider the following Virasoro blocks with insertion of both $V_{\drep{2}{1}}$ and $V_{\drep{1}{\asnum}}$:
\begin{equation}
    \VB^{[0]} _\sign(z,\xr,\xrapp) \coloneqq 
    \begin{tikzpicture}[scale= .5, baseline = .5ex]
        \coordinate (inf) at (-2,-2);
        \coordinate (one) at (-2,2);
        \coordinate (inf-one) at (0,0);
        \coordinate (xr-zero) at (4,0);
        \coordinate (midint) at (2,0);
        \coordinate (xr) at (6,2);
        \coordinate (zero) at (6,-2);
        \draw (inf) -- (inf-one);
        \draw  (one) -- (inf-one);
        \draw (inf-one) -- (xr-zero);
        \draw (xr-zero) -- (xr);
        \draw (xr-zero) -- (zero);
        \node [below left] at (inf) {\scriptsize $P_\infty$};
        \node [above left] at (one) {\scriptsize $P_1$};
        \node [above right] at (xr) {\scriptsize $P_\xr$};
        \node [below right] at (zero) {\scriptsize $P_0$};

        \node [above right] at ($(midint) + (0.5,0)$) {\scriptsize $P_\cmexp$};
        \node [above left] at ($(midint) - (0,0)$) {\scriptsize $P_\cmexp + \frac{i_\asnum}{b}$};
        
        \coordinate (deg-fusion) at ($ 0.3*(xr-zero) + 0.7*(zero)$);
        \coordinate (deg) at ($ (deg-fusion) + (2,2) $);
        \coordinate (deg-fusion-label) at ($0.5*(xr-zero) + 0.5*(deg-fusion)$);
        \draw [dashed] (deg) -- (deg-fusion);
        \node [above right] at (deg) {\scriptsize $P_{\drep{2}{1}}(z)$}; 
        \node [below left] at (deg-fusion-label) {\scriptsize $P_0 + \frac{\sign\, b}{2}$};
        
        \coordinate (heavy-deg) at ($(midint) + (0,2)$);
        \draw[dash dot] (midint) -- (heavy-deg);
        \node [above] at (heavy-deg) {\scriptsize $P_{\drep{1}{\asnum}}(\xrapp)$};
    \end{tikzpicture}
\end{equation}
\begin{equation}
    \VB^{[\xr]} _\sign(z,\xr,\xrapp) \coloneqq 
    \begin{tikzpicture}[scale= .5, baseline = .5ex]
        \coordinate (inf) at (-2,-2);
        \coordinate (one) at (-2,2);
        \coordinate (inf-one) at (0,0);
        \coordinate (xr-zero) at (4,0);
        \coordinate (midint) at (2,0);
        \coordinate (xr) at (6,2);
        \coordinate (zero) at (6,-2);
        \draw (inf) -- (inf-one);
        \draw  (one) -- (inf-one);
        \draw (inf-one) -- (xr-zero);
        \draw (xr-zero) -- (xr);
        \draw (xr-zero) -- (zero);
        \node [below left] at (inf) {\scriptsize $P_\infty$};
        \node [above left] at (one) {\scriptsize $P_1$};
        \node [above right] at (xr) {\scriptsize $P_\xr$};
        \node [below right] at (zero) {\scriptsize $P_0$};

        \node [below right] at ($(midint) + (0.5,0)$) {\scriptsize $P_\cmexp$};
        \node [below left] at ($(midint) - (0,0)$) {\scriptsize $P_\cmexp + \frac{i_\asnum}{b}$};
        
        \coordinate (deg-fusion) at ($ 0.3*(xr-zero) + 0.7*(xr)$);
        \coordinate (deg) at ($ (deg-fusion) + (2,-2) $);
        \coordinate (deg-fusion-label) at ($0.5*(xr-zero) + 0.5*(deg-fusion)$);
        \draw [dashed] (deg) -- (deg-fusion);
        \node [below right] at (deg) {\scriptsize $P_{\drep{2}{1}}(z)$}; 
        \node [above left] at (deg-fusion-label) {\scriptsize $P_\xr + \frac{\sign\, b}{2}$};

        \coordinate (heavy-deg) at ($(midint) + (0,2)$);
        \draw[dash dot] (midint) -- (heavy-deg);
        \node [above] at (heavy-deg) {\scriptsize $P_{\drep{1}{\asnum}}(\xrapp)$};
    \end{tikzpicture}
\end{equation}
Here $i_\asnum$ takes $s$ possible values $i_\asnum = -\frac{\asnum-1}{2}, \cdots, \frac{\asnum-1}{2}$ due to fusion rule of $V_{\drep{1}{\asnum}}$. Thanks to the locality of fusion transformation, the two Virasoro blocks are related by the same fusion matrix as before, viz.,
\begin{equation}
\label{eq-fusion-transformation-apparent-singularity}
    \VB^{[0]} _\sign(z,\xr,\xrapp) = \sum_{\sign^\prime = \pm} \Fus_{\sign\sign^\prime}\prn{P_0,P_x,P_\sigma} \, \VB^{[x]} _{\sign^\prime}(z,\xr,\xrapp)\,,
\end{equation}
where the fusion matrix $\Fus$ is the same as in \eqref{eq-deg-block-fusion-transformation}. The BPZ equation for $V_{\drep{2}{1}}$ now reads
\begin{equation}\label{eq-BPZ-as}
    \prnbig{b^{-2} \,\partial^2_z + \ldiff_{-2}} \VB(z,\xr,\xrapp) = 0\,,
\end{equation}
with the differential operator $\ldiff_{-2}$ now taking the form
\begin{equation}
\begin{split}
\ldiff_{-2} 
&= 
    \frac{h_0}{z^2} + \frac{h_1}{(z-1)^2} + \frac{h_\xr}{(z-\xr)^2} + \frac{h_\xrapp}{(z-\xrapp)^2}+ \frac{h_\infty - h_{\drep{2}{1}} - h_0 - h_1 - h_\xr -h_\xrapp}{z(z-1)} \\
& \qquad 
    + \frac{\xr\, (\xr -1)}{z\,(z-1)\,(z-\xr)}\partial_\xr + \frac{\xrapp\, (\xrapp -1)}{z\,(z-1)\,(z-\xrapp)}\partial_\xrapp 
    \quad - \prn{ \frac{1}{z} + \frac{1}{z-1}} \partial_z \,.
\end{split}
\end{equation}

\paragraph{The connection formula from the semiclassical limit:}  We can now take the classical limit of~\eqref{eq-fusion-transformation-apparent-singularity} to derive the connection formula between $0, \xr$. The exponentiation of Virasoro block and heavy-light factorization is expected to hold here, even with the additional insertion of heavy degenerate operator (as in the standard case). The classical limit of the blocks now reads
\begin{equation} \label{eq-deg-block-CL-as}
\VB^{[i]} _\sign(z,\xr,\xrapp) \;\; \xrightarrow{\CL} \;\; \wfnorm^{[i]}_\sign \,\wf^{[i]}_{\sign}(z) \, \exp\brk{b^{-2}\, \VBcl(\xr,\xrapp)}, \qquad i=0,\,x\,. 
\end{equation}
where the block in question is itself 
\begin{equation} 
 \VBcl(\xr,\xrapp) \coloneqq 
    \begin{tikzpicture}[scale= .5, baseline = .5ex]
        \coordinate (inf) at (-2,-2);
        \coordinate (one) at (-2,2);
        \coordinate (inf-one) at (0,0);
        \coordinate (xr-zero) at (4,0);
        \coordinate (midint) at (2,0);
        \coordinate (xr) at (6,2);
        \coordinate (zero) at (6,-2);
        \draw (inf) -- (inf-one);
        \draw  (one) -- (inf-one);
        \draw (inf-one) -- (xr-zero);
        \draw (xr-zero) -- (xr);
        \draw (xr-zero) -- (zero);
        \node [below left] at (inf) {\scriptsize $\mexp_\infty$};
        \node [above left] at (one) {\scriptsize $\mexp_1$};
        \node [above right] at (xr) {\scriptsize $\mexp_\xr$};
        \node [below right] at (zero) {\scriptsize $\mexp_0$};
        \node [below right] at ($(midint) + (0.5,0)$) {\scriptsize $\cmexp$};
        \node [below left] at ($(midint) - (0,0)$) {\scriptsize $\cmexp + i_\asnum$};
        \coordinate (heavy-deg) at ($(midint) + (0,2)$);
        \draw[dash dot] (midint) -- (heavy-deg);
        \node [above] at (heavy-deg) {\scriptsize $\mexp_{\drep{1}{\asnum}}(\xrapp)$};
    \end{tikzpicture} 
\end{equation}
The normalization factor, as before, is 
\begin{equation}
\wfnorm^{[i]}_\sign = \exp\brk{\frac{\sign}{2}\, \partial_{\theta_i} \VBcl(\xr,\xrapp)} \,.
\end{equation}
The identification of the wavefunctions $\wf^{[i]}_\sign$ with the normalized Frobenius solutions of the generalized Heun's equation~\eqref{eq-heun-as} continues to hold, and the Zamolodchikov relation is generalized to
\begin{equation}
    \acc_\xr = \xr \,\partial_\xr \VBcl(\xr,\xrapp), \qquad 
    \acc_\xrapp = \xrapp\, \partial_\xrapp \VBcl(\xr,\xrapp).
\end{equation}
The connection formula is then read off from the classical limit of~\eqref{eq-fusion-transformation-apparent-singularity} to be
\begin{empheq}[box=\fbox]{equation}\label{eq:conn-app-sing}
\begin{split}
\connm_{\sign \sign^\prime} 
&=  
    \Fuscl_{\sign\sign^\prime}\prn{\mexp_0,\mexp_x,\cmexp} \,\frac{\wfnorm^{[x]}_{\sign^\prime}}{\wfnorm^{[0]}_{\sign}} \\
&= 
    \frac{\Gamma\prn{1-2 \,\sign\, \mexp_0} \;\Gamma\prn{2 \,\sign^\prime \mexp_\xr}}{\Gamma\prn{\half - \sign\, \mexp_0 + \sign^\prime\, \mexp_\xr \pm \cmexp}} \exp\brk{\half \prn{\sign^\prime \,\partial_{\mexp_x} - \sign\, \partial_{\mexp_0} } \VBcl(\xr,\xrapp) } 
\end{split}
\end{empheq}
Now both $\cmexp$ and $i_\asnum$ need to be determined from the generalized Zamolodchikov relation.

\section{Exact results for thermal 2-point functions and QNMs in \texorpdfstring{\AdS{5}}{AdS5}}\label{sec:exactAdS5}

We are now ready to apply the technology reviewed in~\cref{sec-connection-formula-CFT} to obtain exact results for thermal 2-point functions of 4d holographic CFTs. These theories have a dual gravitational description in terms of Einstein-Hilbert gravity \AdS{5}. As a concrete example, one can view the results as being relevant for the thermal correlators of $\mathcal{N}=4$ SYM in the strong 't Hooft coupling regime. 

While we will consider the correlation functions of scalar operators of such CFTs, our primary interest is in understanding correlation functions of the energy-momentum tensor.  The stress tensor is dual to the bulk graviton. The latter has a universal low-energy description in terms of pure Einstein-Hilbert gravity in five dimensions, thanks to consistent truncation in string or supergravity compactifications. The results we derive for the stress tensor correlator are therefore are universally valid for the large class of (super)conformal holographic field theories. In this section, we will summarize the equations of interest, and relate them to the Fuchsian equations studied using 2d CFT techniques in~\cref{sec-connection-formula-CFT}. We illustrate the results with some examples along the way.

\subsection{Black hole wave equations}\label{sec:bhwave}

Since we reviewed the technology for carrying out real-time computations in holography, we will present the wave equations in a form adapted to that discussion. We will later convert them to the form amenable to using the connection formulae.  We also record these in general \AdS{d+1} black hole spacetimes, despite our focus on the $d=4$ case.

The basic wave equation is a variation of the minimally coupled scalar wave equation, given by,
\begin{equation}\label{eq:Mdesign}
\begin{split}
\frac{1}{r^\markov} \Dz_+  \left( r^\markov\, \Dz_+\varphi\right)+\left(\omega^2-\lambda_\Sigma\,f -  m^2\,r^2 f \right)\, \varphi =0\,.
\end{split}
\end{equation}
Here $\lambda_\Sigma$ is either $\ell(\ell+d-2)$ on $\mathbf{S}^{d-1}$ or $k^2$ on $\mathbb{R}^{d-1}$. For $\markov =d-1$ and $m^2 \neq 0$, we have a massive Klein-Gordon scalar, dual to a conformal primary of weight $\Delta = \frac{d}{2} + \sqrt{\frac{d^2}{4} +m^2\, \lads^2}$ of the boundary CFT.

The equation also captures polarizations of the stress tensor dual to components of the linearized graviton wave equations. 
\begin{itemize}[wide,left=0pt]
\item  The $\frac{1}{2}\, d\, (d-3)$ transverse tensor polarizations of stress tensor are obtained for $\markov = d-1$ and  $m^2=0$. 
\item The $d-2$ transverse vector polarizations of the stress tensor are obtained for $\markov = 1-d$ and  $m^2=0$. 
\end{itemize}
We also note that abelian conserved currents dual to bulk Maxwell fields are described by the equation with $m^2=0$; transverse vector polarizations have $\markov =d-3$, while scalar polarization has $\markov = 3-d$~\cite{Ghosh:2020lel}.
In the analysis below, we refer to the minimally coupled massive scalar as the Klein-Gordon scalar, and designate $\markov \neq d-1$ as the designer massless scalar.

The scalar polarization of the stress tensor (the energy density operator), on the other hand, satisfies on $\mathbb{R}^{d-1,1}$ the following equation~\cite{He:2022jnc}:
\begin{equation}\label{eq:Zsound}
\begin{split}
&r^{d-3}\,\Lk(r)^2 \, \Dz_+
        \left( \frac{1}{r^{d-3}\,\Lk(r)^2} \, \Dz_+  \MZ \right)    
            + \left(\omega^2 - k^2 f  \left[1-\frac{d\,(d-2)\, \rp^d}{r^{d-2}\, \Lk(r)}\right]\right) \MZ  =0 \,,\\
&\hspace{3cm}
\Lk(r) = 
    k^2 + \frac{d-1}{2}\, r^3\, f'(r)\,.
\end{split}
\end{equation}  
The corresponding equation for $\Sigma_{d-1} = \mathbf{S}^{d-1}$ is a bit more involved and can be found in~\cite{Kodama:2003jz}. For analogous equations in the case of charged black holes, see Appendix A of~\cite{Loganayagam:2022teq} in the case of $\mathbb{R}^{d-1,1}$ and~\cite{Kodama:2003kk} for the spherically symmetric case. 

\subsection{Heun's oper from radial wave equation}

The radial wave equation \eqref{eq:Mdesign} in $\text{AdS}_5$, with complexified radial coordinate, can be generically mapped to Heun's oper on Riemann sphere $\mathbb{P}^1$ with four regular punctures, reproduced below for convenience:
\begin{equation}
\begin{split}
     &\wf''(z) + \Tcl(z) \,\wf(z) = 0 \,,\\
     &\Tcl(z) = \frac{\delta_0}{z^2} + \frac{\delta_1}{(z-1)^2} + \frac{\delta_\xr}{(z-\xr)^2} +  \frac{\delta_{\infty} - \delta_0 - \delta_1 - \delta_\xr}{z\,(z-1)} + \frac{(\xr-1) \,\acc}{z\,(z-1)\,(z-\xr)} \,,\\
     & \delta_i = \frac{1}{4} - \mexp^2_i\,.
\end{split}
\end{equation}  
The punctures correspond to physical locations in the black hole geometry, such as the AdS boundary, the inner or outer horizons, curvature singularity, and additionally complex horizon loci due to complexification of radial coordinate. For a classification of the punctures, in the case of \SAdS{d+1} and Reissner-Nordstr\"om-\AdS{d+1} black holes, we refer the reader to~\cite{Loganayagam:2022teq}. In dimensions $d>4$ one would encounter a larger number of punctures, but the CFT method should in principle apply as it relies solely on locality of fusion transformation. As noted, for concreteness and simplicity, we will focus on the \AdS{5} case. 

To recast the radial equation \eqref{eq:Mdesign} to Heun's oper, we first pass to the static coordinates outside the horizon (viz., $t = v - \int \frac{dr}{r^2 \, f(r)}$). We then perform the following coordinate transformation:
\begin{equation}\label{eq:zxrmaps}
    z = \xr\; \frac{r^2 - \rp^2}{r^2 - \rm^2}, \qquad 
    \xr = \frac{\rcomplex^2 - \rm^2}{\rcomplex^2 - \rp^2}.
\end{equation}
Here $\rcomplex \in i \real$ is the complex horizon satisfying $f(\rcomplex) = 0$, and $\rpm$ are the outer and inner  horizons, respectively. The transformation applies to both the neutral and charged black holes with spherical or planar horizons. The uncharged case corresponds to $\rm =0$. The transformation recasts the radial equation \eqref{eq:Mdesign} into a second-order ODE in $z$ coordinate, $\Psi''(z) + p(z)\, \Psi'(z) + q(z) \,\Psi(z)$, with four regular singularities. The ODE is then transformed into the canonical Schrödinger form of Heun's oper by the standard transformation $\Psi(z) = \psi(z) \,e^{-\half \int^z p(z^\prime)\, dz^\prime}$, 
with $\Tcl(z) = q(z) - \frac{p^2(z)}{4} - \frac{p^\prime(z)}{2}$.


\paragraph{General features:} Before giving explicit expressions for Heun's parameters in various cases, let us take stock of some features. The physical meaning of the regular punctures differs between the charged and neutral black holes, but is independent of the horizon topology. For simplicity, in the case of the Klein-Gordon scalar with $\markov = 3$, one has   
\begin{equation}
    \{0, x, 1, \infty \} \sim \begin{cases}
        \{\rp,\, \bdy, \, \rcomplex, \,\curv \} & \text{uncharged black holes} \\  
        \{\rp ,\, \bdy, \,\rcomplex, \,\rm \} & \text{charged black holes}
    \end{cases}
\end{equation}
where $\curv$ denotes the curvature singularity. In particular, the curvature singularity does not appear as a puncture in Heun's oper for Klein-Gordon scalar around a charged \AdS{} black hole. However, for $\markov \neq 3$, we will indeed encounter an additional puncture in the charged case at the curvature singularity. 

With a choice of sign conventions, the local monodromy exponents or classical Liouville momenta can be shown to have the following general features:
\begin{equation}
    \mexp_{\rpm} \in i \omega \real_{\geq0}, \qquad \mexp_\bdy \in \real_{\geq0}, \qquad \mexp_{\rcomplex} \in \omega \real_{\geq0}, \qquad \mexp_\curv \in \real_{\geq0}\,.
\end{equation}
In the following, we will also denote $\mexp_{\rp} \equiv \mexp_\hor$. In particular,
\begin{equation}
    \mexp_\hor = \frac{i \,\omega}{4 \pi\, T}\,,
\end{equation}
where $T$ is the temperature of the black hole. The local solution with characteristic exponents $\half \pm \mexp_\hor$ corresponds to outgoing and ingoing waves at the (outer) horizon. The local solutions with $\half \pm \mexp_\bdy$ correspond to normalizable and non-normalizable modes at the boundary, respectively. For certain choices of matter, the  conformal dimension of the dual operator $\Delta$ can be an integer, in which case our assumption of $\theta_i \notin \mathbb{Z}/2$ breaks down. We will treat these cases with a suitable limit procedure (this will of import for energy-momentum tensor correlators). 

The angular or spatial momentum appears in the accessory parameter $\acc$, which is generically a degree two polynomial in frequency and momentum. The cross-ratio lies in the range
\begin{equation}
    x\in \left[ 1/2, 1 \right)
\end{equation}
For the neutral \SAdS{5} black hole $x=\half$. The $t$-channel OPE limit $x \to 1$ turns out to correspond to the small black hole or near-extremal limit. In particular, the $s$-channel OPE limit $x \to 0$ doesn't appear as a physical limit in the black hole perturbation problem.

In the following, we give explicit Heun's parameters in a few representative examples, to illustrate distinct features in the resulting Heun's opers to which the 2d CFT method will be universally applied. It is straightforward to obtain the Heun's parameters in more examples not listed below.

\paragraph{Planar black holes:}
For the planar black hole, i.e., in the large black hole limit, it is convenient to define the following dimensionless frequency and spatial momentum
\begin{equation}
    \freq = \frac{\omega}{2\,\rp}, \quad \mom = \frac{|\vb{k}|}{2\,\rp}.
\end{equation}

\noindent\emph{Neutral planar black hole:} In this case, the metric function $f(r) = 1 - \frac{\rp^4}{r^4}$ has two pairs of roots at $\rp$ and $\rcomplex = i \rp$, giving $x = \half$. The parameters for the associated Heun's opers of Klein-Gordon  and  the designer massless scalar are listed in~\cref{tab-heun-parameters-BB}. In particular, they have identical accessory parameters, only differing in $\mexp_\bdy(x)$ and $\mexp_\curv(\infty)$. We use the notation $\mexp(\cdot)$ to denote the location of the puncture with Liouville momentum $\mexp$.

\begin{table}[H]
    \centering
    \begin{tabular}{|c|c|c|c|c|c|c|}
         \hline
           & $\mexp_\hor(0)$ & $\mexp_\bdy(\xr)$ & $\mexp_{\rcomplex}(1)$ & $\mexp_\curv(\infty)$ & $x$ & $\acc$
         \\
         \hline
         Klein-Gordon scalar & $\frac{i \,\freq}{2}$ & $\frac{\Delta-2}{2}$ & $\frac{\freq}{2}$ & $0$ & $\frac{1}{2}$ & $\mom^2 - \freq^2$ \\
         \hline
         Designer massless scalar & - & $\frac{\abs{\markov+1}}{4}$ & - & $\frac{\abs{\markov-3}}{4}$ & - & - \\
         \hline
          
    \end{tabular}
    \caption{Heun's parameters for a massive Klein-Gordon scalar  and a designer massless scalar in an uncharged planar Schwarzschild-\AdS{5} black hole, with $\mexp(\cdot)$ used to denote the location of the puncture with Liouville momentum $\mexp$.}
    \label{tab-heun-parameters-BB}
\end{table}

\noindent\emph{Charged planar black hole:} The metric function $f(r)$ for the planar Reissner-Nordstr\"om-\AdS{5} has three pairs of roots. We can parameterize it as 
\begin{equation}
    f(r) = \frac{(r^2 - \rp^2)\,(r^2 - \rm^2)\,(r^2 - \rcomplex^2)}{r^6}, \qquad \rcomplex = i \,\ri \,, 
\end{equation}
with $\rpm, \ri \in \real_{\geq 0}$ related by
\begin{equation}
    \frac{\rm}{\rp} = \sqrt{- \frac{1}{2}+\sqrt{\frac{1}{4}+Q^2}}, \qquad 
    \frac{\ri}{\rp} = \sqrt{\frac{1}{2}+\sqrt{\frac{1}{4}+Q^2}} \,.
\end{equation}
Here $Q$ is a dimensionless parameter measuring the charge, and  extremal limit is attained for $\Qext = \sqrt{2}$. 

The cross-ratio $x$ is now given by
\begin{equation}
    \xr = \frac{\ri^2 + \rm^2}{\ri^2 + \rp^2} 
    \;\in\; \left[ 1/2, 1 \right) .
\end{equation}
We see that the near-extremal limit $Q \to \Qext$, therefore, corresponds to the $t$-channel OPE limit $\xr \to 1$.

For completeness, we list below the Heun's parameters for a probe massive Klein-Gordon scalar dual to a boundary operator of dimension $\Delta$ (through the usual relation $m^2 = \Delta(\Delta-4)$) 
\begin{equation}
\begin{split}
\mexp_\hor(0) 
&= 
    \abs{\frac{1}{\rp^2\, f^\prime(\rp)}} \, i\, \omega 
    = \frac{r^4_+}{(r^2_i + r^2_+)(r^2_+ - r^2_-)} \, i\,\freq \,, \\
\mexp_\bdy(\xr) 
&= 
    \frac{\Delta-2}{2}  \,, \\
\mexp_{\rcomplex}(1) 
&= 
    \abs{\frac{1}{\rcomplex^2  f^\prime(\rcomplex)}} \, \omega 
    = \frac{r_i^3\, \rp}{(r^2_i + r^2_+)\,(r^2_i + r^2_-)}\, \freq \,,\\
\mexp_{\rm}(\infty) 
&= 
    \abs{\frac{1}{\rm^2 \,f^\prime(\rm)}} \, i \,\omega 
    =  \frac{\rm^3 \,\rp}{(r^2_i + r^2_-)\,(r^2_+ - r^2_-)} \,
    i\,\freq \,,\\
\acc 
&= 
    \frac{1}{4\, (\rp^2 - \rm^2)} \brkbig{\rp^2\, \prn{4 \,\mom^2 - 4\, \freq^2 + m^2 +2} - \rm^2 \,(\Delta-2)^2 - \ri^2\, \prn{2 + m^2}} \,. 
\end{split}
\end{equation}
One can indeed verify that setting $Q = 0$ reproduces the Heun's parameters in neutral case. The singular behavior of $\acc$ as $Q \to \Qext$ is an artifact of the convention in Heun's oper \eqref{eq-heun-normal-form} as $x \to 1$.

\paragraph{Global black holes:} We can similarly recast the equations for global (i.e., spherically symmetric) black holes in \AdS{5}. For simplicity, we mainly focus on the Schwarzschild-\AdS{5} case.

\noindent\emph{Neutral spherical black hole.} In this case, the metric function reads
\begin{equation}
    f(r) = \frac{\lads^2}{r^2} \brk{ 1 + \frac{r^2}{\lads^2} - \frac{\rp^{2}}{r^{2}}
         \left( 1+ \frac{\rp^2}{\lads^2} \right) }.
\end{equation}
The $\lads \to 0$ limit recovers the planar black hole metric function. We set $\lads=1$ hereafter for convenience. The complex horizon is located at 
$\rcomplex = i\, \sqrt{1+ \rp^2}$, giving the cross-ratio
\begin{equation}
    \xr = \frac{1+\rp^2}{1+2\,\rp^2} \;\in\; \left[1/2, 1 \right).
\end{equation}
The small black hole limit $\rp \to 0$ therefore corresponds to $t$-channel OPE limit $\xr \to 1$, and the large black hole limit $\rp \to \infty$ corresponds to the planar black hole value $\xr = \half$.

Explicit expressions for the Heun's parameters for a  massive Klein-Gordon scalar are 
\begin{equation}
\begin{split}
\mexp_\hor(0) 
&= 
    \frac{\rp}{2+4\,\rp^2} \, i\,\omega \,,\\
\mexp_\bdy(\xr) 
&= 
    \frac{\Delta-2}{2}  \,,\\
\mexp_{\rcomplex}(1) 
&= 
    \frac{\sqrt{1+\rp^2}}{2+4\,\rp^2} \, \omega\,,\\
\mexp_\curv(\infty) 
&= 
    0\,,\\
\acc 
&= 
    \frac{\lambda_{\mathbf{S}^3} -\omega^2 - \Delta\,(\Delta -4) - 2}{4\,\rp^2}\,.
\end{split}
\end{equation}
The spherical harmonic eigenvalue is captured $ \lambda_{\mathbf{S}^3} = l(l+2)$, and the mass has been traded for the conformal dimension $\Delta$.  The $\rp \to \infty$ limit indeed recovers the planar black hole values. The singular behavior of $\acc$ as $\rp \to 0$ is once again an artifact of the convention in Heun's oper \eqref{eq-heun-normal-form} as $x \to 1$. 

\subsection{Universal exact expressions for QNMs and holographic thermal 2-point functions}

We are now ready to present an exact result for the holographic correlator at finite temperature. To this end, we recall from our review in~\cref{sec:QNMs} that the boundary 2-point function $\Gret(\omega)$ is given by the solution to a connection problem~\eqref{eq:Kdef}. This version of the result, which we have argued is justified by a more careful contour prescription, is well-adapted to our differential equation analysis. 

Let is start by noting that local Frobenius solutions at $z_\hor=0$ corresponds to ingoing and outgoing solutions, respectively, which behave as 
\begin{align}
    &\wf_{\ingo/\outgo}(z) = \wf^{[0]}_{\pm} (z) = 
    z^{\half \,\mp\, \mexp_\hor} \,\prn{1 + \cdots}.
\end{align}
On the other hand, the local Frobenius solutions at $z_\bdy = x$ correspond to the normalizable/non-normalizable solutions, and take the form
\begin{align}
    &\wf_{\nor/\nnor}(z) = \wf^{[\xr]}_{\mp} (z) = 
    (z-\xr)^{\half \,\pm\, \mexp_\bdy} \,\prn{1 + \cdots}
\end{align}
We have assumed $\mexp_\bdy \not\in \intz/2$ to keep the discussion generic.\footnote{For $\mexp_\bdy \in \intz/2$ we would have a logarithmic branch to the solution. It will prove convenient to arrive at the result in this case by deforming $\Delta$ or equivalently $\mexp_\bdy$.} 

Consider the connection problem between the (outer) horizon and boundary. Fixing the ingoing solutions at the horizon, and  employing the notation for connection matrix in~\cref{sec-connection-formula-CFT},  we have
\begin{equation}
    \wf_{\ingo}(z) 
    = \connm_{\ingo,\nor}\, \wf_{\nor}(z) + \connm_{\ingo,\nnor} \,\wf_{\nnor}(z) 
    = \connm_{+-} \,\wf_{\nor}(z) + \connm_{++} \,\wf_{\nnor}(z)   
\end{equation}
The retarded Greens function for holographic thermal two-point function and the quantization condition for QNMs are then defined by
\begin{equation}\label{eq:QNM-two-pt-func-def}
\begin{split}
    \Gret(\omega) 
    &\coloneqq 
    \holonorm\prn{\mexp_\bdy} 
    \frac{\connm_{\ingo,\nor}}{\connm_{\ingo,\nnor}} = \holonorm\prn{\mexp_\bdy} \, 
    \frac{\connm_{+-}}{\connm_{++}} \,, \\
    \text{QNMs:}& \quad \connm_{\ingo,\nnor} = \connm_{++}=0\,.
\end{split}   
\end{equation}
We explicitly are using the observation that the QNMs are poles in the thermal retarded Green's function. The prefactor $\holonorm\prn{\mexp_\bdy}$ needs to be determined by holographic renormalization. Since it depends solely on the scaling dimension $\Delta$ it has no effect on the analytic structure of the thermal correlation function.

\paragraph{Exact QNM and correlator in the $s$-channel:} Having identified the relation between the black hole wave equation, which determines the thermal 2-point function and the connection formulae determined using 2d CFT techniques, we can now give the result for the thermal correlator.  One can directly use $s$-channel connection formula explained in~\cref{sec-connection-formula-CFT} to obtain
\begin{empheq}[box=\fbox]{equation}\label{eq-Gret-s}
\Gret(\omega) = 
    \holonorm\prn{\mexp_\bdy} \dfrac{\Gamma\prn{-2\,\theta_\bdy}}{\Gamma\prn{2\,\theta_\bdy}} \, \frac{ \Gamma\prn{\frac{1}{2}  - \theta_\hor + \theta_\bdy \pm \sigma}}{\Gamma\prn{\frac{1}{2}  - \theta_\hor - \theta_\bdy \pm \sigma}} \;
    \exp\brk{-\partial_{\theta_\bdy} \VBcl^s(x)}\,. 
\end{empheq}
We emphasize that the result holds when $ \mexp_\bdy \not\in \frac{\intz}{2} $. We also are implicitly using the Zamolodchikov relation~\eqref{eq:Zamrel} for the accessory parameter. 
Finally, to be concrete, the $s$-channel semiclassical Virasoro block in this context itself is given by
\begin{equation}\label{eq:Wsblock-qnm}
    \VBcl^s(x) = \begin{tikzpicture}[scale= .5, baseline = .5ex]
        \coordinate (inf) at (-2,-2);
        \coordinate (one) at (-2,2);
        \coordinate (inf-one) at (0,0);
        \coordinate (xr-zero) at (4,0);
        \coordinate (midint) at (2,0);
        \coordinate (xr) at (6,2);
        \coordinate (zero) at (6,-2);
        \draw (inf) -- (inf-one);
        \draw  (one) -- (inf-one);
        \draw (inf-one) -- (xr-zero);
        \draw (xr-zero) -- (xr);
        \draw (xr-zero) -- (zero);
        \node [below left] at (inf) {\scriptsize $\mexp_\curv/\mexp_{\rm}(\infty)$};
        \node [above left] at (one) {\scriptsize $\mexp_{\rcomplex}(1)$};
        \node [below] at (midint) {\scriptsize $\cmexp$};
        \node [above right] at (xr) {\scriptsize $\mexp_\bdy(\xr)$};
        \node [below right] at (zero) {\scriptsize $\mexp_\hor(0)$};
    \end{tikzpicture}
\end{equation}
As explained earlier, the internal classical Liouville momentum $\cmexp$ needs to be determined from the Zamolodchikov relation~\eqref{eq:Zamrel}. Using the inversion of Zamolodchikov relation, cf. ~\cref{sec-connection-formula-CFT}, we are led to the following cross-ratio expansion for internal Liouville momentum $\cmexp$:
\begin{equation}\label{eq-cmexp-expansion-gravity}
\begin{split}
    &\cmexp\prn{\vb*{\mexp},\acc} = \cmexpope + \sum^\infty_{k=1} \cmexp_k\prn{\vb*{\mexp},\acc} \,\xr^k \,,\\
    &\cmexpope^2 = \mexp^2_\hor + \mexp^2_\bdy - \acc - \frac{1}{4} 
\end{split}   
\end{equation}
The expansion coefficients $\cmexp_k\prn{\vb*{\mexp},\acc}$ are fixed by the expansion coefficients $\VBcl^s_k\prn{\vb*{\mexp},\cmexp}$ in classical Virasoro block. The first two terms can be found in~\eqref{eq:sigmasol12}.
From the result, or directly from the definition of QNMs in terms of connection coefficient, we deduce an exact quantization condition for QNMs of the black hole. To wit,\footnote{We are assuming that the frequency and momenta are generic. As is now well appreciated, at some special kinematic points $(\freq_*,\mom_*)$ both the ingoing and outgoing solutions are analytic. In such a  situation, the horizon is rendered to be an ordinary point. It has been noted that at such loci, the pole in the numerator of~\eqref{eq-Gret-s} is compensated by a corresponding pole of the denominator factor, leading to the notion of `pole-skipping'~\cite{Blake:2017ris}. For reasons explained in~\cite{Loganayagam:2022zmq} it is more accurate to view these as apparent quasinormal modes, for at these special kinematic points the horizon is an apparent singular point of the oper~\cite{Blake:2019otz,Natsuume:2019xcy,Loganayagam:2022teq}.}
\begin{empheq}[box=\fbox]{equation}\label{eq-QNM-quantization}
\text{QNMs} = \cbrk{\omega \;\bigg|\;\half - \mexp_\hor + \mexp_\bdy \pm \cmexp\prn{\vb*{\mexp},\acc} = -n, \quad n \in \intz_{\geq 0}} 
\end{empheq}

The universality of the answers~\eqref{eq-Gret-s} and thence~\eqref{eq-QNM-quantization} is manifest.  Different types of holographic thermal correlators correspond to variations in the dependence of external classical Liouville momenta $\mexp_i$ and accessory parameter $\acc$. But this simply 
translates to deducing how to relate this data to the black hole parameters, which once done, solves the problem completely.

\paragraph{The logarithmic cases with $\mexp_{\bdy} \in \intz/2$:}
So far we have only defined thermal two-point functions and QNMs via~\eqref{eq:QNM-two-pt-func-def} for $\mexp_{\bdy} \notin \intz/2$, and the results~\eqref{eq-Gret-s} and~\eqref{eq-QNM-quantization} are derived in this regime. Now we address the case $\mexp_{\bdy} \in \intz/2$, which in fact corresponds to most of the physical examples. The general mathematical properties in this situation with logarithmic solution were discussed in~\cref{subsec:conn-prob-frobenius}. In this situation, the normalizable/non-normalizable solutions at AdS boundary are the non-logarithmic/logarithmic solutions, respectively, viz.,
\begin{equation}
        \wf_{\nor}(z) = \wf_{\nlog}(z)\,, \qquad \wf_{\nnor}(z) = \wf_{\log}(z)\,,
\end{equation}
where the non-logarithmic and logarithmic solutions are as defined in~\cref{subsec:conn-prob-frobenius}, with two distinct cases. The connection problem relevant for our purpose is then,
\begin{equation}
     \wf_{\ingo}(z) 
    = \connm_{\ingo,\nor}\, \wf_{\nor}(z) + \connm_{\ingo,\nnor} \,\wf_{\nnor}(z) 
    = \connm_{+\nlog} \,\wf_{\nlog}(z) + \connm_{+\log} \,\wf_{\log}(z)\,.  
\end{equation}
As explained in~\cref{subsec:conn-prob-frobenius}, the connection coefficient $\connm_{+\log}$ is unambiguously defined in both cases, whereas $\connm_{+\nlog}$ has ambiguity in one of the cases. The QNMs can then be defined, in both logarithmic cases, as 
\begin{equation}\label{eq:QNM-def-log}
    \text{QNMs:} \quad \connm_{\ingo,\nnor} = \connm_{+
    \log}=0\,.
\end{equation}

\begin{claim}\label{claim:QNM-log-quant}
    The QNMs in logarithmic cases ($\mexp_{\bdy} \in \intz/2$), defined by~\eqref{eq:QNM-def-log}, obey the same exact quantization condition~\eqref{eq-QNM-quantization} as in the non-logarithmic case ($\mexp_{\bdy} \notin \intz/2$).
\end{claim}
\noindent We have the following reasoning and evidence supporting this claim:
\begin{itemize}
    \item \emph{The analogue of the claim holds in the BTZ case (cf.~\cref{subsec:BTZ})}. In the familiar example of minimally coupled scalar in BTZ background, the logarithmic cases correspond to $\Delta \in \intz$. The analogue of the exact quantization condition~\eqref{eq-QNM-quantization} is a simple algebraic equation (cf.~\eqref{eq-BTZ-answer}) that holds in both non-logarithmic and logarithmic cases.
    \item \emph{The exact quantization condition~\eqref{eq-QNM-quantization} is continuous in $\mexp_\bdy$, and so is the QNM spectrum.} While the expression for exact two-point function~\eqref{eq-Gret-s} is singular at $\mexp_{\bdy} \in \intz/2$ due to the prefactor from degenerate fusion matrix, the associated semiclassical Virasoro block~\eqref{eq:Wsblock-qnm} is not, since in general Virasoro block is expected to be analytic in external weights. Therefore, the internal Liouville momentum $\cmexp$ appearing in the exact quantization condition~\eqref{eq-QNM-quantization} can still be defined from inverting Zamolodchikov relation, and hence the continuity of the condition~\eqref{eq-QNM-quantization} in $\mexp_\bdy$. Physically, $\mexp_\bdy$ corresponds to a mass parameter, in which the QNM spectrum should be continuous.
    \item \emph{The claim can be verified by explicit computation of QNMs.} In the following, we will perform explicit computation of QNMs using the exact quantization condition~\eqref{eq-QNM-quantization} in various examples with $\mexp_{\bdy} \in \intz/2$, and find good agreement with numerical data.
\end{itemize}

\noindent The situation for two-point function is more subtle in the logarithmic cases:
\begin{itemize}
    \item \textbf{Case II ($\mexp_{\bdy} = 0$):} In this case there is no ambiguity in $\connm_{+\,\nlog}$, and the two-point function can be defined as
    \begin{equation}\label{eq:two-point-def-log-case-II}
        \Gret(\omega) \coloneqq \holonorm\prn{\mexp_\bdy} \frac{\connm_{\ingo,\nor}}{\connm_{\ingo,\nnor}} = \holonorm\prn{\mexp_\bdy} \, 
    \frac{\connm_{+\,\nlog}}{\connm_{+\,\log}}
    \end{equation}
    \item \textbf{Case III ($\mexp_{\bdy} = \frac{n}{2}$, $n \in \mathbb{Z}_{>0}$):} In this case $\connm_{+\,\nlog}$ can no longer be unambiguously defined. The two-point function should instead be defined in terms of the fall-off coefficients defined in~\cref{subsec:fall-off-coeff}:
    \begin{equation}\label{eq:two-point-def-log-case-III}
        \Gret(\omega) \coloneqq \holonorm\prn{\mexp_\bdy} \frac{\falloffc_{\ingo,\nor}}{\falloffc_{\ingo,\nnor}} = \holonorm\prn{\mexp_\bdy} \frac{\connm_{+\,{\nlog}} + \frob_{n}\prn{\chexpmn} \connm_{+\,{\log}}}{\connm_{+\,{\log}}}
    \end{equation}
    Here we remind that the combination $\connm_{+\,{\nlog}} + \frob_{n}\prn{\chexpmn} \connm_{+\,{\log}}$ has no ambiguity.
\end{itemize}

\noindent As was done for QNMs, we would like to obtain the two-point functions in the logarithmic cases, defined by~\eqref{eq:two-point-def-log-case-II} and~\eqref{eq:two-point-def-log-case-III}, from the non-logarithmic exact expression~\eqref{eq-Gret-s}. The extra subtlety compared to the QNM case is that, unlike the exact quantization condition~\eqref{eq-QNM-quantization}, the expression~\eqref{eq-Gret-s} is no longer continuous in $\mexp_{\bdy}$. 
\begin{claim}\label{claim:TwoPt-log}
    The exact holographic thermal two-point functions in logarithmic cases ($\mexp_{\bdy} \in \intz/2$), defined by~\eqref{eq:two-point-def-log-case-II} and~\eqref{eq:two-point-def-log-case-III}, can be obtained from the exact two-point function in non-logarithmic case ($\mexp_{\bdy} \notin \intz/2$)~\eqref{eq-Gret-s}, by extracting the regular part in the singular limit $\mexp_{\bdy} \to \intz/2$.
\end{claim}
\noindent Our main evidence for the claim is that the analogous statement holds in the BTZ case, and that the additional semiclassical Virasoro block factor in~\eqref{eq-Gret-s} would not give extra singular behavior apart from the Gamma function factor as long as the internal Liouville momentum doesn't correspond to a degenerate representation, viz., $\cmexp \in \intz/2$, which we assume to be the case. 

\begin{remark}\label{remark:logCFT}
    It should be possible, and is desirable, to give a direct justification of~\cref{claim:QNM-log-quant} by generalizing the CFT method discussed here to \emph{logarithmic CFT}. Such a generalization could also facilitate direct evaluation of exact two-point functions in the logarithmic cases without the need for taking the singular limit as in~\cref{claim:TwoPt-log}.
\end{remark}

\paragraph{QNM expansion in $s$-channel:}The $s$-channel exact quantization condition~\eqref{eq-QNM-quantization} can be used to solve QNMs as an expansion in the cross-ratio parameter $x$:
\begin{equation}\label{eq-QNM-expansion}
    \omega_n = \sum^\infty_{k=0} \,\omega^{(k)}_n\, x^k, \qquad 
    n \in \intz_{\geq 0}.
\end{equation}
This is done by substituting the cross-ratio expansion~\eqref{eq-cmexp-expansion-gravity} for $\sigma$ and the QNM expansion~\eqref{eq-QNM-expansion} to the quantization condition~\eqref{eq-QNM-quantization} to solve $\omega^{(k)}_n$ order by order in $x$. In general, the leading coefficient $\omega^{(0)}_n$ is fixed by the OPE limit, and the higher-order coefficients are recursively determined in terms of $\omega^{(0)}_n$. The QNM expansion is thus fixed by semiclassical Virasoro block data.
 
\begin{figure}[!htbp]
    \centering
    \includegraphics[width=0.8\textwidth]{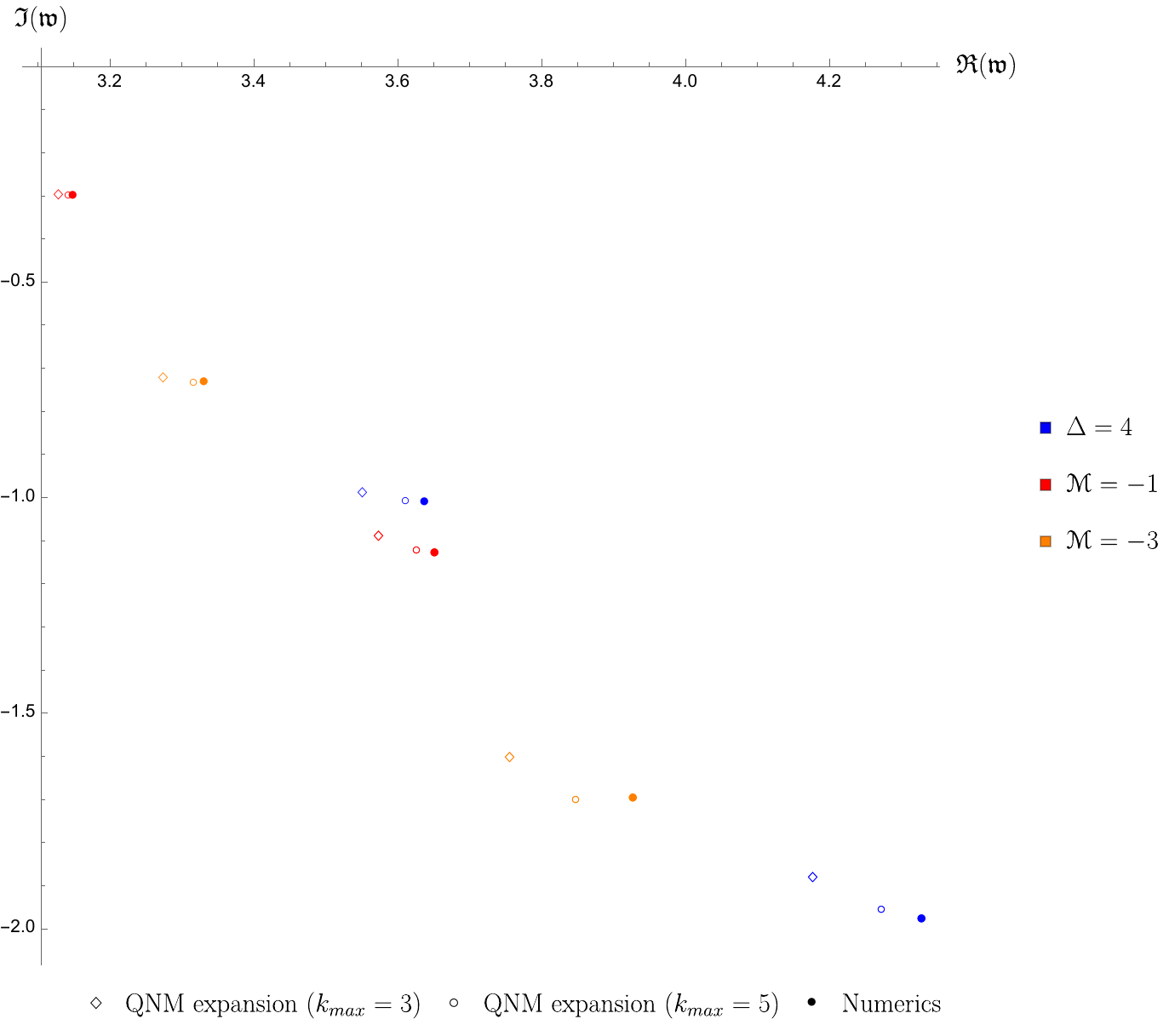}
    \caption{Comparison between QNM expansion and numerics for the first two QNMs ($\freq_0, \freq_1$) at $\mom=3$ for Klein-Gordon  scalar with $\Delta=4$ (massless scalar perturbation), and designer scalars with $\markov=-1$ (scalar polarization of gauge field perturbation) and $\markov = -3$ (vector polarization of metric perturbation). Here $k_{max}$ is the order of truncation in the cross-ratio expansion~\eqref{eq-QNM-expansion}. The $\Re(\freq)\geq0$ part of the QNM spectrum is shown. This data again supports~\cref{claim:QNM-log-quant} that the quantization condition~\eqref{eq-QNM-quantization} continues to holds for $\mexp_\bdy \in \intz /2$.}
    \label{fig:q3QNM}
\end{figure}

\begin{example}[Scalar fields in the planar Schwarzschild-\AdS{5} black hole]\label{ex:QNM-expansion-BB}
    For concreteness, we present explicit forms of the first two terms of the QNM expansion for both Klein-Gordon and designer scalar fields propagating in the neutral black hole spacetime. For planar horizons, we have $\mexp_\hor = \frac{i\, \freq}{2}$, $\mexp_{\rcomplex} = \frac{\freq}{2}$, and $\acc = \mom^2 - \freq^2$. The expansion coefficients as a function of overtone number $n$ read  
\begin{equation}\label{eq:planaromegaexp}
\begin{split}
\freq^{(0)}_n 
&= 
    \half \brk{\pm \sqrt{1+ 3\, \nhalf^2 +  4 \,\mom^2 + 6\, \nhalf\, \mexp_\bdy -\mexp^2_\bdy} - \prn{\nhalf + \mexp_\bdy}i }, \qquad \nhalf = n + \half \,,\\
\freq^{(1)}_n 
&= \
    \frac{i \prn{1+2\,\mom^2-2\,\prn{\freq^{(0)}_n}^2 -4\,\mexp^2_\bdy} \prn{1+2\,\mom^2-2\,\prn{\freq^{(0)}_n}^2 -2\,\mexp^2_\bdy + 2\, \mexp^2_\curv} }{\prn{2+4\,\mom^2-3\,\prn{\freq^{(0)}_n}^2 -4\,\mexp^2_\bdy} \prn{-3\,i\,\freq_n^{(0)} +
    \sqrt{-1-4\,\mom^2+3\,\freq_n^{(0)} + 4\,\mexp^2_\bdy} }} \,.
\end{split}        
\end{equation}
As is the case for $\freq^{(1)}_n$, all higher-order expansion coefficients are recursively determined from the leading term $\freq^{(0)}_n$, which is fixed by OPE limit.

The $s$-channel OPE limit, per se, doesn't correspond to a physical limit in the QNM problem. Nevertheless,  the QNM-type spectral problem  is still well-defined if we take the limit $x \to 0$ in the Heun's oper associated with these scalar probes. The first two terms in the QNM expansion above can indeed be checked to agree with the numerical spectrum in the $\xr \to 0$ limit. 

At the physical value $\xr = \half$, the comparison between the QNM expansion and numerics is shown for 
\begin{itemize}[wide,left=0pt]
\item  A Klein-Gordon scalar with $\Delta=4$ at $\mom=0$ in~\cref{tab-QNM-numerics}.
\item Both Klein-Gordon scalar with $\Delta=4$ and designer scalars with $\markov=-1$ and $\markov=-3$, respectively, at $\mom=3$ in~\cref{fig:q3QNM}.
\end{itemize}
In those cases, we find good agreement with numerical results for the QNMs.

There, however, appears to be a convergence issue when applying the QNM expansion to designer scalars at small $\mom$, the regime where hydrodynamic modes enter the low-lying part of the QNM spectrum. This issue deserves further investigation.
\end{example}

\begin{table}[H]
    \centering
    \begin{tabular}{|c||c|c|c||c|}
    \hline
        $n$ & $k_{\max} = 2$ & $k_{\max} = 4$ &  $k_{\max} = 6$ & Num \\
        \hline
         0 &  $\pm 1.37 -1.15i $ & $\pm 1.47 - 1.29i$ & $\pm 1.51 - 1.34i$ & $\pm 1.56 - 1.37i$ \\
         \hline
         1 & $\pm 2.41 - 1.90i$ & $\pm 2.49 - 2.14i$ & $\pm 2.51 - 2.24i$ & $\pm 2.58 - 2.38i$\\
         \hline
    \end{tabular}
    \caption{Comparison between QNMs $\freq_n$ computed using QNM expansion~\eqref{eq-QNM-expansion} and known numerical values for a Klein-Gordon scalar in the planar Schwarzschild-\AdS{5} background. We have chosen $\Delta = 4, \mom = 0$ for illustration. Here $k_{max}$ is the order of truncation in the cross-ratio expansion~\eqref{eq-QNM-expansion}. This data supports~\cref{claim:QNM-log-quant} that the quantization condition~\eqref{eq-QNM-quantization} continues to holds for $\mexp_\bdy \in \intz /2$. The numerical values can be found in, e.g., Table 1 of \cite{Nunez:2003eq}, or readily computed using numerical packages such as \cite{Jansen:2017oag}. }
    \label{tab-QNM-numerics}
\end{table}

\begin{remark}
    Using the expansion of Virasoro block in terms of elliptic nome $q = e^{-\pi K(1-x)/K(x)}$ should allow faster convergence for the QNM expansion. It would also be useful to further study the convergence property of the QNM expansion. 
\end{remark}

\paragraph{Exact QNM and correlator: $t$-channel:} We can equivalently use the  $t$-channel connection formula obtained in~\cref{sec-connection-formula-CFT}. This  yields 
\begin{empheq}[box=\fbox]{equation}\label{eq-Gret-t}
\begin{split}
\Gret(\omega)
&=
    \holonorm\prn{\mexp_\bdy} \,
    \frac{\mathsf{A}\prn{-\mexp_\bdy}}{\mathsf{A}\prn{  \mexp_\bdy}} \exp\brk{-\partial_{\theta_\bdy} \VBcl^t(x)} \,,
\end{split}
\end{empheq}
where we defined 
\begin{equation}\label{eq:Adef}
\mathsf{A}\prn{\mexp_\bdy} 
:= 
    \sum\limits_{\signconv = \pm} \frac{\Gamma(1-2\, \mexp_{\hor}) \Gamma(2 \,\signconv \,\cmexp)}{\Gamma\prn{\half - \mexp_{\hor} + \signconv \,\cmexp \pm \mexp_{\bullet}}} \frac{\Gamma(1+2\, \signconv \,\cmexp)\Gamma(2 \, \mexp_{\bdy})}{ \Gamma\prn{\half + \signconv\,\cmexp + \mexp_{\bdy} \pm \mexp_{\rcomplex}}} \exp\brk{\frac{\signconv}{2} \,\partial_\cmexp \VBcl^t(x)} \,.
\end{equation}
Here $\mexp_\bullet = \mexp_{\curv}/\mexp_{\rm}$ and we continue to assume that 
$\mexp_\bdy \not\in \frac{\intz}{2}$. The internal Liouville momentum $\cmexp$ is again implicitly defined via Zamolodchikov relation $ \acc = x \,\partial_x \VBcl^t(x) $. The $t$-channel semiclassical Virasoro block here is
\begin{equation}
    \VBcl^t(x) = 
    \begin{tikzpicture}[scale= .5, baseline = -5.5ex]
        \coordinate (xr) at (2,2);
        \coordinate (one) at (-2,2);
        \coordinate (one-xr) at (0,0);
        \coordinate (inf-zero) at (0,-4);
        \coordinate (midint) at (0,-2);
        \coordinate (inf) at (-2,-6);
        \coordinate (zero) at (2,-6);
        \draw (one) -- (one-xr);
        \draw  (xr) -- (one-xr);
        \draw (one-xr) -- (inf-zero);
        \draw (inf-zero) -- (inf);
        \draw (inf-zero) -- (zero);
        \node [below left] at (inf) {\scriptsize $\mexp_\bullet(\infty)$};
        \node [above left] at (one) {\scriptsize $\mexp_{\rcomplex}(1)$};
        \node [left] at (midint) {\scriptsize $\cmexp$};
        \node [above right] at (xr) {\scriptsize $\mexp_\bdy(\xr)$};
        \node [below right] at (zero) {\scriptsize $\mexp_\hor(0)$};   
    \end{tikzpicture}.
\end{equation}
From here, or directly from definition of QNMs in terms of connection coefficient, we deduce that in the $t$-channel exact quantization condition for QNMs is given by $\mathsf{A}\prn{\mexp_\bdy} = 0$, which can be written more explicitly as 
\begin{empheq}[box=\fbox]{equation}\label{eq-QNM-quantization-t}
\begin{split}
\text{QNMs} 
&= 
    \cbrk{\omega \bigg| \;\frac{\mathfrak{A}(\cmexp)}{\mathfrak{A}(-\cmexp)} = - \exp\brk{-\partial_\cmexp \VBcl^t(x)} } \,,\\ 
\mathfrak{A}(\cmexp)  
&=  
    \Gamma(2\,\cmexp)\,\Gamma(1+2\,\cmexp)\, \Gamma\prn{\half - \mexp_\hor -\cmexp \pm \mexp_\bullet} \, \Gamma\prn{\half -\cmexp + \mexp_\bdy \pm \mexp_\bullet} .
\end{split}
\end{empheq}
We are leaving the dependence of internal Liouville momentum $\cmexp$ on $\vb*{\mexp}, \acc$ implicit for simplicity. 

\paragraph{$t$-channel OPE limit:}The $t$-channel expression for the exact quantization takes a more complicated form than its $s$-channel counterpart, but considerably simplifies in the OPE limit $\xr \to 1$. In fact, the limit is physically more relevant for the black hole perturbation problem than the $s$-channel one. In this limit, using the solution to the internal momentum $\cmexp$, cf.~\eqref{eq:topecmexp},
\begin{equation}
    \cmexp \to \cmexpope, \quad \cmexpope^2 = \mexp^2_{\rcomplex} + \mexp^2_\bdy + (1-\xr)\,\acc - \frac{1}{4}.
\end{equation}
As discussed in~\cref{sec-connection-formula-CFT}, if $\Re\prn{\cmexpope} \neq 0$, one of the terms in the sum over $\signconv$ in connection matrix ($\connm_{++}$ in this case) dominates in the OPE limit. With the sign convention $\Re(\cmexpope) > 0$, the $\signconv = +$ term dominates. Alternatively, one can also directly observe that r.h.s.\ of the exact quantization condition in~\eqref{eq-QNM-quantization-t} scales as $(1-x)^\cmexpope \to 0$ in the OPE limit. The QNM spectrum in the OPE limit is thus determined to be 
\begin{equation}\label{eq-QNMope-t}
    \mathrm{QNMs} \to \cbrk{\omega \bigg|\half - \mexp_\hor + \cmexpope \pm \mexp_\bullet = -n \; \;\text{or} \;\; \half + \cmexpope + \mexp_\bdy \pm \mexp_{\rcomplex} = -n, \quad n \in \intz_{\geq 0}, \;  \Re(\cmexpope) > 0}. 
\end{equation}

There are two simple examples which illustrate these general considerations, the small neutral black hole limit, and the near-extremal limit. In both cases, we will examine the wave equation for a Klein-Gordon scalar with mass set by the dimension $\Delta$. 

\begin{example}[Small black hole limit \cite{Dodelson:2022yvn}]
In the $\xr \to 1$ ($\rp \to 0$) limit,
\begin{equation}
        \mexp_\hor \to 0, \quad \mexp_{\rcomplex} \to \frac{\omega}{2}, \quad (1-\xr)\,\acc \to \frac{\lambda_{\mathbf{S}^3} -\omega^2 - \Delta\, (\Delta -4) - 2}{4}
\end{equation}
with $\mexp_\bdy = \frac{\Delta-2}{2}, \mexp_\curv = 0$. One then finds
\begin{equation}
        \cmexpope = \frac{l+1}{2}.
\end{equation}
The first condition in~\eqref{eq-QNMope-t} turns out not to have a solution. The second  recovers the normal mode spectrum in~\AdS{5} at leading order, 
\begin{equation}
        \omega_n = \pm (2n+l+\Delta).
\end{equation}
\end{example}

\begin{example}[Near-extremal planar black hole]\label{ex:near-extremal-QNM} Consider the Klein-Gordon equation for a massive uncharged scalar  in the planar Reissner-Nordstr\"om-\AdS{5} geometry. In the near-extremal limit ($Q \to \Qext = \sqrt{2}$), the Heun parameters are 
\begin{equation}
\begin{split}
        \mexp_\hor &\to \frac{i\, \freq}{2\sqrt{2} \,(\Qext - Q)} + \frac{i \,\freq}{8} \,,\\
        \mexp_{\rcomplex} &\to \frac{2\sqrt{2}}{9} \,\freq \,,\\
        \mexp_{\rm} &\to \frac{i\, \freq}{2\sqrt{2}\, (\Qext - Q)} - \frac{19}{72} \,i\,\freq \,,\\
        (1-\xr)\,\acc &\to \frac{\mom^2}{3}
        -\frac{\freq^2}{3}-\frac{\Delta\,(\Delta -4)}{6}-\half\,.
\end{split}        
\end{equation}
with $\mexp_\bdy = \frac{\Delta-2}{2}$. 
At finite $\freq$, this is a confluent limit leading to irregular puncture instead of a degenerate/OPE limit due to the divergence of $\mexp_\hor,\mexp_{\rm}$. In particular, one may check that in this limit the first few terms of the block coefficients scale as $\VBcl^t_k \sim \prn{\Qext-Q}^{-k} \sim (1-x)^{-k}$ and therefore cannot be ignored when solving the Zamolodchikov relation. However, we may take the $Q \to \Qext$ limit at small frequency $\freq = \order{\Qext - Q}$ so that all Liouville momenta are finite and the OPE analysis still applies.

One then finds
\begin{equation}
    \cmexpope^2 = -\frac{19}{81}\,\freq^2 + \frac{\mom^2}{3} + \frac{m^2}{12} + \frac{1}{4} \simeq \frac{\mom^2}{3} + \frac{\Delta\,(\Delta -4)}{12} + \frac{1}{4}.
\end{equation}
where $\simeq$ denotes to leading order in $\Qext - Q$.
Solving the first condition in~\eqref{eq-QNMope-t} with a negative sign to the leading order in $\Qext-Q$, we find,
\begin{equation}\label{eq:QNMNearExt}
    \freq_n = -i \frac{\Qext - Q}{\sqrt{2}} \prn{2\,n+1 + \sqrt{1 + \frac{\Delta \,(\Delta-4)}{3} + \frac{4\,\mom^2}{3}} },
\end{equation}
while other conditions in~\eqref{eq-QNMope-t} don't admit solutions. To the best of our knowledge, the analytic form~\eqref{eq:QNMNearExt} of the purely decaying modes in near-extremal limit is new in the literature, and we find good agreement with numerics. The oscillation/Christmas-tree-type modes (cf.~\cref{fig:NearExtQNM}) are not visible in this OPE analysis, as they are at finite $\freq$ in the $Q \to \Qext$ limit.

\begin{figure}[!htbp]
    \centering
    \includegraphics[width=0.8\textwidth]{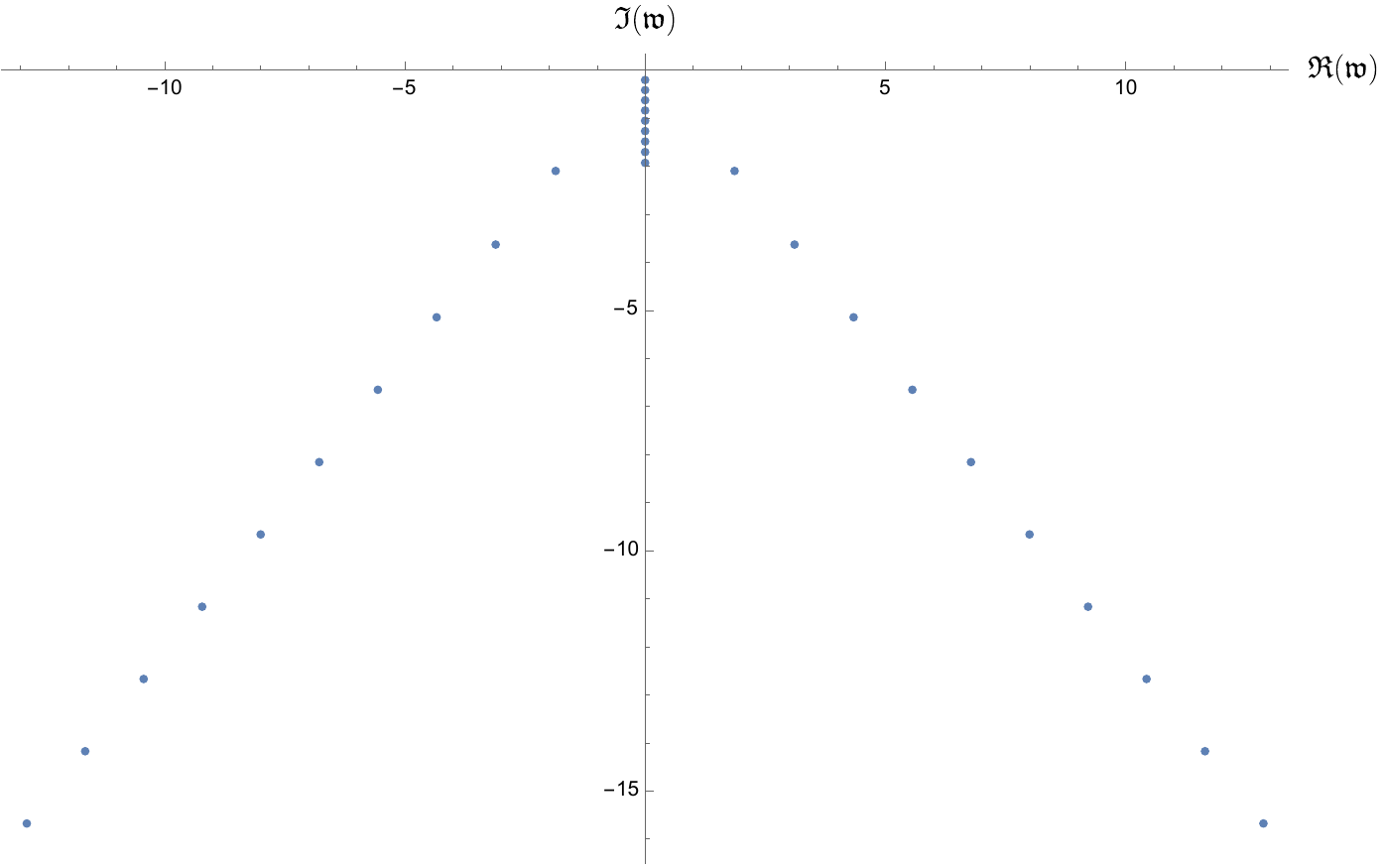}
    \caption{Numerically computed QNM spectrum for massless uncharged scalar at zero momentum in near-extremal planar black hole background with $Q/\Qext = .9$. The purely decaying modes are well-described by \eqref{eq:QNMNearExt} obtained via OPE analysis, equally spaced with gap $\sim \Qext - Q \sim T$. The oscillation/Christmas-tree-type modes at finite $\freq$ are not visible in the OPE analysis; see main text for  more explanations. We also find good agreement between \eqref{eq:QNMNearExt} and numerics for purely decaying modes in near-extremal limit at more generic masses and momenta. The numerical spectra were computed using the numerical package of \cite{Jansen:2017oag}. } 
    \label{fig:NearExtQNM}
\end{figure}

\end{example}

\subsection{An example with five punctures: conserved currents at finite density}\label{subsec:five-punc}

As discussed in~\cref{sec:bhwave} the designer scalar equations in Reissner-Nordstr\"om-\AdS{5} capture the dynamics of conserved currents. In particular, the  scalar and vector polarization of gauge field perturbations, which are dual to the boundary global current,  and the vector polarization of metric perturbation dual to the momentum flux operator, correspond to the radial wave equation~\eqref{eq:Mdesign} with $\markov = -1, 1, -3$, respectively. The tensor polarization of the energy-momentum tensor is equivalent to a massless Klein-Gordon scalar  with $\markov=3$. Since there is a background charge, the equations capture the response of such currents at finite density.\footnote{In the case of gauge dynamics, the global current of the holographic CFT is different from the one that has background charge density. The equations for $\markov = \pm1$ should be viewed as that for a probe Maxwell field in the Reissner-Nordstr\"om-\AdS{5} background, unrelated to the one sourcing the solution.}

The radial wave equation~\eqref{eq:Mdesign} for designer scalar with $\markov \neq 3$ on the planar Reissner-Nordstr\"om-\AdS{5} black hole background is mapped to a (generalized) Heun's oper with five punctures, with aid of the transformation~\eqref{eq:zxrmaps}. The normal form of $\Tcl(z)$ for this five punctured case can be found in~\eqref{eq:T5punctures}. 
 
The additional puncture at $\xrp$ turns out to correspond to the curvature singularity. The two cross-ratios in this case lie in the range
\begin{equation}
\xr = \frac{\ri^2 + \rm^2}{\ri^2 + \rp^2} \in \left[ 1/2, 1 \right) , \qquad 
\xrp = \frac{\ri^2 + \rm^2}{\ri^2 + \rp^2} \; \frac{\rp^2}{\rm^2} \in (1,\infty).
\end{equation}
The near-extremal limit corresponds to both approaching unity, $\xr, \xrp \to 1$. The generalized Heun's parameters can be determined to be 
\begin{equation}
\begin{split}
\mexp_\hor(0) 
&= 
    \abs{\frac{1}{\rp^2 f^\prime(\rp)}} \, i \,\omega = \frac{r^4_+}{(r^2_i + r^2_+)\,(r^2_+ - r^2_-)} \, i\,\freq \,,\\
\mexp_\bdy(\xr) 
&= 
    \frac{\abs{\markov + 1}}{4}  \nonumber\\
\mexp_{\rcomplex}(1) 
&= 
    \abs{\frac{1}{\rcomplex^2  \,f^\prime(\rcomplex)}} \, \omega = \frac{r_i^3 \,\rp}{(r^2_i + r^2_+)\,(r^2_i + r^2_-)} \, \freq \,,\\
\mexp_\curv(\xrp) 
&= 
    \frac{\abs{\markov - 5}}{4}  \,,\\
\mexp_{\rm}(\infty) 
&= 
    \abs{\frac{1}{\rm^2 f^\prime(\rm)}} \, i \,\omega = 
    \frac{\rm^3 \rp}{(r^2_i + r^2_-)\,(r^2_+ - r^2_-)} \, i\,\freq \,,\\
\acc_\xr 
&= 
    \frac{1}{8 \,(\rp^2 - \rm^2)} \brkbig{2\,\rp^2 \prn{4 \mom^2 - 4 \freq^2 + \markov-1} - \rm^2 \prn{\markov^2-1} - 2\, \ri^2 \prn{\markov-1}} \,,\\
\acc_\xrp 
&= 
    \frac{\markov-3}{8 \,\ri^2\prn{\rp^2-\rm^2}} \prn{-2 \,\rp^2\,\rm^2 + 2\,\rm^2\,\ri^2 + (\markov-5)\,\rp^2\,\ri^2 } .
\end{split}        
\end{equation}
As a simple sanity check, we note that for $\markov=3$, which corresponds to a Klein-Gordon scalar, the puncture at $\xrp$ is removed.

The physical cases with $\markov = -1, 1, -3$ correspond to $\mexp_\bdy \in \intz/2$, and therefore the connection formula for five punctures in~\cref{sec-connection-formula-CFT} cannot be directly applied. Nonetheless, we can first consider generic $\markov$ so that the connection formula can be applied to yield
\begin{equation}
\begin{split}
    \Gret = \holonorm\prn{\mexp_\bdy} \frac{\Gamma\prn{-2\,\theta_\bdy}}{\Gamma\prn{2\,\theta_\bdy}} \frac{\prod_\pm \Gamma\prn{\frac{1}{2}  - \theta_\hor + \theta_\bdy \pm \sigma}}{\prod_\pm \Gamma\prn{\frac{1}{2}  - \theta_\hor - \theta_\bdy \pm \sigma}} \exp\brk{-\partial_{\theta_\bdy} \VBcl(\xr,\xrp)}, \quad \mexp_\bdy \not\in \frac{\intz}{2} .
\end{split} 
\end{equation}
Here the semiclassical Virasoro block is in the following channel
\begin{equation}
    \VBcl(\xr,\xrp) = 
    \begin{tikzpicture}[scale= .5, baseline = .5ex]
        \coordinate (inf) at (-2,-2);
        \coordinate (one) at (-2,2);
        \coordinate (inf-one) at (0,0);
        \coordinate (xr-zero) at (4,0);
        \coordinate (midint) at (2,0);
        \coordinate (xr) at (6,2);
        \coordinate (zero) at (6,-2);
        \draw (inf) -- (inf-one);
        \draw  (one) -- (inf-one);
        \draw (inf-one) -- (xr-zero);
        \draw (xr-zero) -- (xr);
        \draw (xr-zero) -- (zero);
        \node [below left] at (inf) {\scriptsize $\mexp_{\rm}(\infty)$};
        \node [above left] at (one) {\scriptsize $\mexp_\curv(\xrp)$};
        \node [above right] at (xr) {\scriptsize $\mexp_\bdy(\xr)$};
        \node [below right] at (zero) {\scriptsize $\mexp_\hor(0)$};
        \node [below right] at ($(midint) + (0.5,-.2)$) {\scriptsize $\cmexp$};
        \node [below left] at ($(midint) - (.5,0)$) {\scriptsize $\cmexpp$};
        \coordinate (heavy-deg) at ($(midint) + (0,2)$);
        \draw (midint) -- (heavy-deg);
        \node [above] at (heavy-deg) {\scriptsize $\mexp_{\rcomplex}(1)$};
    \end{tikzpicture},
\end{equation}
with $\cmexp,\cmexpp$ defined implicitly by
\begin{equation}
    \acc_\xr = \xr \,\partial_\xr \VBcl(\xr,\xrp), \quad \acc_\xrp = \xrp\, \partial_\xrp \VBcl(\xr,\xrp).
\end{equation}
We again anticipate that the following exact quantization condition for QNMs does hold for physical values of $\markov$ with $\mexp_\bdy \in \intz/2$. In any event, we predict the spectrum to be determined by
\begin{empheq}[box=\fbox]{equation}\label{eq-QNM-quantization-five-punc}
   \text{QNMs} = \cbrk{\omega \bigg|\half - \mexp_\hor + \mexp_\bdy \pm \cmexp\prn{\vb*{\mexp},\vb*{\acc}} = -n, \quad n \in \intz_{\geq 0}}
\end{empheq}
One again the dependence of $\cmexp$ on the parameters $\vb*{\mexp},\vb*{\acc}$ is to be determined using the Zamolodchikov relations.

\begin{remark}
    As in the case of scalar perturbation of the planar charged black hole, the near-extremal limit $\xr, \xrp \to 1$ is easier to study with a difference choice of channel than the current one, where the punctures at $1, \xr, \xrp$ sequentially fuse. We expect the near-extremal limit of purely decaying modes can be studied in such OPE limit as in the scalar case.
\end{remark}

\subsection{An example with apparent singularity: energy density correlators}\label{subsec:app-sing}

The correlators of the energy density operator, which corresponds to scalar polarizations of the stress tensor, are determined from the equation~\eqref{eq:Zsound}. This differential equation was obtained by working with suitable gauge invariant variables to account for the diffeomorphism redundancies (see~\cite{Kodama:2003jz,He:2022jnc} for details). Since the fluctuations of energy density correspond to sound propagation, this is also known as the sound channel equation.

The new feature in~\eqref{eq:Zsound} is that it has an apparent singularity with $\asnum = 3$. Its generalized Heun's parameters are listed in~\cref{tab-heun-parameters-SC}. An added complication is that the exponent at the boundary vanishes, $\mexp_\bdy = 0$ (the asymptotic solutions are $\MZ \sim c_1 + c_2\, \log r$).  
Owing to this, the connection formula cannot be directly applied. As a simple trick, we will introduce, as in the designer scalar case,  a one-parameter generalization of~\eqref{eq:Zsound} where the connection formula does apply.
We will then argue that the exact quantization condition for QNMs derived from the connection formula continues to hold as we limit to  the $\mexp_\bdy = 0$ case. 

One cannot naively deform the equation by simply changing $\mexp_\bdy$ alone. The reason being that the apparent singularity condition involves a delicate cancelation of residues, which can fail if the equation is tweaked.  Working with real $d\neq 4$ also doesn't address the issue.  Instead, inspired by the structure in designer scalar case, we choose to deform both $\mexp_\bdy$ and $\mexp_\curv$ while keeping all other parameters fixed. The apparent singularity condition for $s=3$, which is relevant in this case, evaluates to
\begin{equation}
    \asc_3 \propto 3\, (4 \,\mom^4 +9)\, \mexp^2_\bdy - 4 \,\mom^4\, (4\, \mom^4 -3) \,\mexp^2_\curv
\end{equation}
where the proportionality factor depends solely on $\mom$. We therefore introduce the following one-parameter generalization of the sound channel equation:
\begin{equation}\label{eq:Zmodg}
\begin{split}
\mexp_\bdy  = \scpara \,, \quad & \quad 
    \qquad \mexp_\curv = \mathfrak{h}(\mom) \, \scpara \,,\\
\mathfrak{h}^2(\mom) 
&= 
    \frac{3\, (4\, \mom^4 +9)}{4 \,\mom^4 \,(4 \,\mom^4 -3)}\,.
\end{split}    
\end{equation}
In~\cref{tab-heun-parameters-SC} we record the generalized Heun's parameters for both~\eqref{eq:Zsound} and the aforementioned modification.

\begin{table}[H]
    \centering
    \begin{tabular}{|c|c|c|c|c|c|c|c|c|c|}
         \hline
          & $\mexp_\hor$ & $\mexp_\bdy$ & $\mexp_{\rcomplex}$ & $\mexp_\curv$ & $\mexp_\xrapp$ & $\xr$ & $\acc_\xr$ & $\xrapp$ & $\acc_\xrapp$ 
         \\
         \hline
         SC & $\frac{i \,\freq}{2}$ & $0$ & $\frac{\freq}{2}$ & $0$ & $\frac{3}{2}$ & $\frac{1}{2}$ & $\mom^2 - \freq^2 - \frac{3}{2 \,\mom^2}$ & $\frac{1}{2} + \frac{\mom^2}{3}$ & $5 + \frac{3}{2 \mom^2} + \frac{12}{2\, \mom^2 -3}$ \\
         \hline
         SCg & - & $\scpara$ & - & $\mathfrak{h}(\mom) \,\scpara$ & - & - & - & - & - \\
         \hline   
    \end{tabular}
    \caption{Generalized Heun's parameters for sound channel (SC) equation~\eqref{eq:Zsound} and its one-parameter generalization (SCg) as detailed by~\eqref{eq:Zmodg}. Here $\xrapp$ is the apparent singularity occurring at the vanishing locus of the function $\Lk$.}
    \label{tab-heun-parameters-SC}
\end{table}

With this change, we can then readily apply the connection formula for an equation with apparent singularity~\eqref{eq:conn-app-sing}. For the one-parameter deformation of~\eqref{eq:Zsound} introduced herein, we have (assuming $\mexp_\bdy \not\in \frac{\intz}{2} $),
\begin{empheq}[box=\fbox]{equation}
\label{eq-Gret-sc}
\Gret = 
    \holonorm\prn{\mexp_\bdy} \frac{\Gamma\prn{-2\,\mexp_\bdy}}{\Gamma\prn{2\,\mexp_\bdy}} 
    \, \dfrac{\Gamma\prn{\frac{1}{2}  - \mexp_\hor + \mexp_\bdy \pm \sigma}}{\Gamma\prn{\frac{1}{2}  - \mexp_\hor - \mexp_\bdy \pm \sigma}} \exp\brk{-\partial_{\mexp_\bdy} \VBcl(\xr,\xrapp)} \,.
\end{empheq}
The semiclassical Virasoro block here is
\begin{equation}
    \VBcl(\xr,\xrapp) =
    \begin{tikzpicture}[scale= .5, baseline = .5ex]
        \coordinate (inf) at (-2,-2);
        \coordinate (one) at (-2,2);
        \coordinate (inf-one) at (0,0);
        \coordinate (xr-zero) at (4,0);
        \coordinate (midint) at (2,0);
        \coordinate (xr) at (6,2);
        \coordinate (zero) at (6,-2);
        \draw (inf) -- (inf-one);
        \draw  (one) -- (inf-one);
        \draw (inf-one) -- (xr-zero);
        \draw (xr-zero) -- (xr);
        \draw (xr-zero) -- (zero);
        \node [below left] at (inf) {\scriptsize $\mexp_\curv(\infty)$};
        \node [above left] at (one) {\scriptsize $\mexp_{\rcomplex}(1)$};
        \node [above right] at (xr) {\scriptsize $\mexp_\bdy(\xr)$};
        \node [below right] at (zero) {\scriptsize $\mexp_\hor(0)$};
        \node [below right] at ($(midint) + (0.5,0)$) {\scriptsize $\cmexp$};
        \node [below left] at ($(midint) - (0,0)$) {\scriptsize $\cmexp + i_3$};
        \coordinate (heavy-deg) at ($(midint) + (0,2)$);
        \draw[dash dot] (midint) -- (heavy-deg);
        \node [above] at (heavy-deg) {\scriptsize $\mexp_{\drep{1}{3}}(\xrapp)$};
    \end{tikzpicture}
\end{equation}
The internal Liouville momenta $\sigma, i_3$ need to be determined using the Zamolodchikov relations,
\begin{equation}
\acc_\xr = \xr\, \partial_\xr \VBcl(\xr,\xrapp)\,, \qquad 
\acc_\xrapp = \xrapp\, \partial_\xrapp \VBcl(\xr,\xrapp)\,.
\end{equation}
The exact quantization condition for QNMs is given by
\begin{equation}
       \text{QNMs} = \cbrk{\omega \bigg|\half - \mexp_\hor + \mexp_\bdy \pm \cmexp\prn{\vb*{\mexp},\vb*{\acc}} = -n, \quad n \in \intz_{\geq 0}}\,.
\end{equation}

\subsection{Comparison with BTZ}\label{subsec:BTZ}
The exact $s$-channel expressions for thermal two-point function \eqref{eq-Gret-s} and QNMs \eqref{eq-QNM-quantization} in the four-puncture case and their five-puncture generalization \eqref{eq-QNM-quantization-five-punc}, have a formal structure analogous to the well-known BTZ answer, which corresponds to the three-puncture hypergeometric oper.

For minimally coupled massive scalar with $m^2 = \Delta (\Delta-2)$ on planar BTZ background, standard manipulation recasts the radial wave equation to hypergeometric oper with
\begin{equation}
    \begin{split}
        &\Tcl(z) = \frac{\delta_0}{z^2} + \frac{\delta_1}{(z-1)^2}  +  \frac{\delta_{\infty} - \delta_0 - \delta_1}{z\,(z-1)} \,,\\
        &\mexp_0 \equiv \mexp_\hor = \frac{i \,\freq}{2}, \quad \mexp_1    \equiv \mexp_\bdy = \frac{\Delta-1}{2}, \quad \mexp_\infty = \frac{i\, \mom}{2}\,.
    \end{split}
\end{equation}
The well-known results for BTZ thermal two-point function and quantization condition for QNMs read\footnote{These expressions were originally derived from analytically continuing the Euclidean 2d CFT result in~\cite{Gubser:1997cm}. The calculation in the AdS/CFT context was originally considered in~\cite{Birmingham:2001pj} and~\cite{Son:2002sd}. For a recent discussion in the context of the grSK geometry including generalization to the designer scalars, see~\cite{Loganayagam:2022zmq}.}
\begin{equation}\label{eq-BTZ-answer}
    \begin{split}
        \Gret &= \holonorm\prn{\mexp_\bdy} \frac{\Gamma\prn{-2\,\theta_\bdy}}{\Gamma\prn{2\,\theta_\bdy}} \;\frac{\Gamma\prn{\frac{1}{2}  - \theta_\hor + \theta_\bdy \pm \mexp_\infty}}{\Gamma\prn{\frac{1}{2}  - \theta_\hor - \theta_\bdy \pm \mexp_\infty}} \,, \\
        \text{QNMs} &= \cbrk{\freq \bigg|\half - \mexp_\hor + \mexp_\bdy \pm \mexp_\infty= -n, \quad n \in \intz_{\geq 0}} \\
        &= \cbrk{\pm \mom - i \,(2\,n+\Delta) \bigg| n \in \intz_{\geq 0}}\,.
    \end{split}
\end{equation}

Therefore, compared with BTZ answer, the exact $s$-channel expressions in \AdS{5} essentially replaces the external Liouville momentum $\mexp_\infty$ by the interval momentum $\cmexp$ and has an additional factor of semiclassical Virasoro block in the thermal two-point function. In particular, the momentum dependence no longer comes from an external Liouville momentum but arises from the accessory parameters that implicitly define the internal momentum $\cmexp$ via the Zamolodchikov relation. This similarity in structure indeed stems from locality of fusion transformation of the (degenerate) Virasoro block.

It is also the case in BTZ that the expression for $\Gret$ in~\eqref{eq-BTZ-answer} only holds for $\mexp_\bdy \notin \intz/2$ or $\Delta \notin \intz$\footnote{In this case, it is straightforward to obtain the result of $\Gret(\omega)$ for integer $\Delta$ by directly taking the limit of the correlator for non-integer conformal dimension.}, while the quantization condition for QMNs holds regardless of $\Delta$ being integer or not. This is analogous to~\cref{claim:QNM-log-quant}.

\section{Relation with WKB period and Seiberg-Witten curve}\label{sec:SW}

It is well-known that there is a WKB regime of QNMs, either at large overtone number~\cite{Motl:2003cd,Natario:2004jd}, or when the mass of the field (or momentum) gets large~\cite{Festuccia:2008zx}. A natural question is therefore how to recover this WKB regime from the exact quantization condition in~\eqref{eq-QNM-quantization}. 

There is a useful way to realize this answer, but this involves using the gauge theory side of AGT relation~\cite{Alday:2009aq}. Recall that this correspondence links Liouville conformal blocks to the supersymmetric (Nekrasov) partition function of four-dimensional $\mathcal{N} =2$ superconformal field theories. For the four punctured case of interest, the WKB limit corresponds to the Seiberg-Witten (SW) limit of quasi-classical Virasoro block, where it reduces to the SW prepotential of $\mathcal{N}=2, N_f = 4$ gauge theory. By the AGT relation, the internal Liouville momentum $\sigma$ is identified with the expectation value of the scalar $a$ in the $\mathcal{N} =2$ multiplet, and therefore given by SW period in the SW limit. The SW period can furthermore be identified as WKB period from the known relation between $\mathcal{N}=2$ class $\mathcal{S}$ theories and WKB analysis \cite{Gaiotto:2009hg,Gaiotto:2012rg,Hollands:2021itj}.

\subsection{SW limit of semiclassical Virasoro block}

To begin with, note that the semiclassical Virasoro block corresponds to the Nekrasov instanton partition in the Nekrasov-Shatashvili (NS) limit, which involves taking the deformation parameters $\epsilon_1 = \hbar$ and $\epsilon_2 = 0$. To attain the SW limit, one  is instructed to further take the large classical Liouville momenta limit in semiclassical Virasoro block and the associated Heun's equation. This is achieved by
\begin{equation} \label{eq-SW-limit}
\text{SW}: \hbar \to 0, \qquad \mexp_i,\cmexp \to \infty, \qquad 
\hbar\, \mexp_i \to m_i, \qquad \hbar\, \cmexp \to a, \qquad \hbar^2 \,\acc \to u\,.
\end{equation}
Doing so, one obtains a “prepotential” from semiclassical Virasoro block as
\begin{equation}
\hbar^2\, \VBcl\prn{x|\vb*{\mexp},\cmexp} \quad \xrightarrow{\text{\tiny{SW}}} \quad \prepot^{\text{Vir}}\prn{x|\vb*{m},a}.
\end{equation}
As indicated the Liouville parameters have been mapped to the corresponding SW data.  In the aforementioned SW limit, the associated Heun's equation~\eqref{eq-heun-normal-form} reduces to a classical spectral curve, which is nothing but the SW curve. This can be seen from the limiting behavior of $\Tcl(z)$
\begin{equation}
\begin{split}
-\hbar^2 \Tcl(z) &\quad\xrightarrow{\text{\tiny{SW}}}\quad \phi_2(z) \\
&= 
\frac{m^2_0}{z^2} + \frac{m^2_\xr}{(z-\xr)^2} + \frac{m^2_1}{(z-1)^2} + \frac{m^2_\infty - m^2_0 - m^2_\xr - m^2_1 }{z\,(z-1)} -
\frac{u\, (\xr-1)}{z\,(z-1)\,(z-\xr)}\,,
\end{split}
\end{equation}
and the definition of the SW spectral curve
\begin{equation}
        w^2 = \phi_2(z) \,.
\end{equation}
From the perspective of WKB analysis, the SW curve is nothing but the WKB curve for Heun's equation. The branched points in SW curve are the turning points in the WKB analysis. Furthermore, the Zamolodchikov relation~\eqref{eq:Zamrel} reduces to Matone-type relation for prepotential~\cite{Matone:1995rx}
\begin{equation}
    u = x\, \partial_x \prepot^{\text{Vir}}(x)\,.
\end{equation}

There is yet another way to obtain the prepotential from the SW curve. One starts with the SW differential,
\begin{equation}
    \swd = w \,dz = \sqrt{\phi_2(z)}\, dz \,.
\end{equation}
Recall that the SW equations can be expressed as the periods of $\lambda$ around the $A$ and $B$ cycles of the torus. Explicitly, we have 
\begin{equation}
    a = \frac{1}{2\pi i} \oint_A \swd, \qquad \partial_a \prepot^{\text{SW}} = \frac{1}{2\pi i} \oint_B \swd.
\end{equation}
In principle, from calculating the $A$-period integral one can invert $a(u)$ as $u(a)$ and then plug it into the $B$-period integral to solve $\prepot^{\text{SW}}$. The SW periods are then identified as WKB periods.

The two apparently different definitions of prepotential mentioned above are, in fact, equivalent: 
\begin{empheq}[box=\fbox]{equation}
\label{eq-prepot-Vir-equals-SW}
    \prepot^{\text{Vir}}\prn{x|\vb*{m},a} = \prepot^{\text{SW}}\prn{x|\vb*{m},a}
\end{empheq}
This is because: i) the SW prepotential can be extracted from Nekrasov instanton partition function in the $\epsilon_{1,2} \to 0$ limit~\cite{Nekrasov:2002qd}, and ii) from AGT relation, the $\epsilon_{1,2} \to 0$ limit of Nekrasov instanton partition function corresponds to taking the semiclassical limit~\eqref{eq-quasiclassical-limit} followed by the SW limit~\eqref{eq-SW-limit} of the Virasoro block. This relation is important for our purposes. It immediately allows one to identify, in the WKB regime of QNMs, the internal Liouville momentum $\cmexp$ appearing in the exact quantization condition~\eqref{eq-QNM-quantization} as a WKB period.

\paragraph{Computing the prepotential from SW curve through a recursion relation:}
It is instructive to verify~\eqref{eq-prepot-Vir-equals-SW} explicitly by directly computing $\prepot^\SW$  and comparing with $\prepot^\Vir$. Instead of directly computing $A$ and $B$ periods as described above, there is a more efficient way of computing SW prepotential. The trick is to expand the SW differential into a rational differential and use Matone-type relation. This not only avoids the need to compute $B$-period, but also reduces the calculation of $A$-period to evaluation of residue. It eventually yields a recursion relation for the expansion coefficients of prepotential. This is explained, e.g., in Appendix A of~\cite{Tachikawa:2013kta} for the $N_f=0$, pure glue SU(2) theory.

We are interested in the somewhat more involved $N_f=4$ case. To attain the result, we perform the following formal expansion in the quadratic differential $\phi_2(z) \,dz^2$: 
\begin{equation}
\begin{split}
\phi_2(z) 
&= 
    \phi_2(z) \bigg|_{x=0} + \frac{1}{z^2} \,
    \sum^\infty_{n=1} \brk{-u + (n+1)\, m^2_x} \prn{\frac{x}{z}}^n \\
&= 
    \frac{-u + m^2_0 + m^2_x}{z^2} + \frac{1}{z^2} \,
    \sum^\infty_{n=1} \, \prn{-u + m^2_0 + n \, m^2_1 + m^2_x - m^2_\infty} \,z^n \\
&\qquad \qquad 
    + \frac{1}{z^2}\, \sum^\infty_{n=1} \brk{-u + (n+1)\, m^2_x} \prn{\frac{x}{z}}^n \\
&= 
    \frac{-u + m^2_0 + m^2_x}{z^2} \prn{1 + \sum^\infty_{n=1} \alpha_n(u,\vb*{m}) \,z^n + \beta_n(u,\vb*{m}) \,\prn{\frac{x}{z}}^n} , 
\end{split}    
\end{equation}
with $\vb*{m}$ collectively denoting the set of exponents $m_i$. The coefficients appearing in the above expansion are 
\begin{equation}
    \alpha_n(u,\vb*{m}) = \frac{-u + m^2_0 + n m^2_1 + m^2_x - m^2_\infty}{-u + m^2_0 + m^2_x}, \qquad 
    \beta_n(u,\vb*{m}) = \frac{-u + (n+1)m^2_x}{-u + m^2_0 + m^2_x}\,.
\end{equation}
Using this series expression we find that the SW differential itself admits the following expansion:
\begin{align}
\swd = \frac{\sqrt{-u + m^2_0 + m^2_x}}{z} \, 
    \sum^\infty_{k=0} \, \binom{\half}{k} \, \brk{ \sum^\infty_{n=1} \alpha_n(u,\vb*{m}) \,z^n + \beta_n(u,\vb*{m}) \prn{\frac{x}{z}}^n }^k dz \,.
\end{align}
Therefore, the A-period is given by the residue at $z=0$, i.e., 
\begin{align}
\label{eq-A-period}
a = \frac{1}{2\pi i} \oint_A \swd 
   \ = \ \sqrt{-u + m^2_0 + m^2_x} \,\sum^\infty_{k=0} \,\binom{\half}{k} \,\gamma_k\prn{x|\vb*{m},u} \,, 
\end{align}
with 
\begin{align}
\gamma_k\prn{x|\vb*{m},u} = \brk{ \sum^\infty_{n=1} 
\alpha_n(u,\vb*{m})\, z^n + \beta_n(u,\vb*{m}) \prn{\frac{x}{z}}^n }^k \Bigg|_{\order{z^0}}.
\end{align}
In order to find $a$ to order $x^{n_{\text{max}}}$ the sum over $k$ can be truncated at $k= 2 n_{\text{max}}$. One thus obtains a recursion relation for the expansion coefficients in the prepotential, viz.,
\begin{equation}
    \prepot^{\SW}\prn{x|\vb*{m},a} = (-a^2 + m^2_0 + m^2_x) \,\log{x} + \sum^{\infty}_{k=1}\, \prepot^{\SW}_k\prn{\vb*{m},a}\, x^k\,,
\end{equation}
This is achieved by substituting the Matone-type relation
\begin{equation}
 u = x \partial_x\prepot^{\SW}\prn{x|\vb*{m},a} = 
 -a^2 + m^2_0 + m^2_x + \sum^{\infty}_{k=1}\, k\; \prepot^{\SW}_k\prn{\vb*{m},a}\, x^k\,,
\end{equation}
into~\eqref{eq-A-period} and solving order by order in $x$. 

Carrying out this exercise, we find the first two coefficients to be given by
\begin{equation}
    \prepot^{\SW}_1\prn{\vb*{m},a} = -\frac{\left(a^2+m_x^2-m_0^2\right) \left(a^2-m_{\infty }^2 + m_1^2\right)}{2 \,a^2}
\end{equation}
and
{\small{
\begin{equation}
\begin{split}
&\prepot^{\SW}_2\prn{\vb*{m},a} 
= 
 \frac{1}{64\, a^6}\bigg[12 \,a^2 \left(a^2 + 2\, m_x^2 - m_0^2\right) \left(a^2 - m_{\infty}^2 + m_1^2\right)^2 
    + 12\, a^2\,   \left(a^2 + m_x^2 - m_0^2\right)^2 \left(a^2 - m_{\infty}^2 + 2\, m_1^2\right) \\
&\qquad 
    - 21 \left(a^2 + m_x^2 - m_0^2\right)^2   \left(a^2-m_{\infty}^2+m_1^2\right)^2       \\
&\qquad 
    - 8 \left(a^2 + m_x^2 - m_0^2\right) \left(a^2 - m_{\infty}^2 + m_1^2\right) \left(a^2 \left(a^2 + m_x^2 - m_0^2\right) 
    -\left(a^2 + 2 \,m_x^2- 2\,m_0^2\right) \left(a^2 - m_{\infty}^2 + m_1^2\right)\right) \\
&\qquad    
    - 16\, a^4 \left(a^2 + 2\,
   m_x^2 - m_0^2\right) \left(a^2 - m_{\infty }^2 + 2\, m_1^2\right)\bigg]  
\end{split}   
\end{equation}
}}
These can indeed be checked to agree with $\prepot^{\Vir}_k\prn{\vb*{m},a}$ from taking SW limit~\eqref{eq-SW-limit} of the expansion coefficients of quasi-classical Virasoro block.

\subsection{The WKB regime of QNMs and  the SW prepotential}

We can now apply the SW prepotential to ascertain the asymptotic form of quasinormal modes. While we would like to do so by directly using the SW prepotential, for the present, we will instead confirm by a backward check that one indeed reproduces the WKS results. We comment at the end on prospects of directly obtaining the WKB asymptotics. 

\paragraph{The asymptotic QNMs for massless scalar fields:} By WKB analysis or direct numerical computation, one can show that the QNMs for both Klein-Gordon and designer scalars around a planar black hole background have the following asymptotic behavior at large overtone number~\cite{Natario:2004jd,Festuccia:2008zx}:
\begin{equation}\label{eq:asymKGn}
    \freq_n = (\pm 1 -i) \,n + \order{n^0}, \qquad n \to \infty\,.
\end{equation}
Let us try to understand it from the exact quantization condition~\eqref{eq-QNM-quantization}. The leading term should be the solution of the following large $\freq$ limit of~\eqref{eq-QNM-quantization}:
\begin{equation}
     -\frac{i\, \freq}{2} \pm \cmexp(\freq) = -n, \quad n,\freq \to \infty
\end{equation}
The asymptotic behavior~\eqref{eq:asymKGn} is obtained if $\cmexp(\freq) = \frac{\freq}{2}$.

In the SW limit of semiclassical Virasoro block, the leading large $\freq$ limit corresponds to the SW curve with two massive punctures $\prn{m_0, m_\xr, m_1, m_\infty} = \prn{i \,\mu, 0, \mu, 0}$ and modulus $u = -4 \,\mu^2$ at coupling $x = \half$. Here $\mu = \frac{\hbar \,\freq}{2}$. To obtain~\eqref{eq:asymKGn}, we therefore need to verify
\begin{equation}
    a \stackrel{?}{=} \mu
\end{equation}
in this set-up. We can check this by verifying the Matone-type relation
\begin{equation}
    u = x \partial_x \prepot\prnbig{x\Big|\{i\,\mu,0,\mu,0\},\mu} \bigg|_{x=\half} \stackrel{?}{=} -4\,\mu^2\,.
\end{equation}

Computing expansion coefficients of prepotential $\prepot_k\prn{\vb*{m},a}$ using either semiclassical Virasoro block or SW curve (up to $k=10$), we observe the following pattern:
\begin{equation}
    \prepot_k\prnbig{\{i\mu,0,\mu,0\},\mu} = \frac{-2\,\mu^2}{k}.
\end{equation}
We thus find
\begin{equation}
    u = x \partial_x \prepot\prnbig{x\Big|\{i\,\mu,0,\mu,0\},\mu} \bigg|_{x=\half} = -2\,\mu^2 (1 +x +x^2 \cdots) \bigg|_{x=\half} = -4\,\mu^2\,.
\end{equation}
This confirms that the asymptotic QNM spectrum is indeed reproduced by the exact quantization condition~\eqref{eq-QNM-quantization} using SW limit of semiclassical Virasoro block.

\paragraph{QNMs at large $\Delta$:} The second limit where one expects a WKB formula to hold is for large dimension operators in the dual CFT. Consider a massive Klein-Gordon scalar in the planar black hole background at finite $\mom$. The QNMs at large $\Delta$ have the following asymptotic behavior~\cite{Festuccia:2008zx}: 
\begin{equation}
    \freq_n = \prn{\pm 1 -i} \prn{\frac{\Delta}{2} + n} + \order{\Delta^0}, \quad \Delta \to \infty\,.
\end{equation}
In terms of the quantization condition~\eqref{eq-QNM-quantization}, the leading term should be the solution of the following large $\freq, \Delta$ limit of 
\begin{equation}
     -\frac{i\, \freq}{2} + \frac{\Delta}{2} \pm \cmexp(\freq, \Delta) = -n, \qquad \freq, \Delta \to \infty\,.
\end{equation}
Once again $\cmexp(\freq, \Delta) = \frac{\freq}{2}$ will reproduce the leading large $\Delta$ behavior.

The corresponding SW curve has three massive punctures $\prn{m_0, m_\xr, m_1, m_\infty} = \prn{i \,\mu, M, \mu, 0}$, with $\mu = \frac{\hbar\, \freq}{2}$ and  $M = \frac{\hbar \Delta}{2}$. The moduli and coupling are still $u = -4 \,\mu^2$ and $x= \half$. We therefore need to verify
\begin{equation}
    a \stackrel{?}{=} \mu\,,
\end{equation}
in this set-up. We can again check this by verifying the Matone-type relation
\begin{equation}
    u = x \partial_x \prepot\prnbig{x\Big|\{i\,\mu,M,\mu,0\},\mu} \bigg|_{x=\half} \stackrel{?}{=} -4\,\mu^2.
\end{equation}
Direct computation of prepotential coefficients (up to $k=8$) now reveals the following pattern
\begin{equation}
     \prepot_k\prnbig{\{im,M,m,0\},m} = \frac{-2\,\mu^2 - M^2}{k}\,.
\end{equation}
Using this we once again confirm the WKB regime of QNM from SW limit of semiclassical Virasoro block, for 
\begin{equation}
\begin{split}    
u &= 
    x\, \partial_x \prepot\prnbig{x\Big|\{i\,\mu,M,\mu,0\},\mu} \bigg|_{x=\half}  \\
    &= -2\,\mu^2 \,(1 +x +x^2 + \cdots) -M^2 \, (-1+x+x^2 \cdots) \bigg|_{x=\half} = -4\,\mu^2\,,
\end{split}
\end{equation}
indeed holds.

\section{Energy-momentum tensor correlation functions in 4d holographic CFTs}\label{sec:emcorr}

We now turn to the correlation functions of the energy-momentum tensor in a 4d holographic CFT. 
The prototype example to keep in mind is the case of $\mathcal{N} = 4$ SYM. We will work with the CFT on $\mathbb{R}^{3,1}$. To describe the results succinctly, we make the following kinematic choice. We pick the spatial momentum to be oriented along $\vb{k} = k \, \vb{\hat{e}}_z$. This allows us to decompose the stress tensor polarizations into physical components:
\begin{itemize}[wide,left=0pt]
\item  The transverse traceless tensor polarizations (which are two in number) are exemplified by components, such as $T_{xy}(\omega,\vb{k})$ in our chosen kinematics. In the dual gravity theory, they are described by a massless Klein-Gordon scalar, i.e., by~\eqref{eq:Mdesign} with $\markov =3$ on the planar-\SAdS{5} background. We write the asymptotic expansion of this scalar as 
\begin{equation}
\varphi(r,\omega, \vb{k})\bigg|_{\markov =3, m^2 =0}  
\sim \bm{\gamma}(\omega,\vb{k}) + \frac{1}{r^4}\, \expval{\mathcal{O}(\omega,\vb{k})}\,, 
\end{equation}
which defines a boundary operator $\mathcal{O}$ of dimension $\Delta =4$, and its source $\bm{\gamma}$ (which are just the transverse traceless tensor components of the boundary metric). This implies that modulo a normalization factor one can write the result in terms of the correlator  $\mathcal{O} $
\begin{equation}\label{eq:TTret}
\begin{split}
\expval{T_{xy}(-\omega, -\vb{k})\, T_{xy}(\omega,\vb{k})}_{\text{ret}}
&=  i\, \mathfrak{N}\, \expval{\mathcal{O}(-\omega, -\vb{k}) \, \mathcal{O}(\omega, \vb{k})}_{\text{ret}} \,. 
\end{split}
\end{equation}
The normalization factor $\mathfrak{N}$ itself can be deduced by comparing with the asymptotic expansion of the Einstein-Hilbert action as in~\cite{Ghosh:2020lel}. It is given in terms of an effective central charge\footnote{
The overall normalization factor of these correlators, consistent with large $N$ scaling, is specified by $c_{\text{eff}} = \frac{\lads^3}{16\pi\, G_N} $.  For the case of $\mathcal{N} = 4$ SYM with gauge group $\mathrm{SU}(N)$ this normalization is 
$c_{\text{eff}} = \frac{N^2}{8\pi^2}$.}
\begin{equation}
    \mathfrak{N}  =  \pi^4\,T^4\, c_{\text{eff}}\,\,.
\end{equation}

Using the expression for the correlator in the s-channel expansion,~\eqref{eq-Gret-s}, and regulating $\Delta = 4 + \epsilon_c$ to ensure that $\mexp_\bdy \notin \mathbb{Z}/2$, 
we have 
\begin{equation}
\expval{\mathcal{O}(-\omega, -\vb{k}) \, \mathcal{O}(\omega, \vb{k})}_{\text{ret}}
= \mathrm{reg.}\lim_{\epsilon_c \to 0}\; 
       \holonorm\prn{1+\epsilon_c} \dfrac{\Gamma\prn{-2-\epsilon_c}}{\Gamma\prn{2+\epsilon_c}} \, 
       \frac{ \Gamma\prn{\frac{3}{2} +\epsilon_c - \frac{i\,\freq}{2}  \pm \sigma}}{\Gamma\prn{\frac{1}{2} -\epsilon_c -  \frac{i\,\freq}{2} \pm \sigma}} \;
    \exp\brk{-\partial_{\mexp_\bdy} \VBcl_{\mathcal{O}}^s(x)}\,.
\end{equation}
Here $\mathrm{reg.}$ denotes taking the regular part of the singular limit.
The semiclassical block that we need in the above expression 
\begin{equation}
    \VBcl_{\mathcal{O}}^s(x) = \begin{tikzpicture}[scale= .5, baseline = .5ex]
        \coordinate (inf) at (-2,-2);
        \coordinate (one) at (-2,2);
        \coordinate (inf-one) at (0,0);
        \coordinate (xr-zero) at (4,0);
        \coordinate (midint) at (2,0);
        \coordinate (xr) at (6,2);
        \coordinate (zero) at (6,-2);
        \draw (inf) -- (inf-one);
        \draw  (one) -- (inf-one);
        \draw (inf-one) -- (xr-zero);
        \draw (xr-zero) -- (xr);
        \draw (xr-zero) -- (zero);
        \node [below left] at (inf) {\scriptsize $\mexp_\curv(\infty) =0$};
        \node [above left] at (one) {\scriptsize $\mexp_{\rcomplex}(1) = \frac{\freq}{2}$};
        \node [below] at (midint) {\scriptsize $\cmexp$};
        \node [above right] at (xr) {\scriptsize $\mexp_\bdy(\xr) =1+\epsilon_c$};
        \node [below right] at (zero) {\scriptsize $\mexp_\hor(0) =  \frac{i\,\freq}{2}$};
    \end{tikzpicture}
\end{equation}
The intermediate Liouville momentum $\cmexp$ is fixed using the block 
\begin{equation}
    \mom^2 - \freq^2 = x\, \partial_x \,\VBcl_{\mathcal{O}}^s(x) \,, 
\end{equation}
and we set $x=\frac{1}{2}$ at the end of the day. For concreteness, we provide the first two terms in the cross-ratio expansion of $\cmexp = \cmexpope + \sum^{\infty}_{k=1} \cmexp_k x^k$ to $\order{\epsilon^2_c}$:
\begin{equation}
\begin{split}
\cmexpope 
&= 
    \frac{1}{2} \,\sqrt{-4\, \mathfrak{q}^2+3\, \mathfrak{w}^2+3} + \frac{2}{\sqrt{-4\, \mathfrak{q}^2+3 \,\mathfrak{w}^2+3}} \,\epsilon_c + \order{\epsilon^2_c} \\
\cmexp_1 
&= 
    -\frac{\left(-4\, \mathfrak{q}^2 + 2\, \mathfrak{w}^2 + 2\right) \left(-4\, \mathfrak{q}^2 + 4\, \mathfrak{w}^2 + 6\right)}{8 \left(-4 \,\mathfrak{q}^2 + 3 \,\mathfrak{w}^2 + 2\right) \sqrt{-4 \,\mathfrak{q}^2 + 3 \,\mathfrak{w}^2 + 3}} \\
&
     -\frac{4 \left(9\, \mathfrak{w}^6 + 2 \left(11 - 18 \,\mathfrak{q}^2\right) \mathfrak{w}^4 + \left(50\, \mathfrak{q}^4 - 55\, \mathfrak{q}^2 + 16\right) \mathfrak{w}^2 - 3 \,\left(2 \mathfrak{q}^2 - 1\right)^3\right)}{\left(-4 \,\mathfrak{q}^2 + 3\, \mathfrak{w}^2 + 2\right)^2 \left(-4 \,\mathfrak{q}^2 + 3 \,\mathfrak{w}^2 + 3\right)^{3/2}} \,\epsilon_c + \order{\epsilon^2_c} \,.
    \end{split}
\end{equation}

\item The transverse vector polarizations, also comprise two components, the momentum density $T_{zx}(\omega,\vb{k})$ and the momentum current $T_{vx}(\omega, \vb{k})$. In the gravitational description they are described by a massless designer scalar with $\markov =-3$, i.e., again through~\eqref{eq:Mdesign} in the planar-\SAdS{5} background. Such a field has an asymptotic behavior, 
\begin{equation}\label{eq:Pmomfluxasym}
\varphi(r,\omega,\vb{k}) \bigg|_{\markov =-3} 
\sim \ \expval{\mathcal{P}(\omega,\vb{k})} + r^2\, \bm{\alpha}(\omega,\vb{k})\,. 
\end{equation}
Notice that the constant mode of the field in this case defines the dual boundary operator's expectation value. The mode that is growing as $r^2$ picks out the transverse vector component of the boundary metric. 

The correlation functions can thus be read off from those of the designer scalar, which we refer to as the shear mode scalar $\mathcal{P}$ are given by 
\begin{equation}\label{eq:TVret}
\begin{split} 
\expval{T_{vx}(-\omega, -\vb{k})\, T_{vx}(\omega,\vb{k})}_{\text{ret}}
&=  -i\,4\,\mathfrak{N}\,  \mom^2\,  \expval{\mathcal{P}(-\omega, -\vb{k}) \, \mathcal{P}(\omega, \vb{k})}_{\text{ret}} \,, \\ 
\expval{T_{vx}(-\omega, -\vb{k})\, T_{zx}(\omega,\vb{k})}_{\text{ret}}
&=  i\, 4\,\mathfrak{N}\, \freq\, \mom\, \expval{\mathcal{P}(-\omega, -\vb{k}) \, \mathcal{P}(\omega, \vb{k})}_{\text{ret}}  \,,\\ 
\expval{T_{xy}(-\omega, -\vb{k})\, T_{xy}(\omega,\vb{k})}_{\text{ret}}
&=  -i\,4\, \mathfrak{N}\, \freq^2\,  \expval{\mathcal{P}(-\omega, -\vb{k}) \, \mathcal{P}(\omega, \vb{k})}_{\text{ret}}  \,.
\end{split}
\end{equation}
The correlation function of $\mathcal{P}$ itself is given by (taking $\markov = -3+2\,\epsilon_c$)
\begin{equation}
\expval{\mathcal{P}(-\omega, -\vb{k}) \, \mathcal{P}(\omega, \vb{k})}_{\text{ret}}
= \mathrm{reg.} \lim_{\epsilon_c \to 0}\; 
       \holonorm\prn{\tfrac{1-\epsilon_c}{2}} \dfrac{\Gamma\prn{-1+\epsilon_c}}{\Gamma\prn{1-\epsilon_c}} \, 
       \frac{ \Gamma\prn{ -\epsilon_c - \frac{i\,\freq}{2}  \pm \sigma}}{\Gamma\prn{\frac{1}{2} -\epsilon_c -  \frac{i\,\freq}{2} \pm \sigma}} \;
    \exp\brk{-\partial_{\mexp_\bdy} \VBcl_{\mathcal{P}}^s(x)}\,.
\end{equation}
The semiclassical block that we need in the above expression 
\begin{equation}
    \VBcl_{\mathcal{P}}^s(x) = \begin{tikzpicture}[scale= .5, baseline = .5ex]
        \coordinate (inf) at (-2,-2);
        \coordinate (one) at (-2,2);
        \coordinate (inf-one) at (0,0);
        \coordinate (xr-zero) at (4,0);
        \coordinate (midint) at (2,0);
        \coordinate (xr) at (6,2);
        \coordinate (zero) at (6,-2);
        \draw (inf) -- (inf-one);
        \draw  (one) -- (inf-one);
        \draw (inf-one) -- (xr-zero);
        \draw (xr-zero) -- (xr);
        \draw (xr-zero) -- (zero);
        \node [below left] at (inf) {\scriptsize $\mexp_\curv(\infty) =\frac{3-\epsilon_c}{2}$};
        \node [above left] at (one) {\scriptsize $\mexp_{\rcomplex}(1) = \frac{\freq}{2}$};
        \node [below] at (midint) {\scriptsize $\cmexp$};
        \node [above right] at (xr) {\scriptsize $\mexp_\bdy(\xr) = \frac{1-\epsilon_c}{2} $};
        \node [below right] at (zero) {\scriptsize $\mexp_\hor(0) =  \frac{i\,\freq}{2}$};
    \end{tikzpicture}
\end{equation}
Now we fix $\cmexp$ by solving 
\begin{equation}
    \mom^2 - \freq^2 = x\, \partial_x \,\VBcl_{\mathcal{P}}^s(x) \,, 
\end{equation}
and set $x=\frac{1}{2}$ at the end. The first two terms in the expansion $\cmexp = \cmexpope + \sum^{\infty}_{k=1} \cmexp_k x^k$ are 
\begin{equation}
\begin{split}
\cmexpope 
&= 
    \frac{1}{2} \,\sqrt{-4 \,\mathfrak{q}^2+3\, \mathfrak{w}^2} - \frac{2}{\sqrt{-4\, \mathfrak{q}^2+3\, \mathfrak{w}^2}} \,\epsilon_c + \order{\epsilon^2_c} \\
\cmexp_1 
&= 
    \frac{(\mathfrak{q} - \mathfrak{w}) (\mathfrak{q} + \mathfrak{w}) \left(2\, \mathfrak{q}^2 - \mathfrak{w}^2 + 5\right)}{
    \left(4 \,\mathfrak{q}^2 - 3\, \mathfrak{w}^2 + 1\right) \sqrt{3\, \mathfrak{w}^2 
    - 4\, \mathfrak{q}^2}} \\
&
    -\frac{
    18 \,\mathfrak{w}^6 
    - \left(63 \,\mathfrak{q}^2 + 4\right) \mathfrak{w}^4 
    + 2 \left(35\, \mathfrak{q}^4 - 4 \,\mathfrak{q}^2 - 5\right) \mathfrak{w}^2 
    + 3 \,\mathfrak{q}^2 \left(-8 \,\mathfrak{q}^4 + 6\, \mathfrak{q}^2 + 5\right)
    }{ 
    \left(4\, \mathfrak{q}^2 - 3\, \mathfrak{w}^2 + 1\right)^2 
    \left(3\, \mathfrak{w}^2 - 4\, \mathfrak{q}^2\right)^{3/2}} \epsilon_c + \order{\epsilon^2_c}\,.
    \end{split}
\end{equation}

A few comments are in order. Firstly, the field $\varphi$ with $\markov =-3$ is quantized in the bulk with Neumann boundary conditions. This is not only natural given the asymptotic fall-offs~\eqref{eq:Pmomfluxasym}, but it also follows the bulk Einstein-Hilbert dynamics. In particular, the parametrization of metric fluctuations in terms of $\varphi_{\markov =-3}$ involves radial derivatives. These are responsible for converting the Dirichlet boundary condition imposed on the bulk metric into Neumann boundary conditions for $\varphi_{\markov =-3}$. The prefactors involving momentum and frequency arise from the map between the physical stress tensor components and the designer scalar used to parameterize the dual gravitational fluctuations.  For details regarding these statements, see~\cite{Ghosh:2020lel}.\footnote{The dimensionless frequencies and momenta used there differ from our current conventions by a factor of $2$; specifically, $\{\freq,\,\mom\}_{\text{there}} = 2\, \{\freq,\mom\}_{\text{here}}$.} We should also note that the aforementioned reference phrase the answer not for the generating function of $\mathcal{P}$ correlators, but in terms of its Legendre transform, the effective action parameterized by $\expval{\mathcal{P}}$ (denoted $\breve{\mathcal{P}}$ there). 

\item Finally, the single scalar polarization, which encompasses the energy density $T_{vv}(\omega,\vb{k})$ and other scalar components, $\{T_{zz}(\omega,\vb{k}), T_{vz}(\omega,\vb{k}), T_{xx} (\omega,\vb{k})+T_{yy}(\omega,\vb{k})\}$ can be mapped to another designer scalar $\MZ$ with $\markov = -1$, albeit with a more involved equation~\eqref{eq:Zsound}. This field has asymptotics\footnote{This is true for non-vanishing spatial momentum. At zero spatial momentum $\MZ$ satisfies a massless Klein-Gordon equation.  Relatedly, the apparent singularity in~\eqref{eq:Zsound} is absent. The physical reason for this change in behavior is due to the enhanced gauge symmetry in the dual gravity description at $k =0$, cf.~\cite{He:2022jnc} for a detailed discussion. \label{fn:no0mom} }
\begin{equation}
    \MZ = \expval{\mathcal{Z}} + \bm{\zeta}\, \log r\,. 
\end{equation}
In particular, the holographic extrapolate dictionary relates the expectation value energy density operator to the operator $\breve{\mathcal{O}}_{\MZ}$ dual to the field $\MZ$ as 
\begin{equation}
T^v_v  \sim -\frac{k^2}{3} \,\mathcal{Z} \,.
\end{equation}

We fix the correlation function of the energy density operator and then use flat spacetime Ward identities (energy-momentum tensor conservation) to fix the correlators of the other polarizations following~\cite{Policastro:2002tn}. The result can be succinctly expressed in terms of a (kinematic) tensor, $\mathfrak{G}_{\mu\nu,\rho\sigma}(\omega,\bk)$ which is polynomial in $\omega$ and $k$ 
\begin{equation}\label{eq:TSret}
\begin{split} 
\expval{T_{\mu\nu}(-\omega, -\vb{k})\, T_{\rho\sigma}(\omega,\vb{k})}_{\text{ret}}
&= \frac{16\,\mom^4}{9}\, \mathfrak{G}^{\mu\nu,\rho\sigma}(\omega,\bk) \, \expval{\mathcal{Z}(-\omega, -\vb{k})\, \mathcal{Z}(\omega,\vb{k})}_{\text{ret}} \,.
\end{split}
\end{equation}
Explicit expressions for the kinematic tensor can be found in ~\cite{He:2022jnc} (modulo a convention change of $\{\freq,\,\mom\}_{\text{there}} =2\, \{\freq,\mom\}_{\text{here}}$).

The new ingredient here is that the $\MZ$ equation has an apparent singular point of order $s=3$. We cure it as described in~\cref{subsec:app-sing} by introducing a regulator deforming the indicial exponents at the horizon and curvature singularity~\eqref{eq:Zmodg}. Once this is done, the correlation function of $\mathcal{Z}$ itself is given by (taking $\Theta = \epsilon_c$ for convenience)
\begin{equation}
\expval{\mathcal{Z}(-\omega, -\vb{k}) \, \mathcal{Z}(\omega, \vb{k})}_{\text{ret}}
= \mathrm{reg.} \lim_{\epsilon_c \to 0 }\; 
       \holonorm\prn{\epsilon_c} \dfrac{\Gamma\prn{-2\,\epsilon_c}}{\Gamma\prn{2\,\epsilon_c}} \, 
       \frac{ \Gamma\prn{ \half +\epsilon_c - \frac{i\,\freq}{2}  \pm \sigma}}{\Gamma\prn{\frac{1}{2} -\epsilon_c -  \frac{i\,\freq}{2} \pm \sigma}} \;
    \exp\brk{-\partial_{\mexp_\bdy} \VBcl_{\mathcal{Z}}^s(x,\xr)}\,.
\end{equation}
The semiclassical block that we need in this case is a five-point block
\begin{equation}
    \VBcl_{\mathcal{Z}}(\xr,\xrapp) =
    \begin{tikzpicture}[scale= .5, baseline = .5ex]
        \coordinate (inf) at (-2,-2);
        \coordinate (one) at (-2,2);
        \coordinate (inf-one) at (0,0);
        \coordinate (xr-zero) at (4,0);
        \coordinate (midint) at (2,0);
        \coordinate (xr) at (6,2);
        \coordinate (zero) at (6,-2);
        \draw (inf) -- (inf-one);
        \draw  (one) -- (inf-one);
        \draw (inf-one) -- (xr-zero);
        \draw (xr-zero) -- (xr);
        \draw (xr-zero) -- (zero);
        \node [below left] at (inf) {\scriptsize $\mexp_\curv(\infty) = \mathfrak{h}(\mom)\,\epsilon_c$};
        \node [above left] at (one) {\scriptsize $\mexp_{\rcomplex}(1) = \frac{\freq}{2}$};
        \node [above right] at (xr) {\scriptsize $\mexp_\bdy(\xr) = \epsilon_c$};
        \node [below right] at (zero) {\scriptsize $\mexp_\hor(0) = \frac{i\,\freq}{2}$};
        \node [below right] at ($(midint) + (0.5,0)$) {\scriptsize $\cmexp$};
        \node [below left] at ($(midint) - (0,0)$) {\scriptsize $\cmexp + i_3$};
        \coordinate (heavy-deg) at ($(midint) + (0,2)$);
        \draw[dash dot] (midint) -- (heavy-deg);
        \node [above] at (heavy-deg) {\scriptsize $\mexp_{\drep{1}{3}}(\xrapp)$};
    \end{tikzpicture}
\end{equation}
The internal Liouville momenta $\sigma, i_3$ are determined by solving
\begin{equation}
\mom^2-\freq^2- \frac{3}{2\,\mom^2} = \xr\, \partial_\xr \VBcl(\xr,\xrapp)\,, \qquad 
5+\frac{3}{2\,\mom^2} + \frac{12}{2\,\mom^2 - 3} = \xrapp\, \partial_\xrapp \VBcl(\xr,\xrapp)\,.
\end{equation}
and setting $\xr=\frac{1}{2}$ and $\xrapp = \frac{1}{2} + \frac{\mom^2}{3}$.

To obtain the result for the $\MZ$ correlation function, we quantize the field with  Neumann boundary condition. The reason is similar to that for $\varphi_{\markov=-3}$ mentioned above, but the details, which can be found in~\cite{He:2022jnc}, are more involved. That work, which was interested in the low-energy limit, $\freq,\mom\ll 1$,   obtained results in a gradient expansion. We need not restrict to this low-energy regime, and do stay away from $\mom\neq 0$ to avoid the complications alluded to in~\cref{fn:no0mom}.

\end{itemize}

\section{Discussion}\label{sec:discuss}

The primary focus of the paper was to exploit the connection between wave equations in \AdS{} black hole backgrounds, and the BPZ equation for degenerate Virasoro conformal blocks, to analyze thermal correlation functions of holographic CFTs. These correlators are meromorphic, with the location of the poles being associated with the quasinormal modes (QNMs) of the black hole. The QNMs themselves are solutions to the connection problem of the aforesaid differential equation, and can be directly obtained using an exact quantization condition. 

While these insights have been discussed in the literature, we have argued for a simpler formula for the thermal 2-point function. This was achieved by working with a different $s$-channel expansion of the semiclassical Virasoro blocks. In particular, our final result for the 2-point function bears striking resemblance to the answers obtained in a thermal \CFT{2}. Specifically, 
\begin{equation}\label{eq:structure2pt}
\begin{split}
\Gret(\omega,k)\Bigg|_{\text{CFT}_4} 
&\propto 
    \dfrac{\Gamma\prn{\frac{1-i\,\freq}{2}  + \frac{\Delta-2}{2} + \cmexp}\, \Gamma\prn{\frac{1-i\,\freq}{2}  + \frac{\Delta-2}{2} - \cmexp}}{\Gamma\prn{\frac{1-i\,\freq}{2} - \frac{\Delta-2}{2} + \cmexp}\,\Gamma\prn{\frac{1-i\,\freq}{2}  - \frac{\Delta-2}{2} - \cmexp}}  \,, \\
\Gret(\omega,k)\Bigg|_{\text{CFT}_2} 
&\propto 
    \dfrac{\Gamma\prn{\frac{1-i\,\freq}{2}  + \frac{\Delta-1}{2} + \frac{i\,\mom}{2}}\, \Gamma\prn{\frac{1-i\,\freq}{2}  + \frac{\Delta-1}{2} -  \frac{i\,\mom}{2}}}{\Gamma\prn{\frac{1-i\,\freq}{2} - \frac{\Delta-1}{2} + \frac{i\,\mom}{2}}\,\Gamma\prn{\frac{1-i\,\freq}{2}  - \frac{\Delta-1}{2} - \frac{i\,\mom}{2}}}  \,. 
\end{split}
\end{equation}
Apart from a shift of the conformal dimension, we see that  $\cmexp = \frac{i\,\mom}{2}$ is elementary in \CFT{2} (holographic or otherwise), but is a more involved function of $\cmexp(\freq,\mom,\Delta)$ in higher dimensional holographic CFTs. In the latter case, one needs to determine $\cmexp$ from the semiclassical Virasoro block. This form of the result not only holds for scalar conformal primaries of the holographic CFT, but also for the conserved currents. In order to determine the latter, we have generalized the formalism to include equations with apparent singular points. In the auxiliary 2d CFT these correspond to heavy degenerate operator insertions. 

One useful result we have been able to derive using this formalism is an exact quantization condition for purely decaying QNMs in a near-extremal black hole background. Schematically, this result takes the form
\begin{equation}\label{eq:nearextTqnm}
    \freq_n = -i\, T \left(2\,n +\Delta_\text{near-horizon}\right) \,, \qquad n\in \mathbb{Z}_{\geq 0} \,.
\end{equation}
Here $\Delta_\text{near-horizon}$ is the conformal dimension of the operator in the near-horizon \AdS{2} throat present in the extremal limit. We have derived this expression for a neutral scalar primary of the holographic CFT, but expect it to hold more generally. One reason for this expectation is based on the structure of correlators in the \AdS{2} throat region~\cite{Faulkner:2009wj}.\footnote{As noted in~\cite{Dodelson:2023vrw} one can alternately interpret these modes as relics of bound states in the effective potential that are trapped between the inner and outer horizon. We thank Matthew Dodelson for alerting us to this possibility.} We comment further on this result below.

There are several directions in which our results can be generalized,  some of which we outline below. 

\paragraph{Conserved current correlators:} A natural extension of our analysis would be to carry out an analogous exercise for correlation functions in a finite density system. For instance, by analyzing the fluctuations of gravitons and photons in the Reissner-Nordstr\"om-\AdS{5} background, one can extract the correlation functions of the energy-momentum and charge current in $\mathcal{N}=4$ SYM. This example also has the advantage of teaching us about thermal correlation functions in the presence of 't Hooft anomalies using holography. This problem has recently been analyzed in a series of works~\cite{He:2021jna,He:2022deg} and there are even results for anomaly induced correlation functions~\cite{Rangamani:2023mok}. Another interesting analysis in this context would be to consider charged fermionic operators along the lines of~\cite{Loganayagam:2020iol,Martin:2024mdm}. 

One aspect we have not discussed is the low energy gradient expansion of the correlation functions. One of the motivations for analyzing the equations for us originated from the recent understanding of open effective field theories for quantum systems coupled to conserved currents of a thermal environment. The presence of long-lived hydrodynamic modes in such thermal environments leads to non-Markovian open system dynamics. This class of problems was analyzed for neutral holographic environments in~\cite{Ghosh:2020lel,He:2022jnc} and for charged holographic environments in~\cite{He:2021jna,He:2022deg}. These works derived the 2-point retarded thermal Green's functions in a low frequency and momentum expansion, both for conserved charged currents and the energy-momentum tensor.  Our results are complementary in that we have not attempted to directly analyze this low-energy regime. It would also be interesting to understand the connection to the recent work~\cite{Aminov:2023jve}, who argue for an expansion in terms of multiple polylogarithms, for a similar structure originates in the gradient expansion in the aforementioned works.

\paragraph{Asymptotic QNMs:} The SW limit of the exact quantization condition allows one to obtain the asymptotic QNMs at large overtone number. As discussed in~\cref{sec:SW} it would be desirable to directly deduce the asymptotic behavior from the SW prepotential. In particular, it would be interesting to apply this to determine the asymptotic gap in QNMs (for both purely decaying and the so-called Christmas-tree type with a non-vanishing real part) around charged black hole backgrounds.  Based on numerical results, we expect the asymptotic gap for the purely decaying mode to go from linear to temperature in the near-extremal limit to infinity in the neutral black hole.
 
Another interesting question to analyze is to analytically determine purely decaying QNMs  for all channels of scalar, gauge field, and metric perturbations in near-extremal black hole backgrounds. As we noted above, we expect a result analogous to~\eqref{eq:nearextTqnm}. The calculation ought to be doable as near-extremal limit at small frequency corresponds to certain OPE or degeneration limit of four or five-punctured sphere to three-punctured sphere. The latter corresponds to the hypergeometric oper, which is natural given the emergence of a long \AdS{2} throat and thus an $\mathrm{SL}(2,\mathbb{R})$ isometry in the extremal limit.

\paragraph{One loop determinants around black holes:} The determination of an analytic formula for purely decaying QNMs around near-extremal black holes, has interesting implications for black hole thermodynamics. It is now well understood that the Bekenstein-Hawking result for the near-extremal thermodynamics receives corrections~\cite{Ghosh:2019rcj,Iliesiu:2020qvm,Iliesiu:2022onk} owing to gapless modes localized in the near-horizon region. This result is derived by computing the one-loop partition function around the black hole background, and realizing that these gapless modes lead to a temperature dependent one-loop determinant. The connection with QNMs arises owing to an elegant formula for the black hole determinant over the set of QNMs (and their conjugate anti-QNMs)~\cite{Denef:2009kn}. We believe it should be possible to explicitly deduce this temperature dependent contribution from these purely decaying QNMs, and hope to report on it in the near-future (for another perspective,  see~\cite{Kolanowski:2024ta}). See also~\cite{Arnaudo:2024rhv} for a recent application of the connection coefficients method discussed here to black hole determinant. 

\paragraph{Higher point functions:} Our discussion has primarily focused on thermal 2-point functions, which have the nice feature of being directly related to the connection problem. However, as reviewed  in~\cref{sec:algorithm}  it is more natural to work with the Schwinger-Keldysh formalism for real-time computations. The essential ingredient in that case is the ingoing boundary-to-bulk Green's function $G_{\text{in}}$, in terms of which we can obtain all the correlation functions using bulk Witten diagrams on the grSK geometry. Therefore,  should one be able to compute $G_{\text{in}}$ from the auxiliary 2d CFT description, we would be in a position to compute any desired holographic thermal correlator that can be obtained with the Schwinger-Keldysh time-ordering. This ingoing Green's function is an on-shell solution to the black hole wave equation subject to non-normalizable boundary conditions at the \AdS{} boundary, and is ingoing  at the horizon. In our language, this is given by the wavefunction $\wf_{\ingo}(z)$, which can be obtained from the Virasoro block $ \VB(z,\xr)$. In particular, taking the ratio of the full degenerate Virasoro block and its heavy part in the semiclassical limit gives us $\wf_{\ingo}(z)$, cf.~\eqref{eq-deg-block-CL-s}. While obtaining the degenerate Virasoro block, and its semiclassical limit, is still a challenging proposition, doing so would allow us to extend the formalism to computing higher point thermal correlation functions.   

\paragraph{Generalization to logarithmic CFT:} Technically, the CFT method discussed here only applies to black hole perturbation problems holographically dual to  CFT operators with non-integer conformal dimensions, with the integer cases involving logarithmic solutions. In fact, the integer cases correspond to most of the physically relevant examples, especially when the dual CFT operators are conserved currents. While, as advocated in~\cref{claim:QNM-log-quant,claim:TwoPt-log}, the answers for the integer cases can in principle be extracted from taking a suitable limit from the non-integer case, the procedure is not particularly straightforward for calculating the two-point function, where regularization of a singular limit is required. It is thus desirable to give a direct description of the integer cases without the need for taking such limit. As already mentioned in~\cref{remark:logCFT}, we expect that this can be achieved by generalizing the current CFT method to logarithmic CFT.

\paragraph{The origins of the auxiliary 2d CFT:}
While one might have guessed at a formula as in~\eqref{eq:structure2pt} for holographic CFTs from the knowledge of the analytic structure, it is intriguing that this form of the answer is naturally suggested by viewing the black hole wave equation as the BPZ Ward identity for a higher-point function in an auxiliary 2d CFT. Thus far, the origin of this auxiliary 2d CFT appears somewhat empirical to us. Knowing that the wave equations in black hole backgrounds are Fuchsian, one can identify this auxiliary CFT as a tool to help solve the equation, or at least deduce the physical features of interest such as the QNMs. However, it is interesting to ask why this relation arises in the first place.

Consider computing the thermal correlator of strongly coupled planar CFT, and imagine that we are unaware of a holographic dual. Could one deduce from the structure of the thermal 2-point function that there is an auxiliary 2d CFT lurking in the background? Note that this question is intimately tied with bulk locality in the holographic context, since the BPZ equation of the auxiliary CFT is the radial wave equation. Therefore, being able to show that a relation to the auxiliary CFT exists would introduce naturally the bulk radial coordinate, a feature that is a priori not visible in just the thermal correlator. This is a fascinating question that we think deserves further attention. For instance, one could ask if a similar result would hold away from the classical gravity regime. Would, for example, thermal correlation functions of $\mathcal{N}=4$ SYM at large $N$, but finite 't Hooft coupling continue to be related to a connection problem, which can be interpreted in terms of an auxiliary 2d CFT? 

\section*{Acknowledgements}

It is a pleasure to thank Matthew Dodelson, Alba Grassi, Tom Hartman, R.~Loganayagam,  Julio Virrueta, and Sasha Zhiboedov for insightful discussions. We would also like to thank Matthew Dodelson and Julio Virrueta for comments on a draft of the paper. 
H.~F.~J.~was supported by U.S. Department of Energy grant DE-SC0020360 under the HEP- QIS QuantISED program. M.R.~was supported by U.S.~Department of Energy grant DE-SC0009999 and funds from the University of California. 
H.~F.~J~and M.R.~would like to thank KITP for hospitality during the program, “What is string theory? Weaving perspectives together”, which was supported by the grant NSF PHY-2309135 to the Kavli Institute for Theoretical Physics (KITP). M.R.~would also  like to thank the Aspen Center for Physics, which is supported by National Science Foundation grant PHY-2210452 during the course of this work.

\appendix

\section{Further details on apparent singularities and their CFT description}
\label{app-apparent-singularity}

Consider a second order Fuchsian equation in normal form, with apparent singularities at $\{w_\alpha\}$:
\begin{equation}
\begin{split}    
& \hspace{3cm} \psi^{\prime \prime}(z) + \Tcl(z) \psi(z) = 0 \,,\\
\Tcl(z) & = \sum_i \prn{ \frac{\delta_i}{(z-z_i)^2} + \frac{c_i}{z - z_i}} + \sum_\alpha \prn{ \frac{ \delta_\alpha}{(z-w_\alpha)^2} + \frac{d_\alpha}{z - w_\alpha}}\,, \\
& \hspace{2cm} \delta_i = \frac{1}{4} - \theta^2_i, \quad \delta_\alpha = \frac{1 - \asnum^2_\alpha}{4}, \quad \asnum_\alpha \in \intz \,.
\end{split}
\end{equation}
The exponents are $\half \pm \theta_i$ at $\{z_i\}$ , and $\half \pm \frac{\asnum_\alpha}{2}$ at $\{w_\alpha\}$. For $\{w_\alpha\}$ to be apparent singularities without logarithmic branch, additional conditions on $\Tcl(z)$ are needed. 

\paragraph{A useful property:} A basis of solution $\psi_\pm(z)$ of the Fuchsian equation satisfies,
\begin{equation}
\label{eq-sol-ratio-sch-der}
    \cbrk{\frac{\psi_+}{\psi_-} , z}  = 2 \,\Tcl(z)
\end{equation}
where $\cbrk{\cdot,z}$ denote Schwarzian derivative w.r.t.~$z$.

\subsection{The apparent singularity condition}

At each $w_\alpha$, define the expansion
\begin{equation}\label{eq-l-coeff-def}
\Tcl(z) = 
    \sum_{m=0} \lop^{(w_\alpha)}_{-m} \,(z-w_\alpha)^{m-2}\,, 
    \qquad
    \lop^{(w_\alpha)}_{-m} = \Res_{z = w_\alpha} \brk{(z-w_\alpha)^{1-m} \,\Tcl(z)}\,.
\end{equation}
The coefficients themselves are
\begin{equation}\label{eq-l-coeff}
\begin{split}
\lop^{(w_\alpha)}_0 
&= 
    \delta_\alpha\,, \qquad 
\lop^{(w_\alpha)}_{-1} = d_\alpha \,, \\
\lop^{(w_\alpha)}_{-m} 
&= 
    \frac{1}{(m-2)!} \, \partial^{m-2}_z \Tcl^{(\Bar{\alpha})}(z)\bigg|_{z=w_\alpha}, \qquad m \geq 2 \\
&= 
    \sum_{i} \frac{(m-1) \,\delta_i}{(z_i - w_\alpha)^m} - \frac{c_i}{(z_i - w_\alpha)^{m-1}} + \sum_{\beta \neq \alpha} \frac{(m-1) \,\delta_\beta}{(w_\beta - w_\alpha)^m} - \frac{d_\beta}{(w_\beta - w_\alpha)^{m-1}}\,, 
\end{split}
\end{equation}
where $\Tcl^{(\Bar{\alpha})}(z)$ denotes $\Tcl(z)$ with singular terms at $w_\alpha$ excluded.  

The apparent singularity condition with exponent difference $n$ is an algebraic equation between $\lop_{-1}, \cdots, \lop_{-n}$. The condition can be extracted from the Schwarzian derivative of $\frac{\psi_+}{\psi_-} = (z-w)^n \,F(z)$.

\begin{example}[n=3] Using~\eqref{eq-sol-ratio-sch-der} and~\eqref{eq-l-coeff-def}, we compute $\lop_{-1}$, $\lop_{-2}$, $\lop_{-3}$ from Schwarzian derivative of $\frac{\psi_+}{\psi_-} = (z-w)^3\, F(z)$:
\begin{equation}
\begin{split}
\lop_{-1} 
&= 
    -\frac{4}{3}\, \frac{F'(w)}{F(w)} \,, \\
\lop_{-2} 
&= 
    \frac{8 \,F'(w)^2 - 15\, F(w)\, F''(w)}{18\, F(w)^2} \\
\lop_{-3} 
&= 
    \frac{32 \, F'(w)^3 - 30\, F(w)\, F'(w)\, F''(w)}{27\, F(w)^3}.
\end{split}
\end{equation}
We note the absence of $F^{(3)}(w)$ in $\lop_{-3}$. This allows one to write $\lop_{-3}$ in terms of $\lop_{-1}$ and $\lop_{-2}$ by first solving $\frac{F'(w)}{F(w)}$ and $\frac{F''(w)}{F(w)}$ using the first two conditions. 
\end{example}

This procedure continues to work for higher $n$ where $F^{(n)}(w)$ is absent from $l_{-n}$. The first few apparent singularity conditions (ASCs) were listed in \cref{tab-apparent-singularity-conditions}.

\subsection{Degenerate Virasoro representations}

To understand what the ASC means, we recall from~\cref{sec:Liouville} that the Liouville theory has a set of degenerate representations $V_{\drep{r}{s}}$ with conformal weights $h_{\drep{r}{s}}$. In the Hilbert space, we have null states $\ket{\chi_{\drep{r}{s}}}$ at level $r s$:
\begin{equation}
    \ket{\chi_{\drep{r}{s}}} = L_{\drep{r}{s}} \ket{h_{\drep{r}{s}}}.
\end{equation}
The null states at the first few levels are collated in~\cref{tab-null-vectors}.

\renewcommand{\arraystretch}{2.2}
\begin{table}[ht]
	\tiny
    \centering
    \begin{tabular}{|c|c|c|}
       \hline
       $rs$  & $\drep{r}{s}$ & $L_{\drep{r}{s}}$ \\
       \hline\hline
       
       1 & $\drep{1}{1}$ & $L_{-1}$ \\
       \hline
       
       \multirow{2}{*}{2} 
        & $\drep{1}{2}$ & $b^{2}\, L^2_{-1} + L_{-2}$ \\ 
        \cline{2-3}
       & $\drep{2}{1}$ & $b^{-2} \, L^2_{-1} + L_{-2}$ \\     
       \hline

       \multirow{2}{*}{3} & $\drep{1}{3}$ & 
       $\dfrac{b^4}{2\,(2 - b^2)} \,L^3_{-1} + \dfrac{2\, b^2}{2 - b^2}\, L_{-1} \, L_{-2} + L_{-3}$ \\ \cline{2-3}
       & $\drep{3}{1}$ & $\dfrac{1}{2\,b^2 (2\,b^2-1)} \, L^3_{-1} + \dfrac{2}{2\,b^2-1} \,L_{-1}\, L_{-2} + L_{-3}$ \\ 
       \hline

       \multirow{3}{*}{4} & $\drep{1}{4}$ & 
       $\dfrac{b^6}{6\,(6-4\,b^2 + b^4)} \,L^4_{-1} 
       + \dfrac{5 \,b^4}{3\, (6- 4\,b^2 + b^4)} \,L^2_{-1}\, L_{-2} 
       + \dfrac{3\,b^2}{2\, (6-4\,b^2 + b^4)}\, L^2_{-2} 
       + \dfrac{b^2\, (12-5\,b^2)}{3\, (6-4\,b^2+b^4)} \,L_{-1}\,L_{-3} + L_{-4}$ \\ \cline{2-3}
       & $\drep{4}{1}$ & 
       $\dfrac{1}{6 \,(b^2 - 4\,b^4 + 6\,b^6)}\, L^4_{-1} 
       + \dfrac{5}{3(1-4\, b^2+ 6\, b^4)}\, L^2_{-1} \,L_{-2} 
       + \dfrac{3\,b^2}{2-8\,b^2+12\,b^4}\, L^2_{-2} 
       + \dfrac{(12\,b^2 - 5)}{3\, (1 - 4\,b^2 + 6\, b^4)}\,L_{-1}\, L_{-3} 
       + L_{-4}$ \\ \cline{2-3}
       & $\drep{2}{2}$ & $\dfrac{b^2}{3\,(-1+b^2)^2}\, L^4_{-1} 
       + \dfrac{2\,(1+b^4)}{3\,(-1+b^2)^2}\, L^2_{-1}\, L_{-2} 
       + \dfrac{(1+b^2)^2}{3\,b^2}\, L^2_{-2} 
       -\dfrac{2\,(1-3\,b^2+b^4)}{3\,(-1+b^2)^2}\,L_{-1}\, L_{-3} + L_{-4} $\\
       \hline    
    \end{tabular}
    \caption{Null vectors at first few levels.}
    \label{tab-null-vectors}
\end{table}
\renewcommand{\arraystretch}{1.6}

\subsection{Apparent singularities from degenerate representations}

Consider the following degenerate Virasoro block in any channel
\begin{equation}
    \pf\prn{z,\vb{z},\vb{w}} = \expval{V_{\drep{2}{1}}(z) \,\prod_i V_{P_i}(z_i) \,\prod_\alpha V_{\drep{r_\alpha}{s_\alpha}}(w_\alpha) }, \quad s_\alpha \neq 1.
\end{equation}
The operator $V_{\drep{2}{1}}(z)$ is singled out, while the other degenerate operators $V_{\drep{r_\alpha}{s_\alpha}}(w_\alpha) $ are 
inserted at locations $w_\alpha$, which we will identify with the location of apparent singularties.  

The degenerate block satisfies BPZ null vector decoupling equations for each of the degenerate representations involved, viz., 
\begin{equation}
\begin{split}
     \ldiff^{(z)}_{\drep{2}{1}} \pf &= 0 \\
    \ldiff^{(w_\alpha)}_{\drep{r_\alpha}{s_\alpha}} \pf &= 0 \quad \forall \alpha.   
\end{split}
\end{equation}
The differential operator $\ldiff^{(\zeta)}_{\drep{r}{s}}$ is induced from $L_{\drep{r}{s}}$, with action of each Virasoro generator given by
\begin{equation}
    \ldiff^{(\zeta)}_{-n} = \sum_I \frac{(n-1)\,h_I}{(\zeta_I - \zeta)^n} - \frac{1}{(\zeta_I - \zeta)^{n-1}}\, \partial_{\zeta_I}
\end{equation}
where the sum is over all other operators in the correlation function. In particular,
\begin{equation}
    \ldiff^{(\zeta)}_{-1} = -\sum_I \partial_{\zeta_I} = \partial_\zeta.
\end{equation}
We now consider the semiclassical limit~\eqref{eq-quasiclassical-limit} of this degenerate block, which we record here for convenience
\begin{equation}
    b \to 0, \quad P_i \to \infty, \quad b P_i \to \theta_i \ (b^2 h_i \to \delta_i)
\end{equation}
In this limit, $V_{\drep{2}{1}}$ is light with $\order{b^0}$ weight, while other operators have $\order{b^{-2}}$ weights, including the other degenerate operators with $s_\alpha \neq 1$. This motivates the heavy-light factorization ansatz:
\begin{equation}
    \pf_\cl(z,\vb{z},\vb{w}) = \psi(z|\vb{z},\vb{w}) \; e^{b^{-2} \,\cb\prn{\vb{z},\vb{w}}}.
\end{equation}

The BPZ equation $\ldiff^{(z)}_{\drep{2}{1}} \pf_\cl = 0$ leads to
\begin{equation}\label{eq:mutiBPZ}
\begin{split}
&\hspace{1.5cm} 
    b^{-2} \brkbig{\psi^{\prime \prime}(z) + \Tcl(z) \psi(z)} + \order{b^0} = 0 \,,\\
\Tcl(z) &= 
    \sum_i \prn{ \frac{\delta_i}{(z-z_i)^2} + \frac{\partial_{z_i} \cb}{z - z_i}} + \sum_\alpha \prn{ \frac{ \delta_\alpha}{(z-w_\alpha)^2} + \frac{\partial_{w_\alpha} \cb}{z - w_\alpha}} \,,\\
&\hspace{2cm} 
    \delta_i = \frac{1}{4} - \theta^2_i, \quad \delta_\alpha = \frac{1 - s^2_\alpha}{4}\,.
\end{split}
\end{equation}
Note that one has the following identification between accessory parameters and derivatives of classical conformal block:
\begin{equation}\label{eq-accessory-param-cb-relation}
c_i = \partial_{z_i} \cb\,, 
\qquad
 d_\alpha = \partial_{w_\alpha} \cb.
\end{equation}

For the other BPZ equations $\ldiff^{(w_\alpha)}_{\drep{r_\alpha}{s_\alpha}} \pf_\cl = 0$, consider the action of each Virasoro generato r at level $m$:
\begin{equation}
\begin{split}
\ldiff^{(w_\alpha)}_{-m} \pf_\cl 
&= 
    b^{-2} \brk{\sum_i \frac{(m-1)\, \delta_i}{(z_i - w_\alpha)^m} 
    - \frac{\partial_{z_i} \cb}{(z_i - w_\alpha)^{m-1}} 
    + \sum_{\beta \neq \alpha} \frac{(m-1)\delta_\beta}{(w_\beta -w_\alpha)^m} - \frac{\partial_{w_\beta} \cb}{(w_\beta - w_\alpha)^{m-1}}} \pf_\cl  \\ 
& \qquad \qquad 
    + \order{b^0}   
\end{split}
\end{equation}
Recalling relations~\eqref{eq-l-coeff} and~\eqref{eq-accessory-param-cb-relation}, we therefore have
\begin{empheq}[box=\fbox]{equation}
    \ldiff^{(w_\alpha)}_{-m} \pf_\cl \ = \ 
    b^{-2}\, \lop^{(w_\alpha)}_{-m} \pf_\cl 
    + \order{b^0}, \qquad m \geq 1 \,.
\end{empheq}
In other words, in the classical limit $\pf_\cl$ acts as a common eigenfunction for all (raising) Virasoro generators. Each BPZ equation $\ldiff^{(w_\alpha)}_{\drep{r_\alpha}{s_\alpha}} \pf_\cl = 0$ then gives the following constraint on $\lop^{(w_\alpha)}_{-m}$:
\begin{empheq}[box=\fbox]{equation}
\lim_{b \to 0} L_{\drep{r}{s}} \bigg|_{ {L_{-m} \,\to\, b^{-2} \,\lop_{-m}}} = 0, \qquad s \neq 1\,.
\end{empheq}

So we have two constraints for $\pf_\cl$, one which is a second order equation~\eqref{eq:mutiBPZ}, which is of the form we seek to solve for holographic correlators, and another which constraints the 
accessory parameters
\begin{equation}
\sum_i \frac{(m-1)\, \delta_i}{(z_i - w_\alpha)^m} 
    - \frac{c_i}{(z_i - w_\alpha)^{m-1}} 
    + \sum_{\beta \neq \alpha} \frac{(m-1)\,\delta_\beta}{(w_\beta -w_\alpha)^m} - \frac{d_\beta}{(w_\beta - w_\alpha)^{m-1}} =0
\end{equation}

From \cref{tab-apparent-singularity-conditions,tab-null-vectors}, we see that, up to $n=4$, the apparent singularity conditions for exponent difference $n$ are precisely recovered from the classical limits of BPZ equations for degenerate representations $V_{\drep{1}{n}}$.

\printbibliography

\end{document}